\documentclass[pdflscape,lscape,iop,apj,numberedappendix,appendixfloats,stfloats,url,microtype]{emulateapj}

\pdfoutput=1
\usepackage{paralist,amsmath}
\usepackage{upgreek}
\usepackage{apjfonts}
\usepackage{textcomp}
\makeatletter
  \newcommand\tinyv{\@setfontsize\tinyv{7.0pt}{7.0}}
\makeatother
\slugcomment{Submitted to The Astrophysical Journal.}
\shorttitle{Faraday Rotation from Magnesium~II Absorbers towards Polarized Background Radio Sources}
\shortauthors{Farnes et al.}

\begin{document}

\title{Faraday Rotation from Magnesium~II Absorbers towards Polarized Background Radio Sources}

\author{J.~S. Farnes\altaffilmark{1,2}, S.~P. O'Sullivan\altaffilmark{1,2}, M.~E. Corrigan\altaffilmark{1}, B.~M. Gaensler\altaffilmark{1,2}}
\altaffiltext{1}{Sydney Institute for Astronomy, School of Physics, The University of Sydney, NSW 2006, Australia.}
\altaffiltext{2}{ARC Centre of Excellence for All-sky Astrophysics (CAASTRO).}
\email{jamie.farnes@sydney.edu.au}

%==============================================================================%
\begin{abstract}
Strong singly-ionized magnesium (MgII) absorption lines in quasar spectra typically serve as a proxy for intervening galaxies along the line of sight. Previous studies have found a correlation between the number of these MgII absorbers and the Faraday rotation measure (RM) at $\approx5$~GHz. We cross-match a sample of 35,752 optically-identified non-intrinsic MgII absorption systems with 25,649 polarized background radio sources for which we have measurements of both the spectral index and RM at 1.4~GHz. We use the spectral index to split the resulting sample of 599 sources into flat-spectrum and steep-spectrum subsamples. We find that our flat-spectrum sample shows significant ($\sim3.5\sigma$) evidence for a correlation between MgII absorption and RM at 1.4~GHz, while our steep-spectrum sample shows no such correlation. We argue that such an effect cannot be explained by either luminosity or other observational effects, by evolution in another confounding variable, by wavelength-dependent polarization structure in an active galactic nucleus, by the Galactic foreground, by cosmological expansion, or by partial coverage models. We conclude that our data are most consistent with intervenors directly contributing to the Faraday rotation along the line of sight, and that the intervening systems must therefore have coherent magnetic fields of substantial strength ($\bar{B}=1.8\pm0.4$~$\upmu$G). Nevertheless, the weak nature of the correlation will require future high-resolution and broadband radio observations in order to place it on a much firmer statistical footing.
\end{abstract}


%==============================================================================%
\section{Introduction}
\label{introduction}
Metal enriched gaseous structures, such as normal star-forming galaxies, can lie along the line of sight between us and a quasar \citep[e.g.][]{1999AAS...195.5201C}. These intervening galaxies are believed to give rise to absorption lines in an observed quasar spectrum, with the magnesium~II (MgII) doublet appearing at $\lambda\lambda$2796,~2803~\AA, in the rest-frame of the absorber. This lies in the optical from $z=0.3$~to~$2.4$ and serves as a probe of low-ionization gas \citep[e.g.][]{1999AAS...195.5201C,2010ApJ...715.1497J}. Detections of MgII absorption lines have therefore been used to infer the presence of an intervening system along the line of sight. In some cases these absorbers are associated with the quasar itself. However, in cases where the absorber is at a lower redshift than that of the quasar, the absorption is most likely taking place in an intervening galaxy between us and the quasar (i.e.\ a non-intrinsic system or an `intervenor').

Faraday rotation is a powerful tool for measuring the magnetic field strength along the line of sight towards astrophysical objects. The combination of cosmic magnetic fields and charged particles causes rotation of the polarization angle of linearly polarized synchrotron emission from background radio sources \citep[e.g.][]{2011hea..book.....L}. Along a line of sight, the observed polarization angle is altered by an amount equal to
\begin{equation}
\Phi = \Phi_{\textrm{0}} + \textrm{RM}\lambda^2 \,,
\end{equation}
where $\lambda$ is the observing wavelength, $\Phi$ and $\Phi_{\textrm{0}}$ are the measured and intrinsic polarization angles respectively, and the constant of proportionality RM, the `rotation measure', is generally related to the integrated product of the electron number density, \(n_{\rm e}\), and the strength of the component of the magnetic field parallel to the line of sight, \(B_{\parallel}\). The observed RM is also related to the redshift at which the Faraday rotating medium is located, but in practice this is typically a non-simple relationship, as there are actually multiple rotating media and it is not known where all of these media are distributed along the line of sight. Nevertheless, measurements of the RM can be used to infer the presence of magnetic fields and ionised gas somewhere along the line of sight between an observer and a source.

It has previously been suggested that there is a correlation between metal-line absorption and the RM of distant polarized sources \citep{1982ApJ...263..518K,1984ApJ...279...19W,1990ApJ...355L..31K,1991MNRAS.248...58W,1995ApJ...445..624O,2008ApJ...676...70K}. More recent studies have extended these previous works by finding a correlation between the magnitude of the RM and the number of strong MgII absorbing systems along the line of sight \citep{2008Natur.454..302B}. This has been used to suggest that these intervening systems are magnetized, and that the magnetic fields in these intervening normal galaxies are of much higher strength than is typically expected in this earlier epoch of the Universe. This adds an extra challenge to our understanding of cosmic magnetism, as it implies the Faraday rotation towards a background quasar consists of a Galactic, intrinsic\footnote{By intrinsic Faraday rotation, we refer to an additional component to the RM that occurs directly within a radio source or within the source's immediate environment, such that the Faraday rotation is directly related to the background quasar itself in some manner.}, and also an additional \emph{intervening} contribution.

\citet{2008Natur.454..302B} inferred a population of intervening magnetized sources from RMs measured at relatively high radio frequencies, i.e.\ at $\approx$5~GHz, using a sample of 71 optical quasar spectra. The correlation claimed between the presence of MgII absorption lines and increased magnitude of RM is relatively weak, with a signal equivalent to a $\approx1.7\sigma$ detection when comparing $N_{\textrm{MgII}}$=0 and $N_{\textrm{MgII}}$>0, where $N_{\textrm{MgII}}$ is the number of MgII absorbers along a line of sight, and a $\approx3.3\sigma$ detection between $N_{\textrm{MgII}}$=0 and $N_{\textrm{MgII}}$=2, albeit with only five sources with $N_{\textrm{MgII}}$=2. This has been suggested as evidence that the intervening systems must increase the RM along the line of sight. Conversely, when using RMs measured at lower frequencies, i.e.\ at 1.4~GHz, the correlation between RM and the presence of MgII absorbers is consistent with no signal \citep{2012ApJ...761..144B}, or with a weakly positive result at the $1.7\sigma$ level \citep{2013MNRAS.434.3566J}. This observed dichotomy between results at 1.4~GHz and 5~GHz has been used to suggest that the intervenors provide `partial coverage', and obscure only a fraction of the background radio source \citep[e.g.][]{2012ApJ...761..144B,2013ApJ...772L..28B}. Under such circumstances, \citet{2012ApJ...761..144B} suggests that MgII absorbers provide partial coverage of the background source with an inhomogeneous Faraday screen, which could perhaps depolarize the high RM component at low radio frequencies -- thereby giving rise to an observed `Faraday complexity', i.e.\ a system with a non-linear relationship between polarization angle and squared wavelength so that RM$\rightarrow$RM($\lambda$). As the sight lines with intervening systems that exhibit Faraday complexity appear associated with low fractional polarization at low frequency \citep{2012ApJ...761..144B}, this has been interpreted as evidence of depolarization due to partial coverage. In addition, the suggested presence of partial coverage has also been inferred from polarized spectral energy distributions (SEDs) \citep[e.g.][]{rosetti08,mantovani08}. 

However, the apparent frequency-dependence of the effect could alternatively be a result of observational selection effects: at high and low radio frequencies we select different source populations, with different morphology and position in relation to the optical counterparts. Such relationships could also be caused by a number of confounding variables that require interpretation of their effect on the data. One possible proxy for overcoming these selection effects is the total intensity spectral index, $\alpha$, defined such that $S\propto\nu^{+\alpha}$. In this paper, we attempt to take these selection effects and confounding variables into account. We re-examine the relationship between MgII absorption and RM from first principles, to determine whether this relationship extends to low observational frequencies. The consequences of such a relationship are important, as the emergence of magnetic fields in normal galaxies plays a strong role in star-formation in galaxy discs, drives the structure of the interstellar medium, influences other astrophysical processes that drive galaxy evolution, has implications for the cosmological growth of magnetic fields, and constrains dynamo mechanisms \citep[e.g.][]{2008RPPh...71d6901K}. Investigating such relationships may also provide the first conclusive empirical discriminant between theories of magnetic field amplification and structure. The standard $\alpha$--$\Omega$ dynamo predicts that a small seed field is amplified by the combined action of differential rotation and turbulence on a large-scale in a galactic disk. These seed fields could be either primordial or have been generated by supernovae and amplified by dynamo action. Primordial fields could also be amplified in the process of the collapse of protogalaxies, or by dynamo action in oblique shocks as a protogalaxy collapses. Observational constraints on these competing models are currently lacking \citep[e.g.][]{1993ApJ...406..407P}.

This paper is structured as follows: we present our observational data in Section~\ref{data}, where we detail and justify how we created our sample and its properties. We detail the quantitative analysis of our main sample in Section~\ref{analysis}, with the analysis of our subsamples being presented in Section~\ref{subsamples}. We discuss our results and the effect of confounding variables in Section~\ref{discussion}, while a summary of the physical implications of our findings are presented in Section~\ref{conclusion}. In Appendix~\ref{theory}, we argue that current mathematical models of partial coverage are incompatible with observational evidence. We refer to `polarization' on multiple occasions, in all cases we are referring to linear radio polarization -- both circular and optical polarization are beyond the scope of this work. All derived uncertainties are calculated using standard error propagation. Unless otherwise specified, all quantities are as measured in the observed-frame.

%==============================================================================%
\section{Observational Data}
\label{data}
\subsection{Cross-Matching}
We use the broadband radio polarization catalog of \citet{farnescatalogue} as our primary data source. This catalog accumulates and cross-matches data from throughout the literature over the last 50 years, taking resolution effects into account through the cross-matching criteria, and incorporating a significant number of major radio surveys including the NRAO VLA Sky Survey (NVSS), AT20G, B3-VLA, WENSS, NORTH6CM, GB6, and Texas \citep[e.g.][]{1980A&AS...40..319S,1980A&AS...39..379T,1981A&AS...43...19S,1982A&AS...48..137S,1991ApJS...75....1B,1996AJ....111.1945D,1996ApJS..103..427G,1997A&AS..124..259R,1998AJ....115.1693C,1999A&AS..135..571Z,2003A&A...406..579K,2003PASJ...55..351T,2009ApJ...702.1230T,2010MNRAS.402.2403M}.

The \citet{farnescatalogue} catalog expands upon the NVSS RM catalog at 1.4~GHz \citep{2009ApJ...702.1230T}, providing total intensity spectral indices, $\alpha$, for 25,649 sources, and polarization spectral indices, $\beta$, for 1,171 sources.\footnote{The polarized spectral index, $\beta$, is defined such that $\Pi \propto \lambda^{\beta}$, where $\Pi$ is the polarized fraction and $\lambda$ is the observing wavelength. Note that $\beta$ is defined in the opposite sense to the total intensity spectral index, $\alpha$, which is defined as $S\propto\nu^{+\alpha}$ and is the exponent of observing frequency rather than wavelength.} Furthermore, the catalog contains 951 polarized SEDs that are defined between 0.4~GHz to 100~GHz, with up to 56 independent polarization measurements per source. \citet{farnescatalogue} use model fitting and an automated classification algorithm based on the Bayesian Information Criterion to distinguish between different models for Faraday depolarization and to constrain total intensity radio spectral indices and curvature. In attempting to fit physical models of depolarization to the data, the assumption is made that the polarization fraction, $\Pi$, is intrinsically a meaningful quantity that is related to the degree of magnetic field ordering in the source. \citet{farnescatalogue} fit to the polarization angle as a function of wavelength, obtaining broadband RM measurements, and also include spectroscopic redshifts for 4,003 linearly polarized radio sources that were identified by \citet{2012arXiv1209.1438H} using various resources including the Sloan Digital Sky Survey (SDSS) \citep[e.g.][]{2009ApJS..182..543A}.

In this paper, we cross-match the data from \citet{farnescatalogue} with the catalog of \citet{2013ApJ...770..130Z}, which presents a sample of 84,534 quasars with a total of 35,752 non-intrinsic MgII absorption systems along their lines of sight, as derived from SDSS spectra. Since the catalog of RM versus redshift \citep{2012arXiv1209.1438H} and the catalog of MgII absorption \citep{2013ApJ...770..130Z} are both based on data from the SDSS, one way to combine the catalogs would be to use an arbitrarily small cross-matching radius. However, \citet{2012arXiv1209.1438H} do not necessarily nominate the nearest SDSS source, as seen in projection on the sky, as the most likely matched candidate. The catalog instead provides a `selected redshift' which is determined from the inclusion of other redshift catalogs and which takes into account the morphology of, for example, double-lobed radio sources. To combine the \citet{2013ApJ...770..130Z} data with the \citet{farnescatalogue} catalog, we therefore use the redshift of the background quasar, $z$, in the cross-matching criteria. Cross-matching was carried out relative to the radio source positions provided by \citet{2009ApJ...702.1230T}, each of which has an associated RM measurement at 1.4~GHz. For a match to be accepted, it must have been listed in the \citet{2012arXiv1209.1438H} catalog, be within an astrometric radius of $90^{\prime\prime}$ of the NVSS source position, and have a maximum redshift difference between the catalogs of \citet{2012arXiv1209.1438H} and \citet{2013ApJ...770..130Z} of $\Delta z \le 0.05$. This additional criterion helps to eliminate false cross-matches by ensuring consistent redshifts throughout both datasets. Note that the $90^{\prime\prime}$ astrometric radius is only used to find associated radio emission, and it is the quasar redshift that is used to minimize the number of false matches. The cross-matching of optical and radio data, including complex morphological effects, has already been done rigorously by \citet{2012arXiv1209.1438H}.

\subsection{Combining Radio and Optical Lines of Sight}
\label{combowavelengths}
A typical model of an extragalactic radio source includes at least two components: (i) the core-region surrounding the active galactic nucleus (AGN) itself, and (ii) the radio lobes and/or jets. Where an extragalactic radio source is also detected at optical wavelengths, the bright optical counterpart is generally associated with the core. Since MgII absorption systems are optically identified, they only provide information on the presence of intervenors towards the core. The presence of an MgII absorber therefore provides no constraints on the presence of absorption towards the radio lobes or jets. Furthermore, at high frequencies we tend to select the flat-spectrum cores of radio sources, while at low frequencies we tend to select the steep-spectrum lobes and jets. As the lobes are physically offset from the cores, it is therefore possible that such an effect is `diluting' the measured relationship between RM and the number of MgII absorbers, when measured at low frequencies. Analysing the polarized fraction and RM of an AGN at radio wavelengths within a finite resolution element will therefore potentially include contaminating polarized emission from the radio lobes and/or jets. Previous studies have not had the data available to investigate the extent of these contaminating effects. A sample that attempts to ensure that the core is probed at both radio and optical wavelengths will therefore assist in minimizing observational biases.

Note that we wish to probe the same emitting region\footnote{We use the term `region' to describe the corresponding surrounding area within the source in which there are similar physical conditions, e.g.\ the core-, jet-, or lobe-regions. The same emitting region, in this case, is unlikely to correspond to emission from the same physical material at both optical and radio wavelengths, although it is likely to be separated by only a very small angle on the sky.} within the source at both optical and radio wavelengths, and not just as seen in projection in the plane of the sky. This is important for three reasons: (i) probing the same emitting region ensures that we are probing very similar lines of sight, (ii) this ensures that we make no assumptions with respect to the physical size of an intervening system, and (iii) it is the only way to guarantee we are not affected by projection effects -- it is possible we could observe the core region at optical wavelengths, and meanwhile probe a lobe/jet at radio wavelengths as seen in projection on the sky. Point (iii) highlights that the requirement is not just for similar lines of sight, but for similar emitting regions. One cannot easily attempt to confirm that the emitting regions are coincident for an unresolved radio source using positional offsets only, i.e.\ offsets in the plane of the sky, as very unresolved sources would incorrectly appear to emit from the same region at both optical and radio wavelengths. This is potentially important at the $\approx45^{\prime\prime}$ resolution of the NVSS as used here. It is also not possible to explore projection effects using positional offsets of data with mismatched resolution -- for example, a steep-spectrum radio lobe can appear to be coaligned with an optical core, despite both physical features not emanating from the same line of sight. We therefore highlight that without the introduction of either multiple simplifying assumptions or very long baseline interferometric data, the same physical line of sight cannot be trivially probed using merely the alignment of radio and optical counterparts.

We therefore suggest an improved measure of the same emitting region, and by extension the same physical line of sight. This can be provided by the total intensity spectral index, $\alpha$. A prototypical model of an extragalactic radio source is one that consists of at least two emitting regions: (i) a flat-spectrum core ($\alpha\approx0$), and (ii) steep-spectrum jets/lobes ($\alpha\approx-0.7$). The spectral index therefore serves as a powerful discriminator of the physical emitting region that is largely independent of both resolution and projection effects. Although unresolved radio sources can contain emission from both the core region and the jets/lobes, the spectral index allows us to determine from which physical region the emission dominates. Consequently, flat-spectrum sources can be used as a proxy for the optical and radio counterparts being aligned (i.e.\ a core-dominated source), and steep-spectrum sources for those not aligned (i.e.\ a lobe-dominated source). This provides a reliable divider between different physical emitting regions and by extension of different lines of sight, while simultaneously reducing the likelihood of selecting regions that are merely aligned through projection or resolution effects. It is likely that the sources would need to be angularly resolved in order to completely eliminate such projection effects, although we believe this is only a very infrequent effect in our sample.

\subsection{Our Sample}
The cross-matching process provides an initial sample in which each source has a measurement of the number of MgII absorbing systems along the line of sight, the redshifts of the background quasar and of the intervening systems, and also a polarized fraction and an RM measurement at 1.4~GHz \citep{2009ApJ...702.1230T,2012arXiv1209.1438H,2013ApJ...770..130Z,farnescatalogue}. We also have supplementary data on the equivalent width of each absorbing system. All of the MgII absorption lines are non-intrinsic (aka intervening) \citep{2013ApJ...770..130Z}, and are blue-shifted from their background quasar by at least $\Delta z = 0.04$. 

To improve the quality of our sample, we exclude sources that are best modeled by a curved spectrum in total intensity \citep{farnescatalogue}, keeping only sources that were best fit by a conventional power law. To avoid poor-quality total intensity spectral indices in our sample, we exclude sources with a reduced-$\chi^2 \ge 4.0$. As the data used to construct the SEDs were taken at different epochs, more variable sources may tend to have an increased reduced-$\chi^2$ -- we are therefore likely selecting the least variable sources, in addition to those with low measurement errors. In order to minimize the effects of the Galactic foreground, we discard sources at Galactic latitudes $|b|\le25^{\circ}$. A full discussion of foreground effects is provided in Section~\ref{foreground}. 

We are also primarily concerned with the \emph{strong} absorbing systems. Based upon statistical arguments, it historically had been suggested that very small rest-frame equivalent width MgII absorption did not exist \citep[e.g.][]{1992ApJS...80....1S}. However, due to high signal--to--noise and high-resolution spectroscopy, the detection thresholds eventually dropped below the previous sensitivity levels of $0.3$~$\AA$ so that this could be tested through observation \citep[e.g.][]{1997ApJS..112....1T, 1999AAS...195.5201C}. The conventional divider between strong and weak MgII absorption is therefore assumed throughout the literature, by definition, to be at a rest-frame equivalent width of $0.3$~$\AA$ \citep[e.g.][]{1998AAS...193.0405R,2009AJ....138.1817B}. We therefore, in this paper, consider the strong absorbers to be those with an equivalent width $W_{\text{r}}\ge0.3$~\AA. Most contemporary studies have found that both strong and weak MgII absorbers are associated with different clouds of material: strong absorbers are typically associated with outflows from star-forming normal galaxies \citep[e.g.][]{2014ApJ...784..108B}, whereas weak absorbers may be related to the outskirts of normal galaxies, dwarf galaxies, material stripped through tidal interactions, and low surface-brightness galaxies \citep[e.g.][]{1999ApJS..120...51C}. Our initial sample contained both strong and weak absorbers, although it was dominated by strong absorbers. We calculate the mean rest-equivalent width, $\overline{W_{\text{r}}} = (W_{2796\AA} + W_{2803\AA})/2$, using the data of \citet{2013ApJ...770..130Z}. Absorbing systems with $\overline{W_{\text{r}}} <0.3$~\AA~($n=31$) were excluded from the rest of our analysis. This final sample contains 599 sources. We shall refer to this as the `main sample'. From \citet{farnescatalogue}, we also have a measurement of the total intensity spectral index, $\alpha$, for 548 of these sources. The main sample contains 398 sources without MgII absorption and 201 sources with strong non-intrinsic MgII absorption. Of the absorbing lines of sight, 152 contain a single MgII absorber, 38 contain two absorbers, 10 contain three absorbers, 0 contain four absorbers, and 1 contains five absorbers. 

The size of our sample is an improvement of almost an order of magnitude upon that of \citet{2008Natur.454..302B}, which contained measurements of MgII absorption and RM for 71 quasar spectra, and is also 10\% larger than the sample of \citet{2013MNRAS.434.3566J}, which contained 539 measurements of quasar spectra and RM but did not have spectral index information available. Our sample is further assisted by the high-reliability of the MgII data, which have both a purity and a completeness of $>95$\% \citep{2013ApJ...770..130Z}.

Using the spectral indices, we further split the sources into two subsamples: `flat-spectrum' ($\alpha\ge-0.3$) and `steep-spectrum' ($\alpha\le-0.7$). The gap in spectral index from $-0.7<\alpha<-0.3$ serves to avoid cross-contamination between the two samples. Our flat-spectrum subsample contains 87 sources with no absorber, 39 with one absorber, and 16 with more than one absorber. Our steep-spectrum subsample contains 154 sources with no absorber, 62 with one absorber, and 16 with more than one absorber.

Estimates of many other parameters are available in the catalog of \citet{farnescatalogue}, but are in a regime of small sample statistics after cross-matching with the MgII catalog. We therefore exclude these other variables from our analysis, and consider only the RMs and polarized fractions at 1.4~GHz, and the depolarization spectral indices. There is insufficient sample size for an analysis of the broadband RMs or weak MgII absorbers.

The main, flat-spectrum, and steep-spectrum samples, are all summarized in Table~\ref{tab:oursamples}. The source coordinates and other properties of our main sample are listed in Appendix~\ref{sampledeets}.

\begin{deluxetable}{ccccccccc}
\tablewidth{0pt}
\tablecaption{The total sample size and the number of lines of sight with a given number of strong MgII absorbing systems ($W_{\text{r}}\ge0.3$~\AA). The sample size is listed for both the main sample and the defined subsamples.}
\tablehead{
\colhead{} & \colhead{} & \multicolumn{7}{c}{$N_{\textrm{MgII}}$}
\\
\cline{3-9}
\\
\colhead{Sample Name} & \colhead{$N_{\textrm{total}}$} & \colhead{0} & \colhead{1}  & \colhead{2}  & \colhead{3}    & \colhead{4} & \colhead{5}    & \colhead{>0}      }
\startdata
Main (All sources)                            & 599   & 398    &  152   & 38 & 10  & 0  &   1 &  201 \\
Flat-spectrum\tablenotemark{a}      & 142     & 87      & 39    & 12  & 4  & 0  &   0 &  55 \\
Steep-spectrum\tablenotemark{b}   & 232   & 154    & 62    & 11  & 4  & 0  &   1 &  78
\enddata
\tablenotetext{a}{$\alpha\ge-0.3$}
\tablenotetext{b}{$\alpha\le-0.7$}
\label{tab:oursamples}
\end{deluxetable}

%==============================================================================%
\section{Main Sample Analysis}
\label{analysis}
We first perform an analysis similar to that of \citet{2013MNRAS.434.3566J}, and use our main sample to look for differences between the RMs and polarized fractions at 1.4~GHz of sources with $N_{\textrm{MgII}} =$0, 1, and $\ge2$ MgII absorbers (irrespective of whether each source has a spectral index measurement). In order to test whether any differences are statistically significant, we calculate the empirical cumulative-distribution functions (ECDFs) and statistical measures for various aspects of our main sample, both with and without non-intrinsic MgII absorption systems along the line of the sight. Our flat- and steep-spectrum subsamples will be presented in Section~\ref{subsamples}. For all of our analyses, we take a frequentist approach and use the two-sample Kolmogorov--Smirnov test (KS-test) for which the null hypothesis is that the ECDFs are calculated from independent samples drawn from the same underlying population. The $p$-values we obtain therefore indicate the probability of getting a result as extreme as or greater than the one obtained, if the null hypothesis is true. Note that the $p$-value only provides the probability with which one would reject the null-hypothesis, if it \emph{were} correct -- it provides no information on the probability that the null hypothesis \emph{is} correct, i.e.\ we have calculated $p(\ge D|H_{0})$ and not $p(H_{0}|D)$. This test is non-parametric, i.e.\ it does not assume that the data are sampled from any particular distribution. We will on occasion refer to the `$p$-value' and the `probability' interchangeably -- unless otherwise specified, we refer to $p(\ge D|H_{0})$, the probability of the two samples being as different as observed, or more so, if drawn from the same distribution.

%\newpage
\subsection{Rotation Measure at 1.4~GHz}
\begin{figure}
\centering
\includegraphics[clip=true, trim=0cm 0cm 0cm 0cm, width=8.7cm]{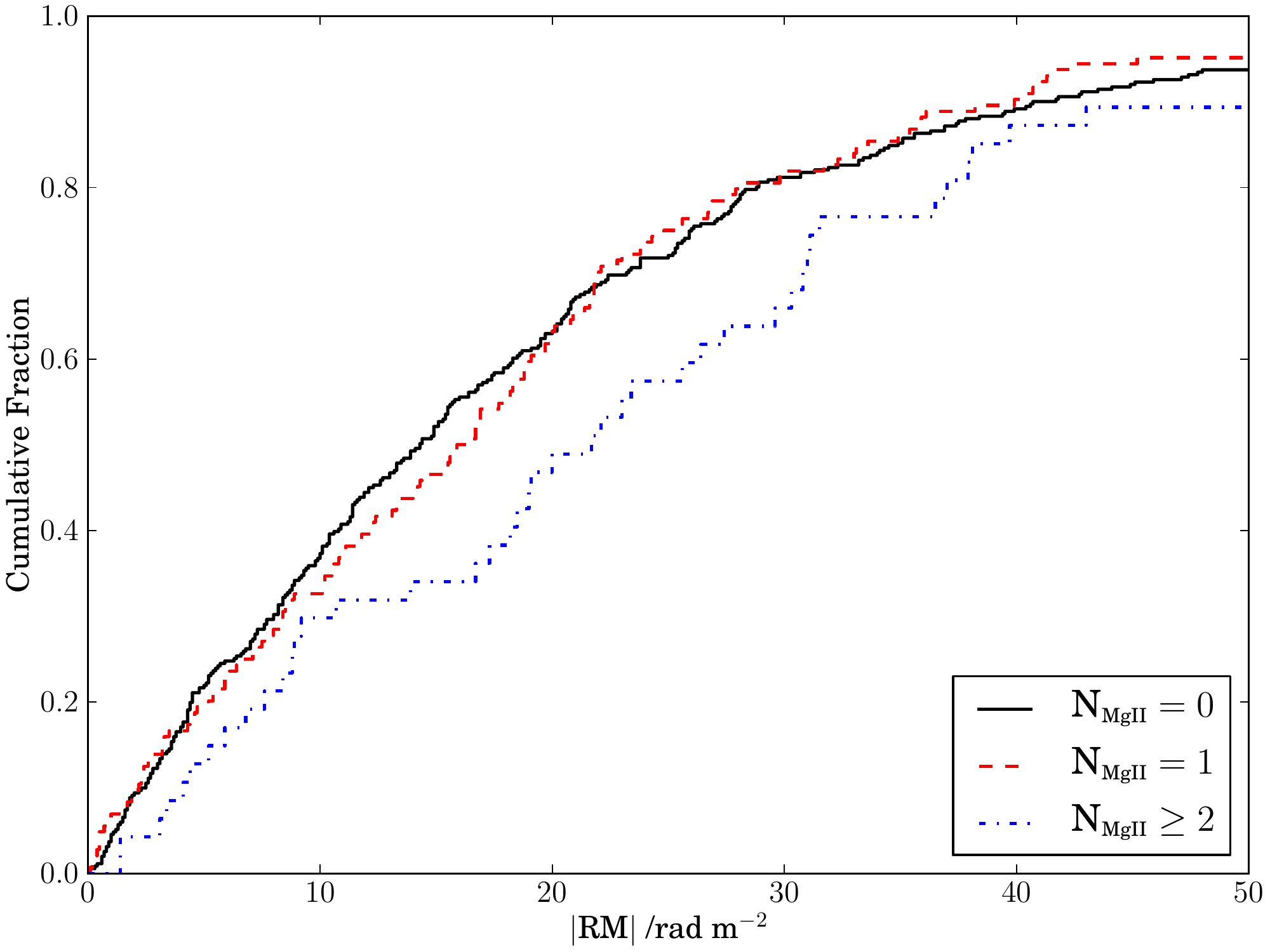}
\caption{ECDFs of the absolute value of the NVSS RMs for all 599 sources in the main sample. The black solid line shows the 398 sources without MgII absorption along the line of sight, the red dashed line shows the 152 sources with $1$ absorbing system, and the blue dotted line shows the 49 sources with $\ge2$ absorbing systems. There is no statistically significant difference between the three samples.}
\label{ecdfs_nosep}
\end{figure}

The ECDFs of the NVSS RMs from the main sample for $N_{\textrm{MgII}} =$0, 1, and $\ge2$ are shown in Fig.~\ref{ecdfs_nosep}. The KS-test provides a $p$-value of 17\% for sources with $N_{\textrm{MgII}}=0$ and $>0$. We also obtain a $p$-value of 77\% between sources with $N_{\textrm{MgII}}=0$ and $1$. For sources with $N_{\textrm{MgII}}=0$ and 2, the $p$-value is 2.9\%. There is no significant difference between any of the ECDFs. This is consistent with the results of \citet{2012ApJ...761..144B} and \citet{2013MNRAS.434.3566J}.

\subsection{Polarized Fraction}
\begin{figure}
\centering
\includegraphics[clip=true, trim=0cm 0cm 0cm 0cm, width=8.7cm]{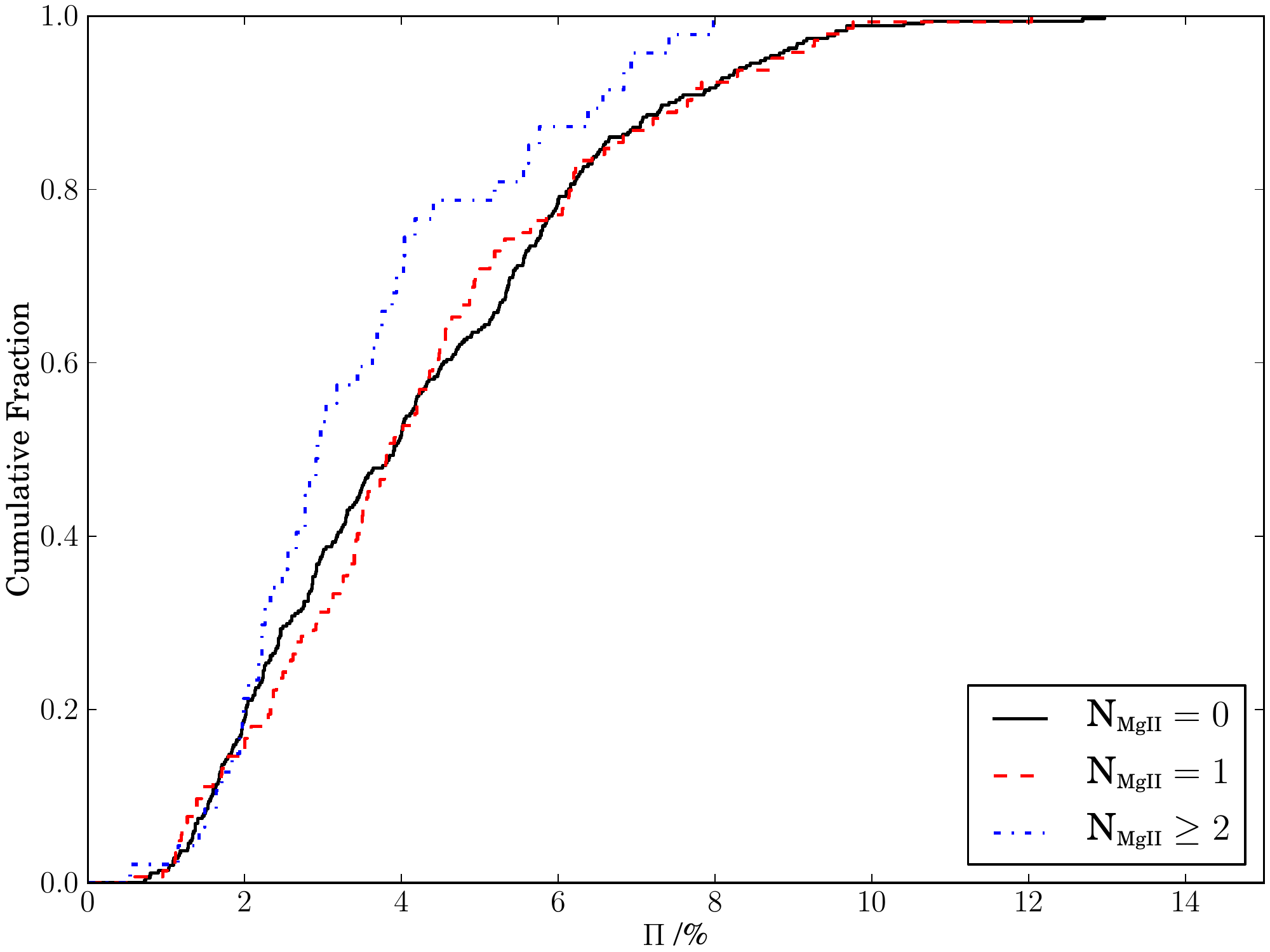}
\caption{ECDFs of the NVSS polarized fractions for all 599 sources in the main sample. The black solid line shows the 398 sources without MgII absorption along the line of sight, the red dashed line shows the 152 sources with $1$ absorbing system, and the blue dotted line shows the 49 sources with $\ge2$ absorbing systems. There is no statistically significant difference between the three samples.}
\label{ecdfs_nosep_polfrac}
\end{figure}

The ECDFs of the NVSS polarized fractions from the main sample are shown in Fig.~\ref{ecdfs_nosep_polfrac}. The KS-test provides a $p$-value of 25\% for sources with $N_{\textrm{MgII}}=0$ and $>0$. We also obtain a $p$-value of 60\% between sources with $N_{\textrm{MgII}}=0$ and $1$, and a $p$-value of 3.5\% between sources with $N_{\textrm{MgII}}=0$ and $2$. There is no significant difference between any two of the ECDFs. This indicates that MgII absorption has no significant effect on the polarized fraction of sources at 1.4~GHz. This is consistent with the results of \citet{2012ApJ...761..144B}.

\subsection{Polarization Spectral Indices}
\begin{figure}
\centering
\includegraphics[clip=true, trim=0cm 0cm 0cm 0cm, width=8.7cm]{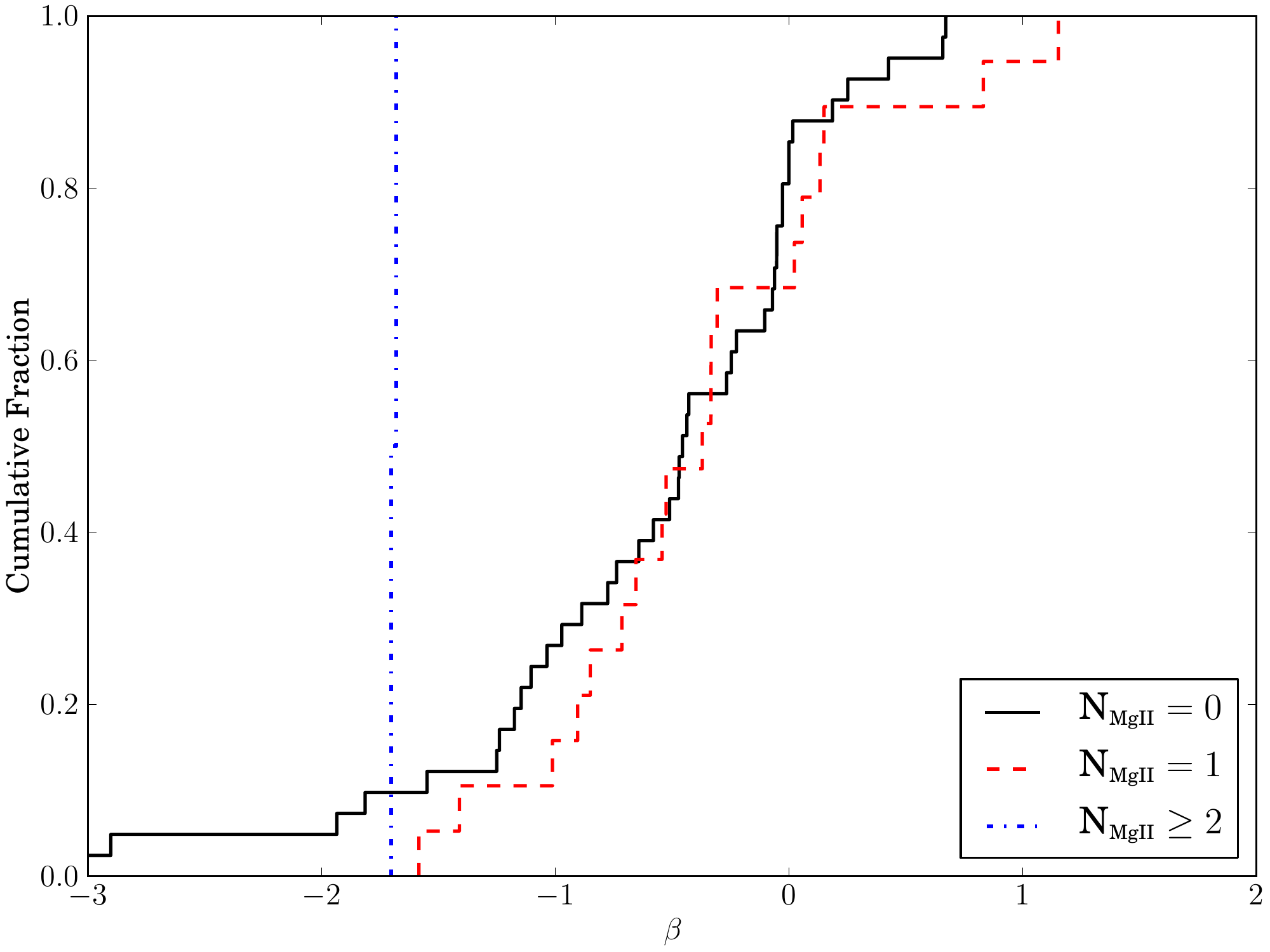}
\caption{ECDFs of the \citet{farnescatalogue} polarization spectral indices for all 62 sources in the main sample with complementary depolarization information. The black solid line shows the 41 sources without MgII absorption along the line of sight, the red dashed line shows the 19 sources with $1$ absorbing system, and the blue dotted line shows the 2 sources with $\ge2$ absorbing systems. There is no statistically significant difference between the three samples.}
\label{ecdfs_nosep_depol}
\end{figure}

The ECDFs of the \citet{farnescatalogue} polarization spectral indices, $\beta$ defined such that $\Pi \propto \lambda^{\beta}$, are shown in Fig.~\ref{ecdfs_nosep_depol}. Of the sources in the main sample, there is complementary depolarization information for 41 sources without an absorber, 19 sources with one absorber, and 2 sources with two or more absorbers. The KS-test provides a $p$-value of 82\% for sources with $N_{\textrm{MgII}}=0$ and $>0$. We also obtain a $p$-value of 66\% between sources with $N_{\textrm{MgII}}=0$ and $1$, and a $p$-value of 3.4\% between sources with $N_{\textrm{MgII}}=0$ and $2$. There is no significant difference between any two of the ECDFs. This suggests that MgII absorption has no significant effect on the depolarization of sources.

%==============================================================================%
\section{Flat- and Steep-Spectrum Subsample Analysis}
\label{subsamples}
We now extend our analysis to the flat- and steep-spectrum subsamples as defined in Section~\ref{data}. As a frequentist `significance' tends to be subjective, we shall provide a summary of our results, and individually consider both the significance and the effects of confounding variables in Section~\ref{discussion}. While we consider the redshift distribution of our sample in Section~\ref{flatsteepredshift}, we are unable to trivially separate our sample based on luminosity, which has been calculated assuming an optically-thin synchrotron emitting region \citep{farnescatalogue}. The physical meaning of such a calculation is unclear due to beaming and in the presence of self-absorption due to an optically-thick emitting region, as might be occurring if flat-spectrum sources are core-dominated AGN. As shall be shown, the flat-spectrum subsample is the most important in which to check for luminosity effects.

%\newpage
\subsection{Rotation Measure at 1.4~GHz}
\label{flatsteepRM}
\begin{figure*}
\centering
\includegraphics[clip=true, trim=0cm 0cm 0cm 0cm, width=8.95cm]{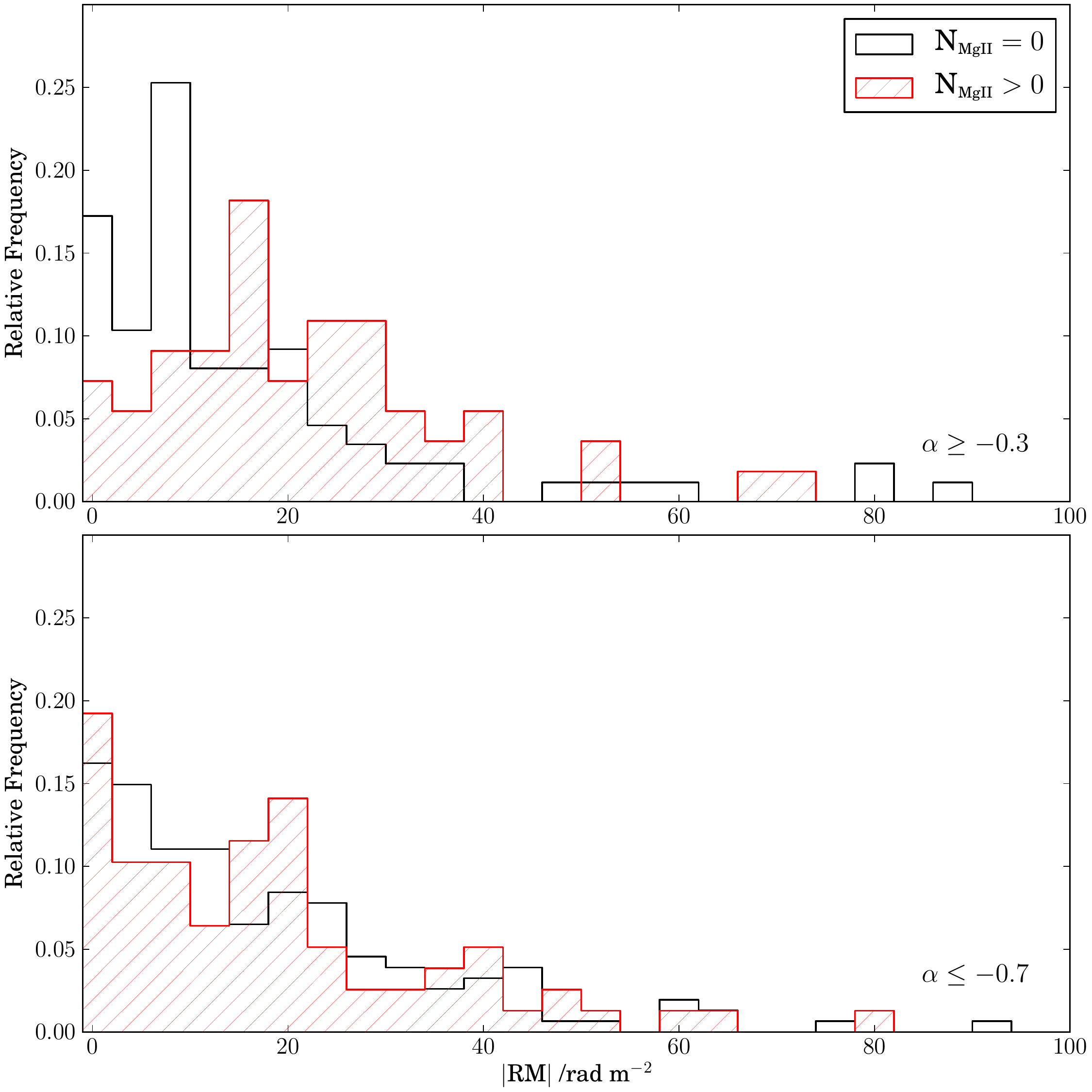}
\includegraphics[clip=true, trim=0cm 0cm 0cm 0cm, width=8.95cm]{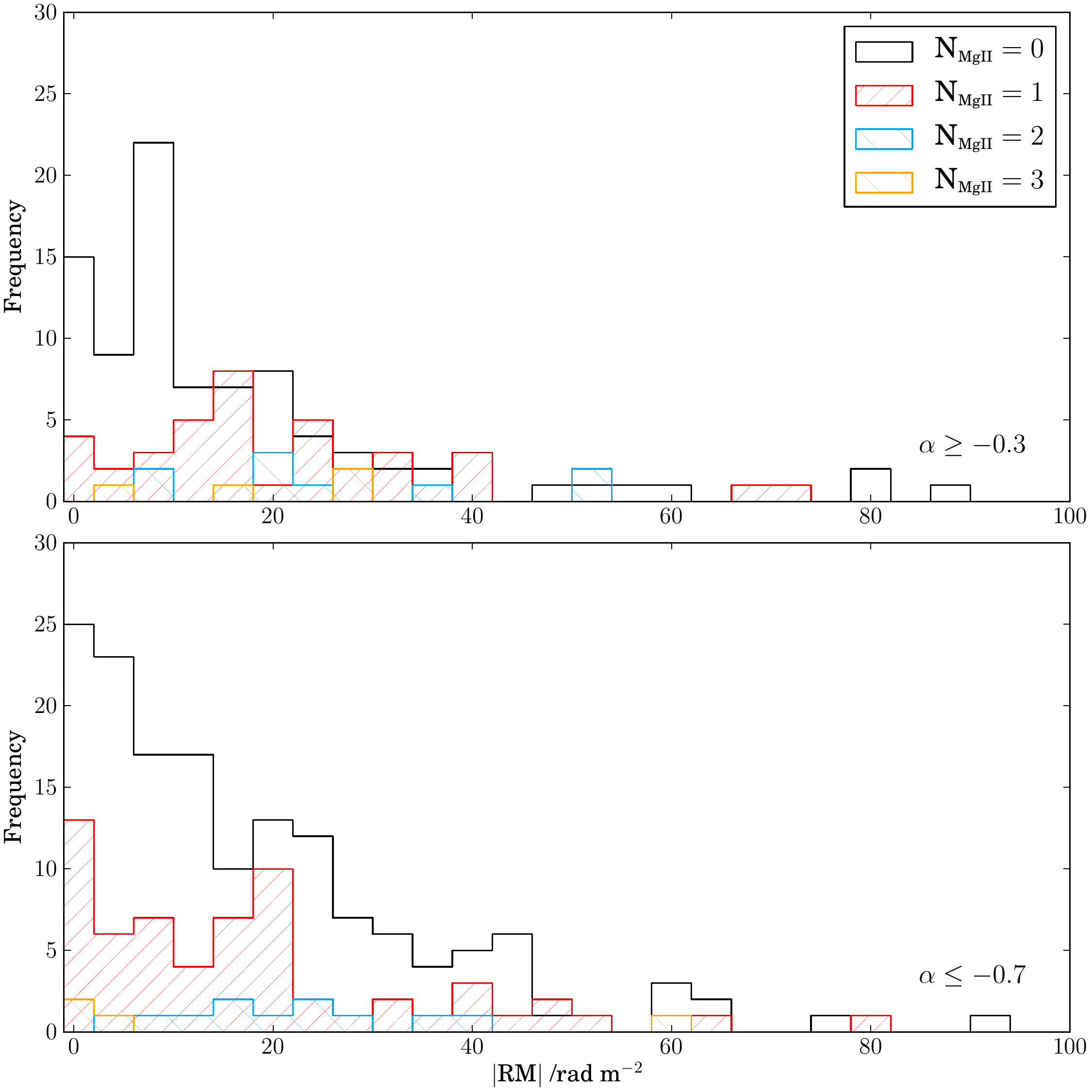}
\caption{Histograms showing the number of lines of sight with a given number of MgII absorbing systems, as a function of |RM|. Both the flat-spectrum (top row) and steep-spectrum (bottom row) subsamples are displayed. Histograms are shown for the cases where the number of MgII absorption systems along the line of sight, $N_{\textrm{MgII}}$, is either equal to zero or greater than zero (left column) and for each individual number of MgII absorbers (right column). Values of $N_{\textrm{MgII}}$ are displayed in the legend to the top-right of the upper plots. Note that the histograms in the left column are normalized; those in the right column are not.} 
\label{MgII_Histo}
\end{figure*}

Our sample is displayed in histograms of the number of MgII absorbers with different RMs at 1.4~GHz in Fig.~\ref{MgII_Histo}. For the flat-spectrum subsample, the sources with $N_{\textrm{MgII}}>0$ appear to have a greater dispersion in the absolute value\footnote{Note that the sign of the RM tells us only about the direction of the magnetic field along the line of sight. Here we are interested only in the field strength, which is best traced by |RM|. For consideration of the Galactic foreground, see Section~\ref{foreground}.} of RM relative to the steep-spectrum sources. We now test this statistically.

\begin{figure}
\centering
\includegraphics[clip=true, trim=0cm 0cm 0cm 0cm, width=8.5cm]{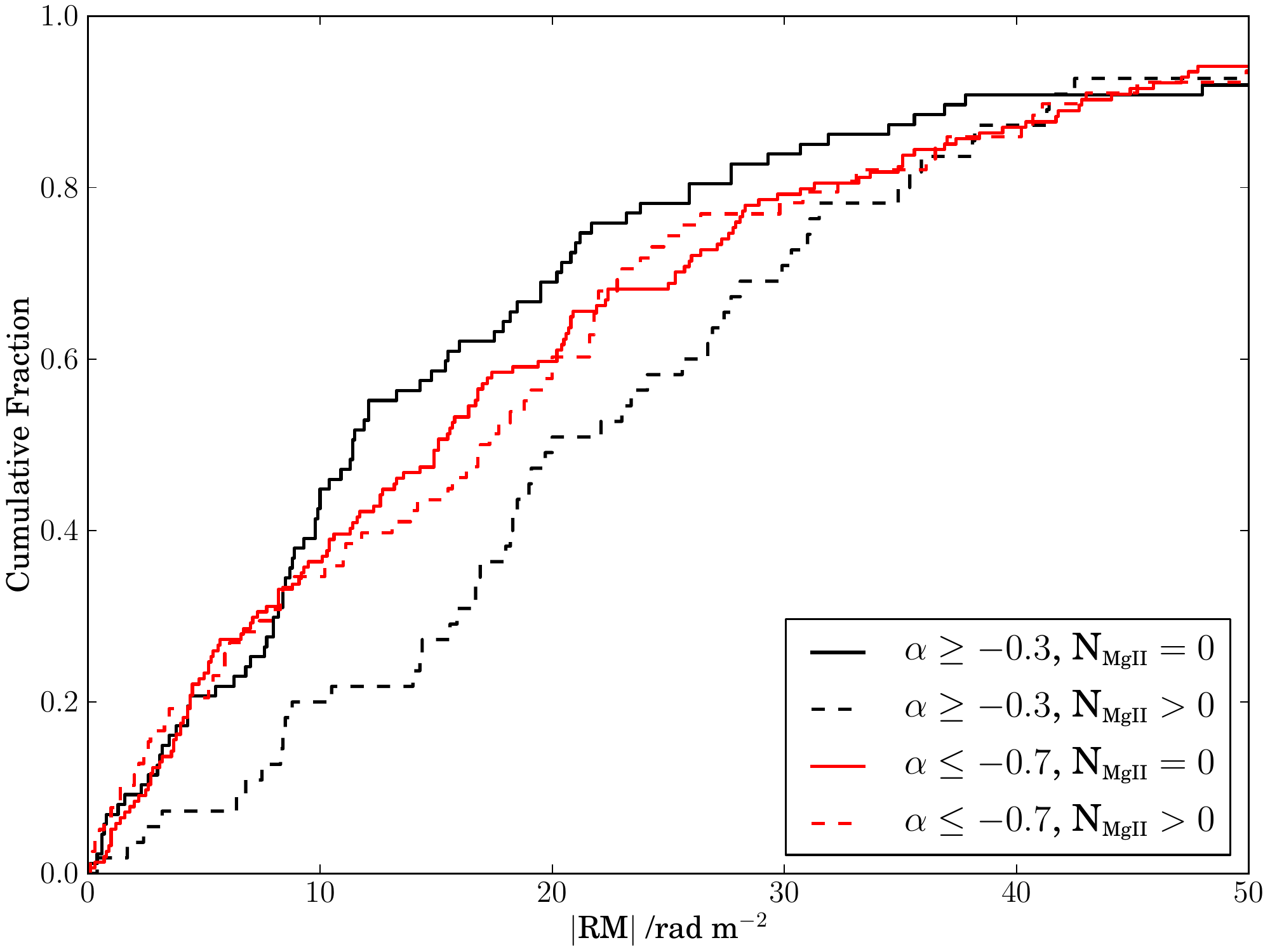}
\bigskip
\includegraphics[clip=true, trim=0cm 0cm 0cm 0cm, width=8.5cm]{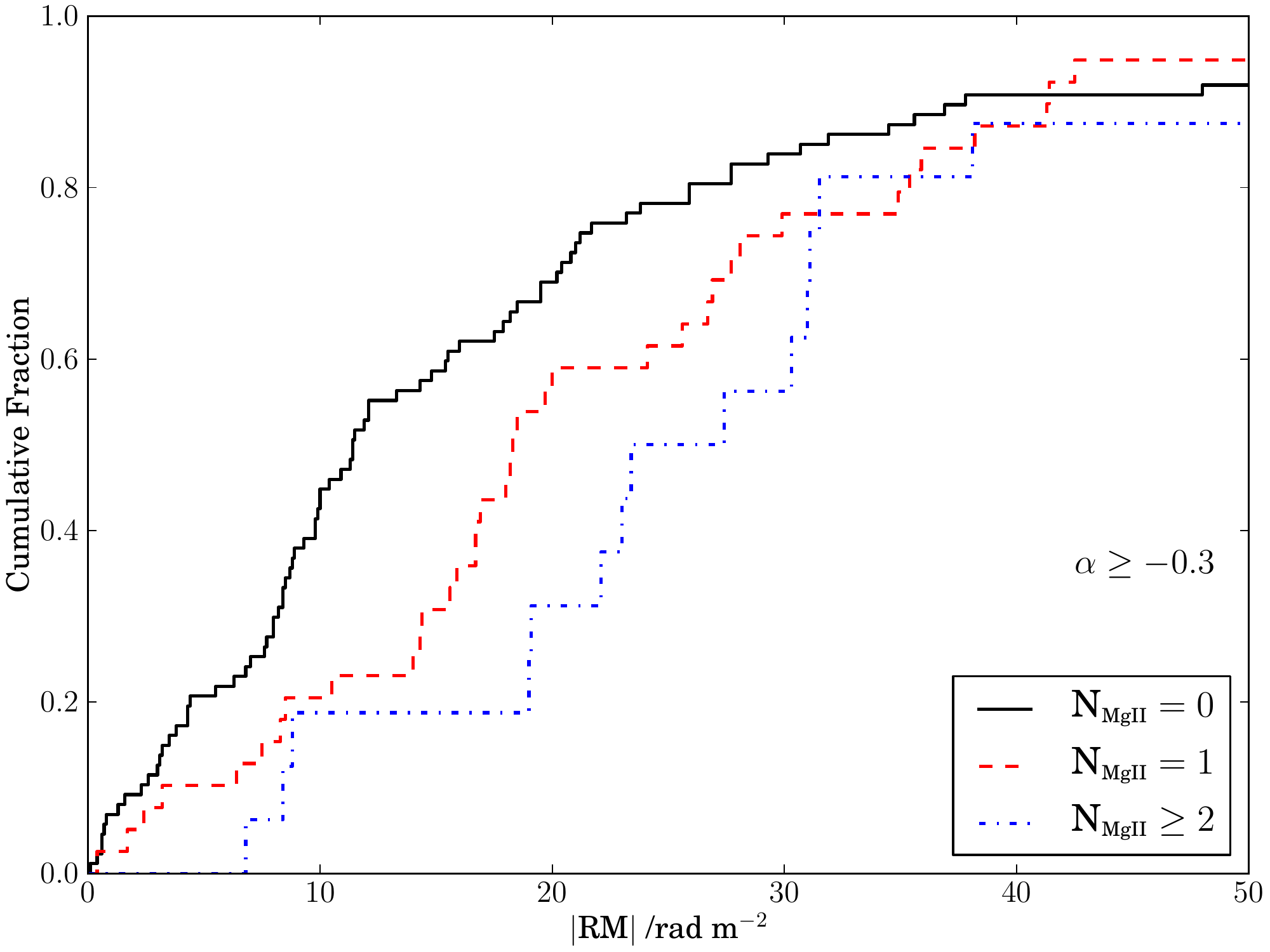}
\bigskip
\includegraphics[clip=true, trim=0cm 0cm 0cm 0cm, width=8.5cm]{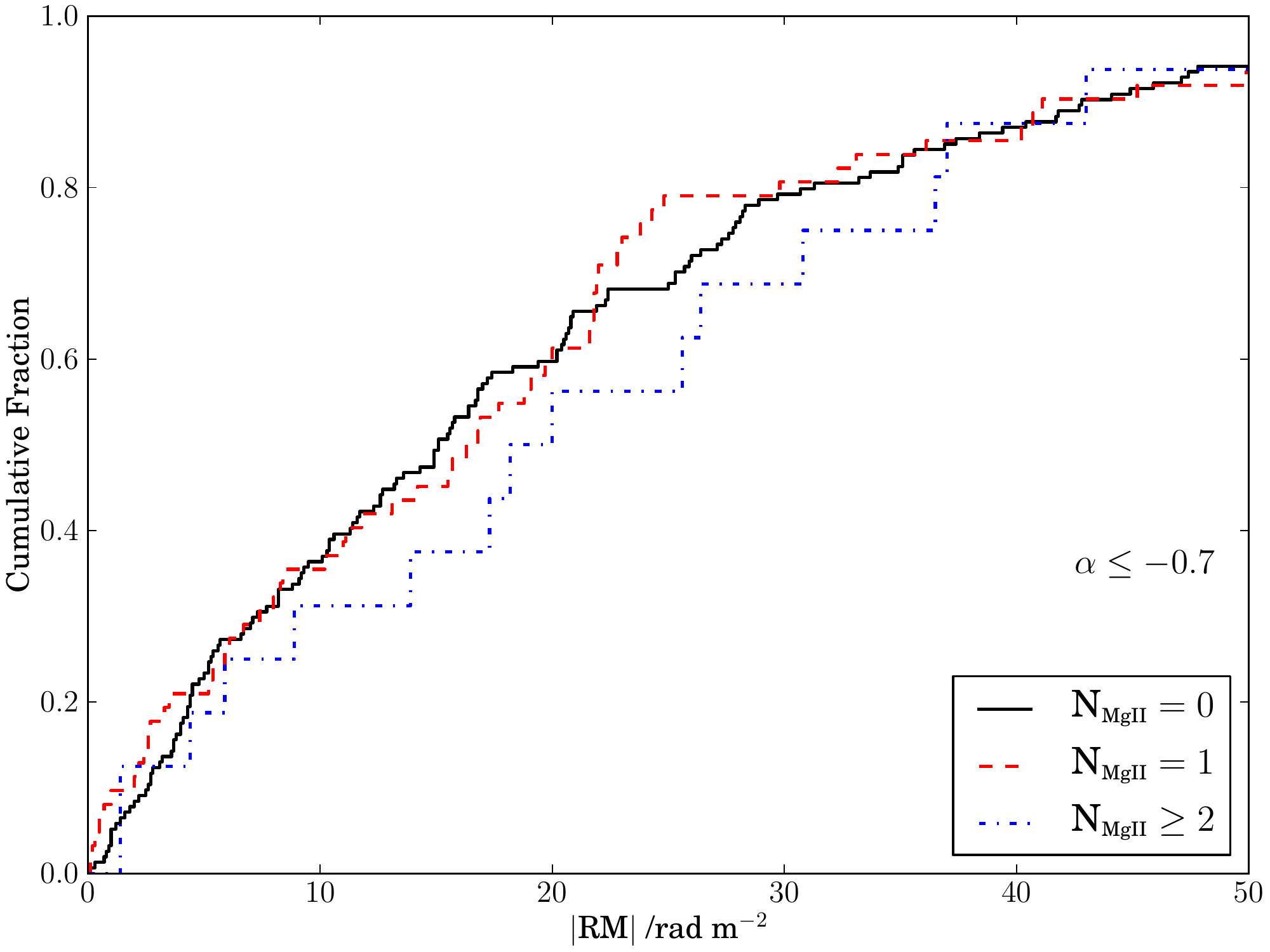}
\caption{ECDFs of the absolute value of the NVSS RMs for (i) Top panel: flat- (black), and steep- (red) spectrum sources. The solid lines show the sources without MgII absorption, while the dashed lines show the sources with $\ge1$ absorbing system along the line of sight, (ii) Middle panel: flat-spectrum sources only, (iii) Bottom panel: steep-spectrum sources only. In (ii) and (iii), the black solid lines show the sources without MgII absorption along the line of sight, the red dashed lines show the sources with $1$ absorbing system, and the blue dotted lines show the sources with $\ge2$ absorbing systems.}
\label{ecdfs_cores_lobes}
\end{figure}

The ECDFs of the RMs of the flat- and steep-spectrum sources are shown in Fig.~\ref{ecdfs_cores_lobes}. From the top panel, the difference between the flat-spectrum sources with versus without MgII absorption has a $p$-value of 0.044\% of being this large or larger if drawn from the same underlying distribution. Meanwhile, the steep-spectrum sources have a $p$-value of 90\%. This difference between the flat- and steep-sources is also identified from the ECDFs displayed in the middle and bottom panels; flat-spectrum sources with 0 and 1 (0 and 2) absorbers have a $p$-value of 0.37\% (0.24\%), while steep-spectrum sources with 0 and 1 (0 and 2) absorbers have a $p$-value of 65\% (54\%). For only sources with no absorption, the difference between the flat- and steep-sources has a $p$-value of 29\%.

\subsection{Polarized Fraction}
\begin{figure}
\centering
\includegraphics[clip=true, trim=0cm 0cm 0cm 0cm, width=8.5cm]{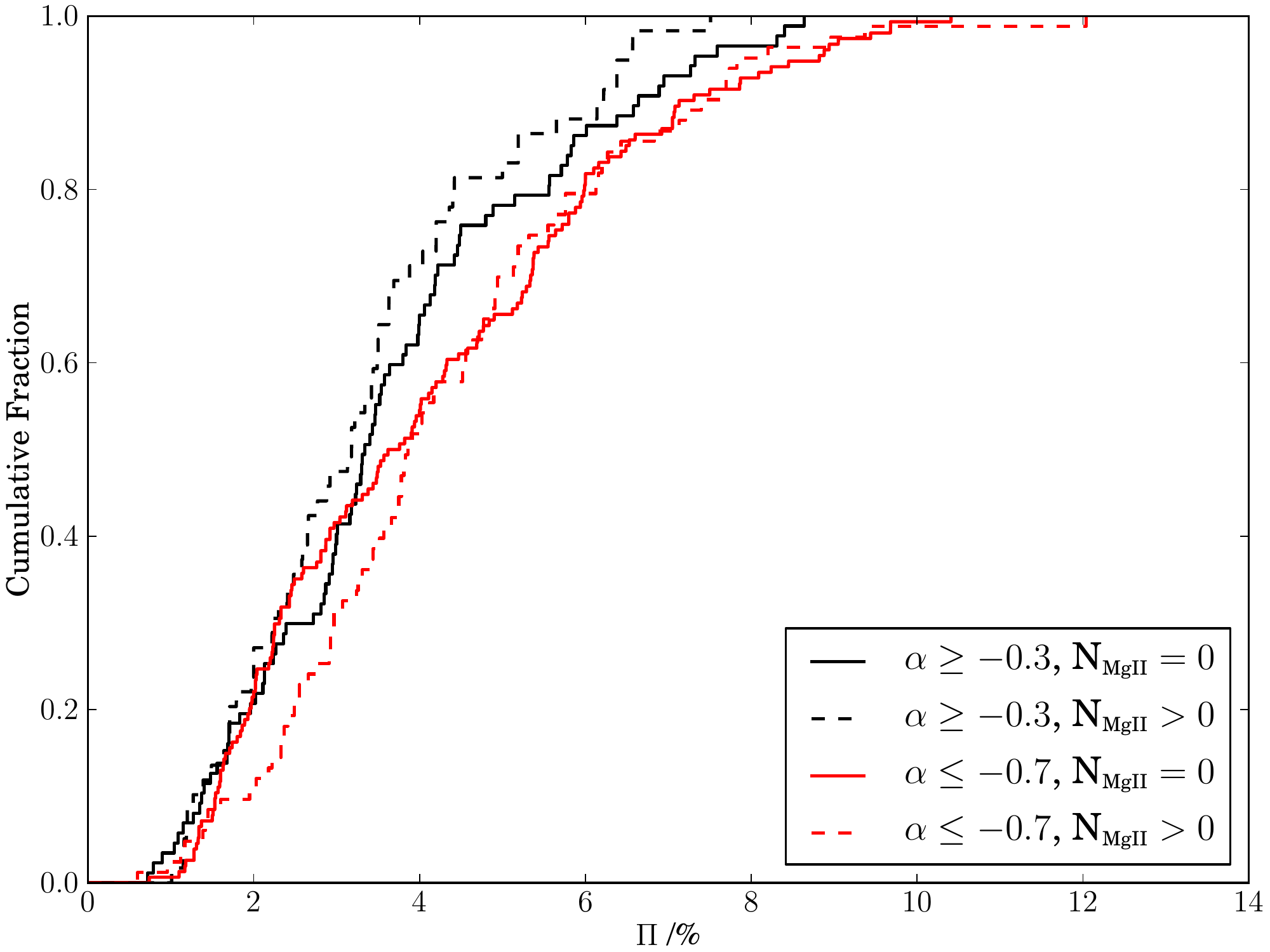}
\caption{ECDFs of the NVSS polarized fraction, $\Pi$, for flat- (black), and steep- (red) spectrum sources. The solid lines show the sources without MgII absorption, while the dashed lines show the sources with $\ge1$ absorbing system along the line of sight.}
\label{ecdfs_polfrac}
\end{figure}

The ECDFs of the polarized fractions, $\Pi$, for flat- and steep-spectrum sources are shown in Fig.~\ref{ecdfs_polfrac}, for sources that are both with and without an absorber. The flat-spectrum sources with/without MgII absorption have a $p$-value of 56\%. Meanwhile, the steep-spectrum sources have a $p$-value of 11\%.

\subsection{Polarization Spectral Indices}
\label{depol_section}
\begin{figure}
\centering
\includegraphics[clip=true, trim=0cm 0cm 0cm 0cm, width=8.5cm]{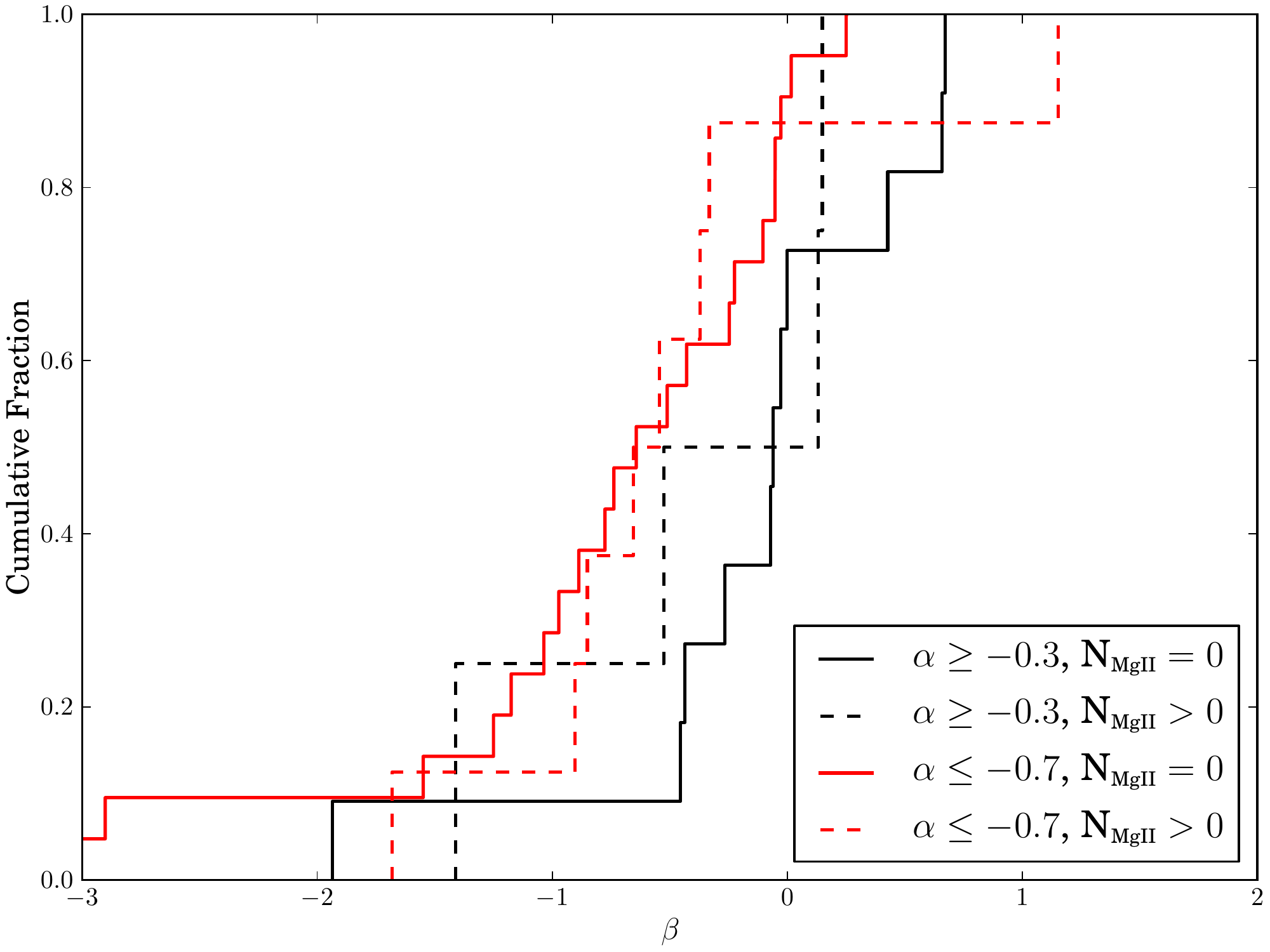}
\caption{ECDFs of the \citet{farnescatalogue} polarization spectral indices, $\beta$, for flat- (black), and steep- (red) spectrum sources. The solid lines show the sources without MgII absorption, while the dashed lines show the sources with $\ge1$ absorbing system along the line of sight.}
\label{ecdfs_depol}
\end{figure}

The ECDFs of the \citet{farnescatalogue} polarization spectral indices, $\beta$, are shown in Fig.~\ref{ecdfs_depol} (By flat- and steep-spectrum, we still refer to the total intensity spectral index, $\alpha$). For the flat-spectrum subsample, there are complementary $\beta$ measurements for 9 sources without an absorber, 5 sources with one absorber, and 1 source with two absorbers. For the steep-spectrum subsample, there are complementary $\beta$ measurements for 15 sources without an absorber, 7 sources with one absorber, 0 sources with two absorbers, and 1 source with three absorbers. The KS-test provides a $p$-value of 58\% between flat-$\alpha$ sources with $N_{\textrm{MgII}}=0$ and $>0$, and a $p$-value of 78\% between steep-sources with $N_{\textrm{MgII}}=0$ and $>0$. There is a known difference between the depolarization of flat- and steep-$\alpha$ sources, as discussed by \citet{farnescatalogue}. Nevertheless, there is no significant difference detected between sources with and without absorbers, regardless of whether the source has flat- or steep-$\alpha$. Our data are consistent with the presence of intervening MgII absorbers having no effect on the depolarization of sources, although we note that the sample size is very small. 

\subsection{Redshift}
\label{flatsteepredshift}
\begin{figure}
\centering
\includegraphics[clip=true, trim=0cm 0cm 0cm 0cm, width=8.5cm]{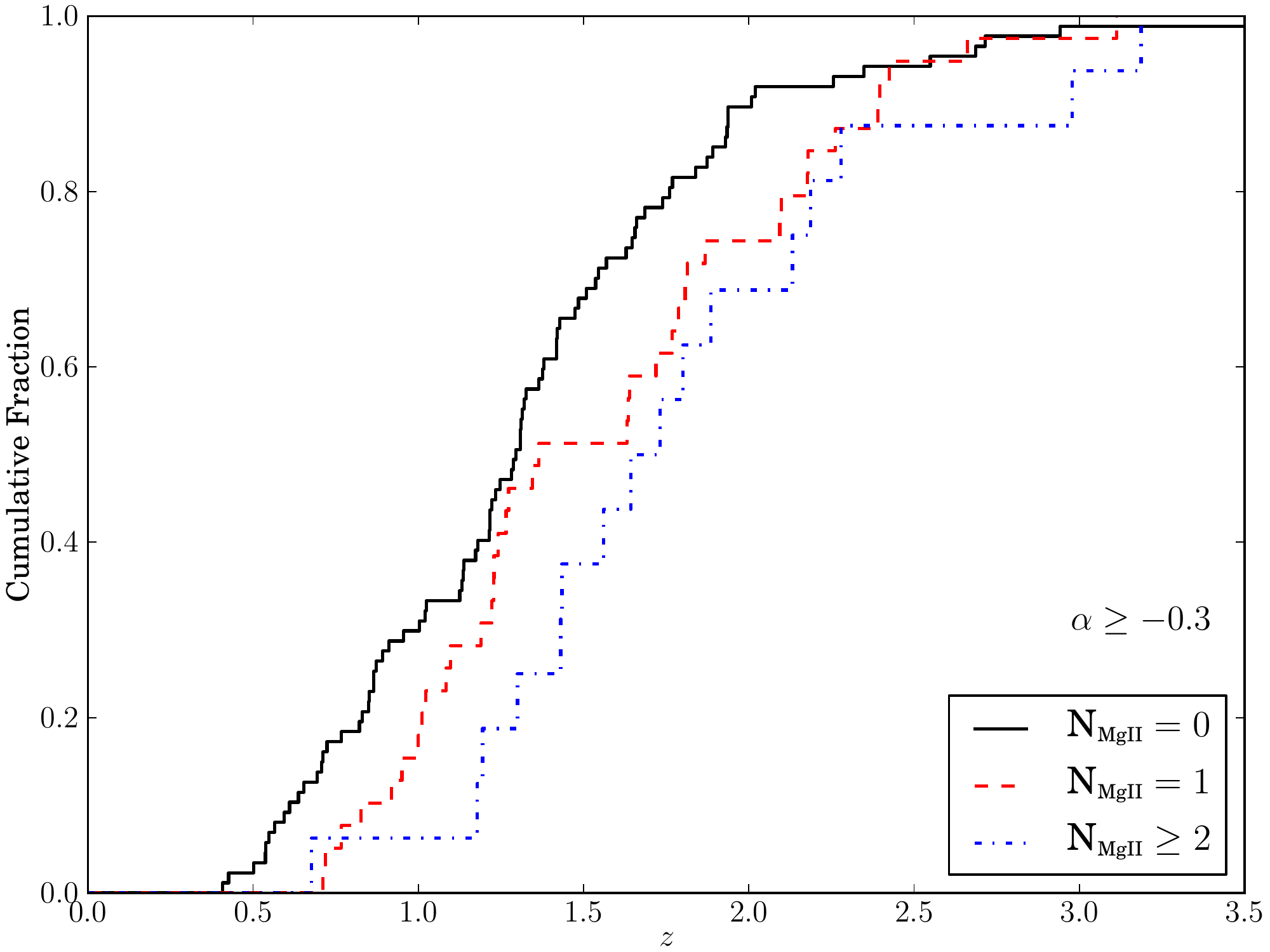}
\bigskip
\includegraphics[clip=true, trim=0cm 0cm 0cm 0cm, width=8.5cm]{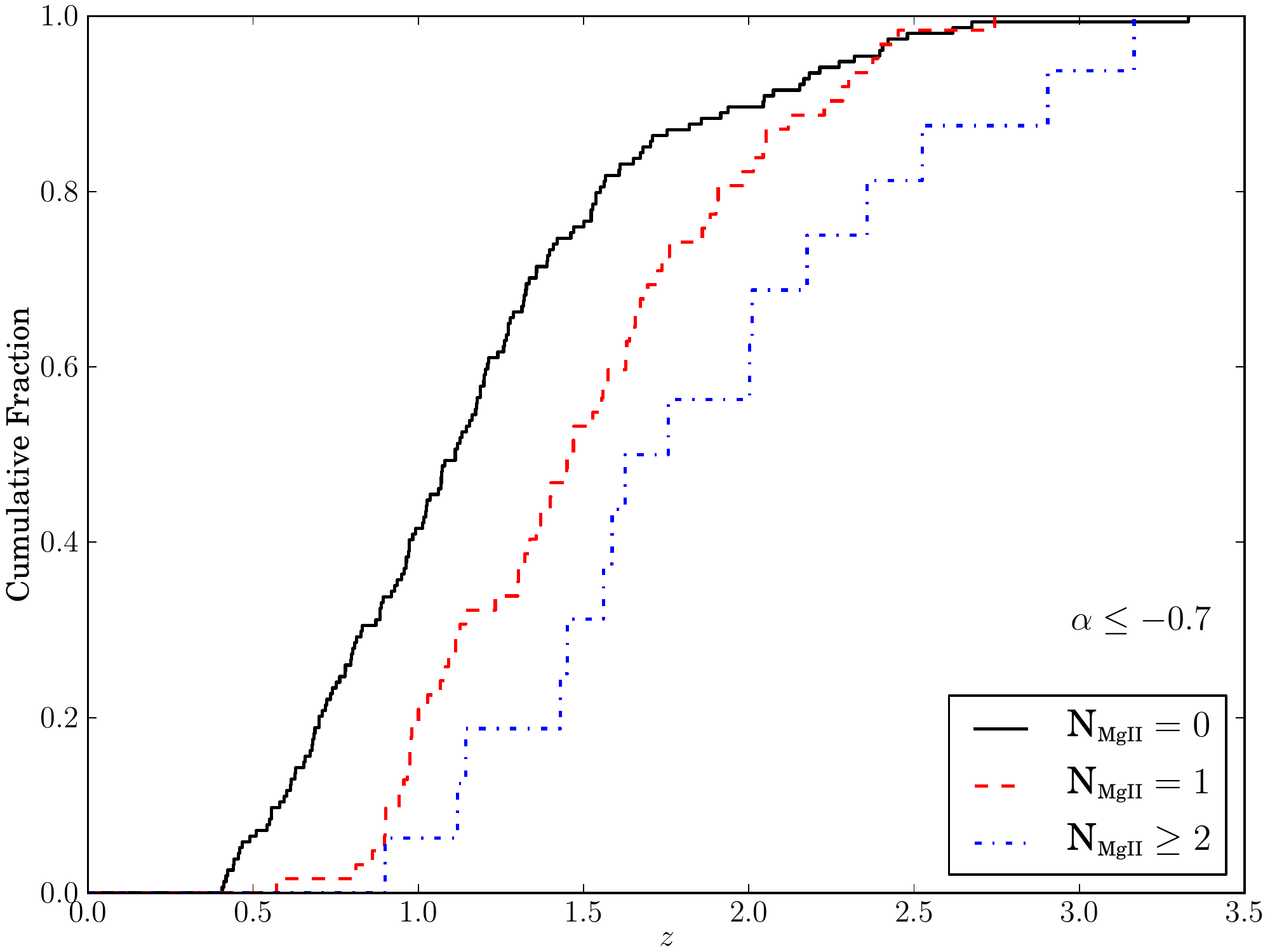}
\caption{ECDFs of the background quasar redshifts for flat- (top), and steep- (bottom) spectrum sources. The black solid line shows the sources without MgII absorption along the line of sight, the red dashed line shows the sources with $1$ absorbing system, and the blue dotted line shows the sources with $\ge2$ absorbing systems. In our sample, the quasars with intervening absorbers tend to be located at higher redshifts. However, there is no statistically significant difference between the flat- and steep-spectrum distributions.}
\label{ecdfs_redshift}
\end{figure}

The ECDFs of the redshifts, $z$, for flat- and steep-spectrum sources are shown in Fig.~\ref{ecdfs_redshift}, for sources with varying numbers of MgII absorbers. The flat-spectrum sources with versus without MgII absorption have a $p$-value of 1.0\%. Meanwhile, the steep-spectrum sources have a $p$-value of 0.00003\%. In our sample, the radio sources with intervening absorbers clearly tend to be located at higher redshifts. However, all flat- and all steep-spectrum sources have a $p$-value of 80.5\% -- there is no statistically significant difference between the flat- and steep-spectrum distributions.

\subsection{The Galactic Foreground}
\label{foreground}
One possibility to explain the apparent difference between flat-spectrum sources both with and without MgII absorption, as detailed in Section~\ref{flatsteepRM}, is that it is the consequence of contributions to the Faraday rotation from the Galactic foreground. It is typically assumed that the observed $\text{RM}=\text{GRM}+\text{RRM}$, where RRM contains contributions from both the background source environment and also the extragalactic line of sight, and GRM is the contribution from the Galaxy \citep[e.g.][]{2012arXiv1209.1438H}. Nevertheless, it has been suggested that while the Galactic foreground \emph{can} be estimated using surveys such as the NVSS, it \emph{cannot} be reliably subtracted to obtain an RRM without knowing the relative uncertainties \citep{2014arXiv1404.3701O}. Regardless, we expect the effect of the Galactic foreground to be low, as we have removed sources at low Galactic latitudes from our sample (see Section~\ref{data}) and we would expect sources with high RMs to be preferentially located in the Galactic plane \citep[e.g.][]{2009ApJ...702.1230T}. For the foreground to be influencing our main results, our sample would have to be anisotropically distributed on the sky such that there was either a difference in the RM of flat- and steep-spectrum sources, or of sources with differing numbers of MgII absorbers. Therefore if the Galactic foreground was causing our result, we could expect a different estimation of GRM between these different samples.

To investigate the possibility of the Galactic foreground affecting our result, we plot the ECDFs of the GRM. As various foreground estimation methods have been previously proposed \citep[e.g.][]{2009ApJ...702.1230T,2012arXiv1209.1438H,2014arXiv1404.3701O,2014arXiv1405.5087X}, we use two independent techniques to ensure there is no dependence on the method used for foreground correction. In both cases, we use the NVSS RMs \citep{2009ApJ...702.1230T} as the input to the reconstruction algorithm.
\begin{compactenum}[i]
\item For the first algorithm, for each point source we find the mean RM of all sources within an $8^{\circ}$ radius of the central source while excluding the central source itself \citep[e.g.][]{1995ApJ...445..624O}. We refer to this as the `mean RM' algorithm.
\item For the second algorithm, we use the reconstruction of \citet{2014arXiv1404.3701O} which uses the extended critical filter formalism that is derived within the framework of information field theory \citep[see e.g.][for further details]{2012A&A...542A..93O,2014arXiv1404.3701O}.
\end{compactenum}

\begin{figure*}
\centering
\includegraphics[clip=true, trim=0cm 0cm 0cm 0cm, width=8.5cm]{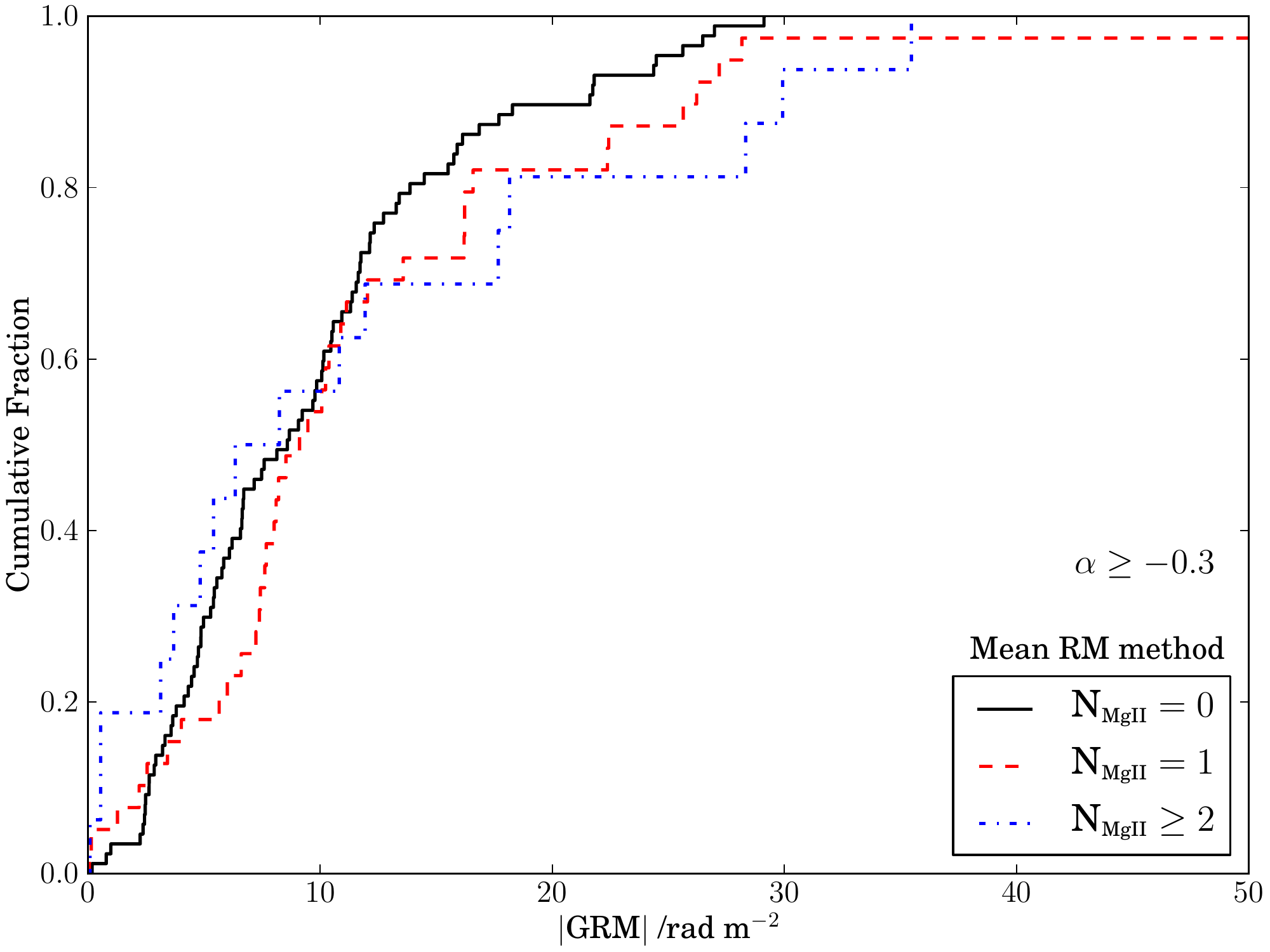}
\includegraphics[clip=true, trim=0cm 0cm 0cm 0cm, width=8.5cm]{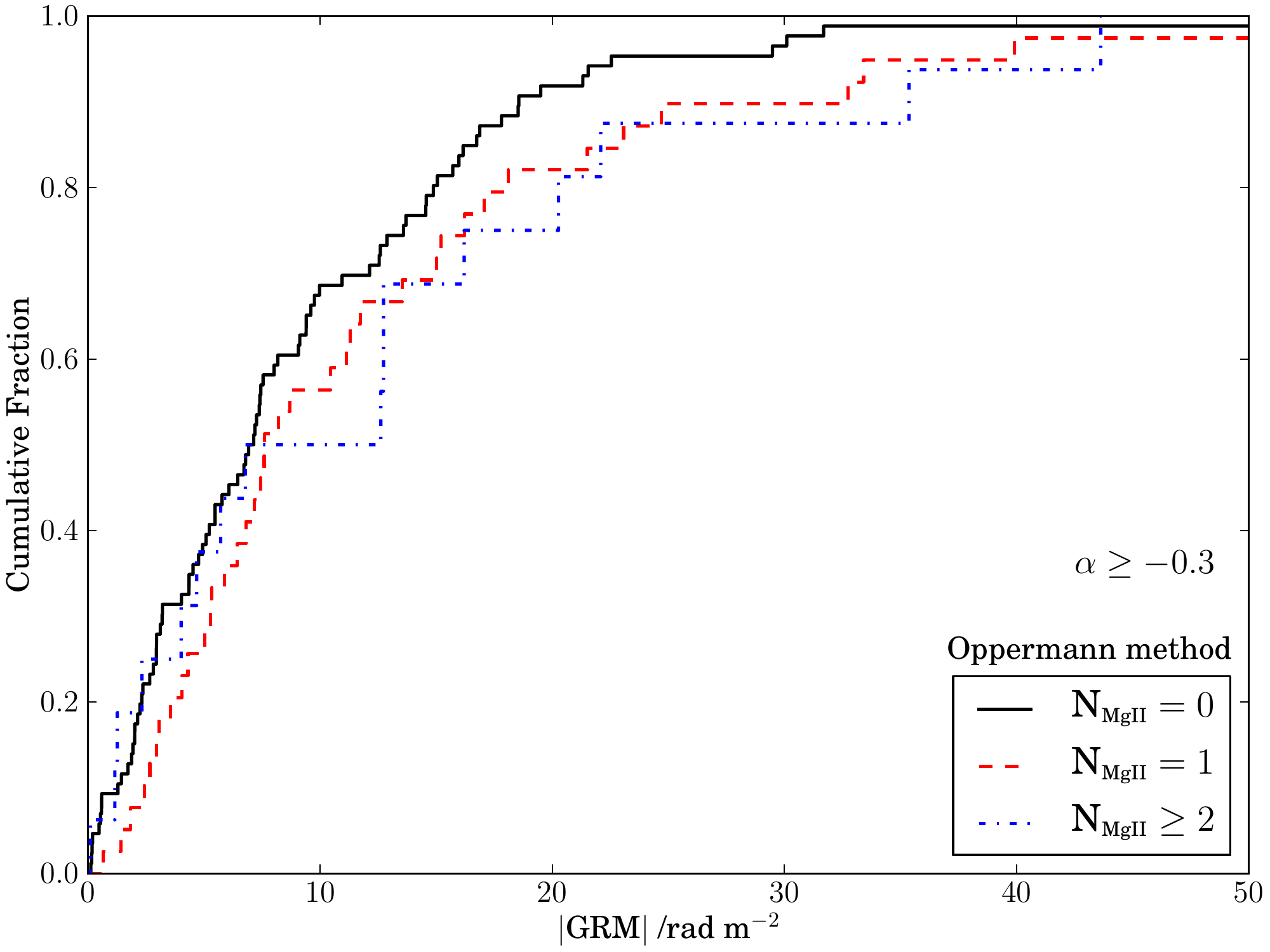}
\bigskip
\includegraphics[clip=true, trim=0cm 0cm 0cm 0cm, width=8.5cm]{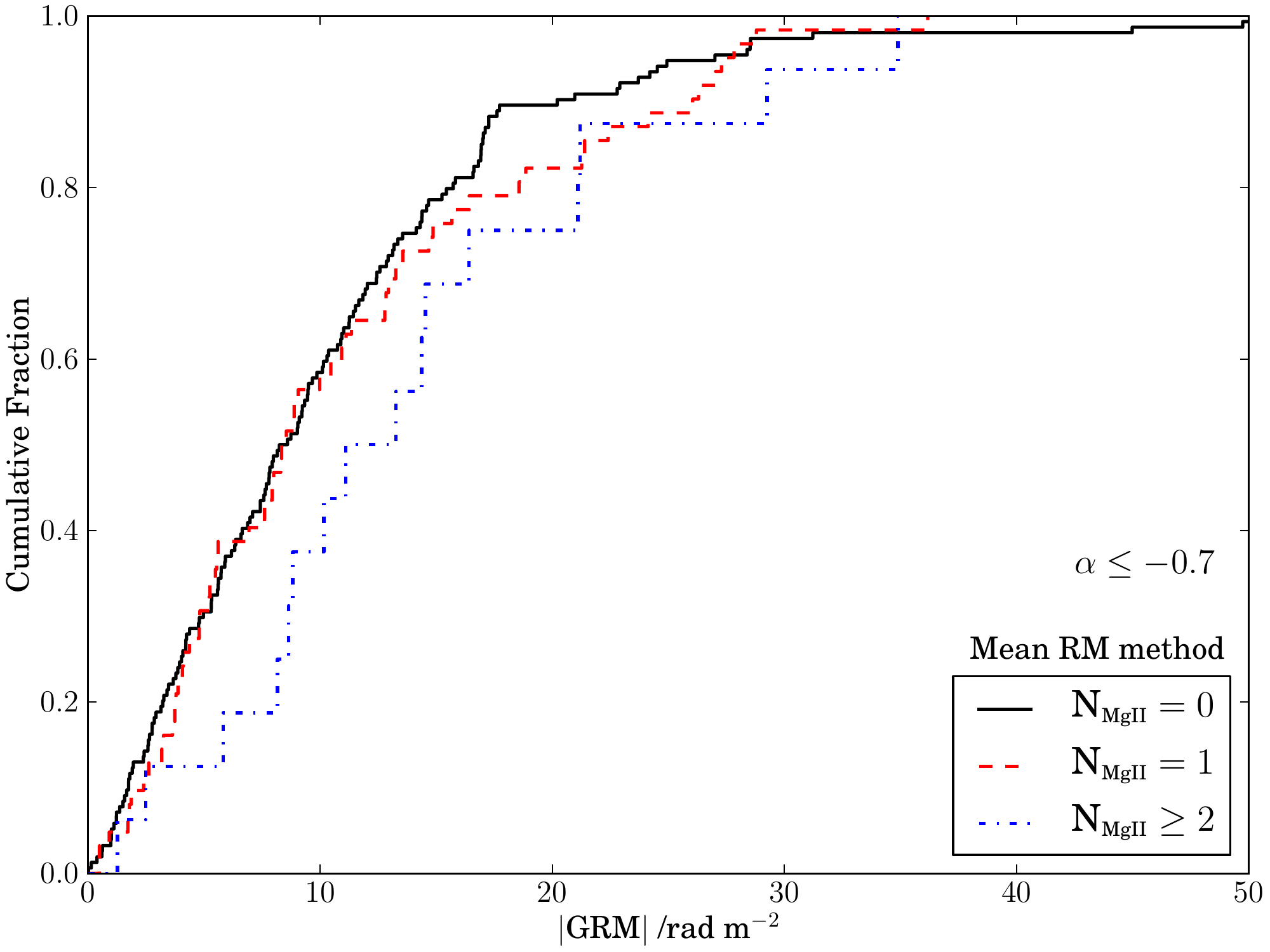}
\includegraphics[clip=true, trim=0cm 0cm 0cm 0cm, width=8.5cm]{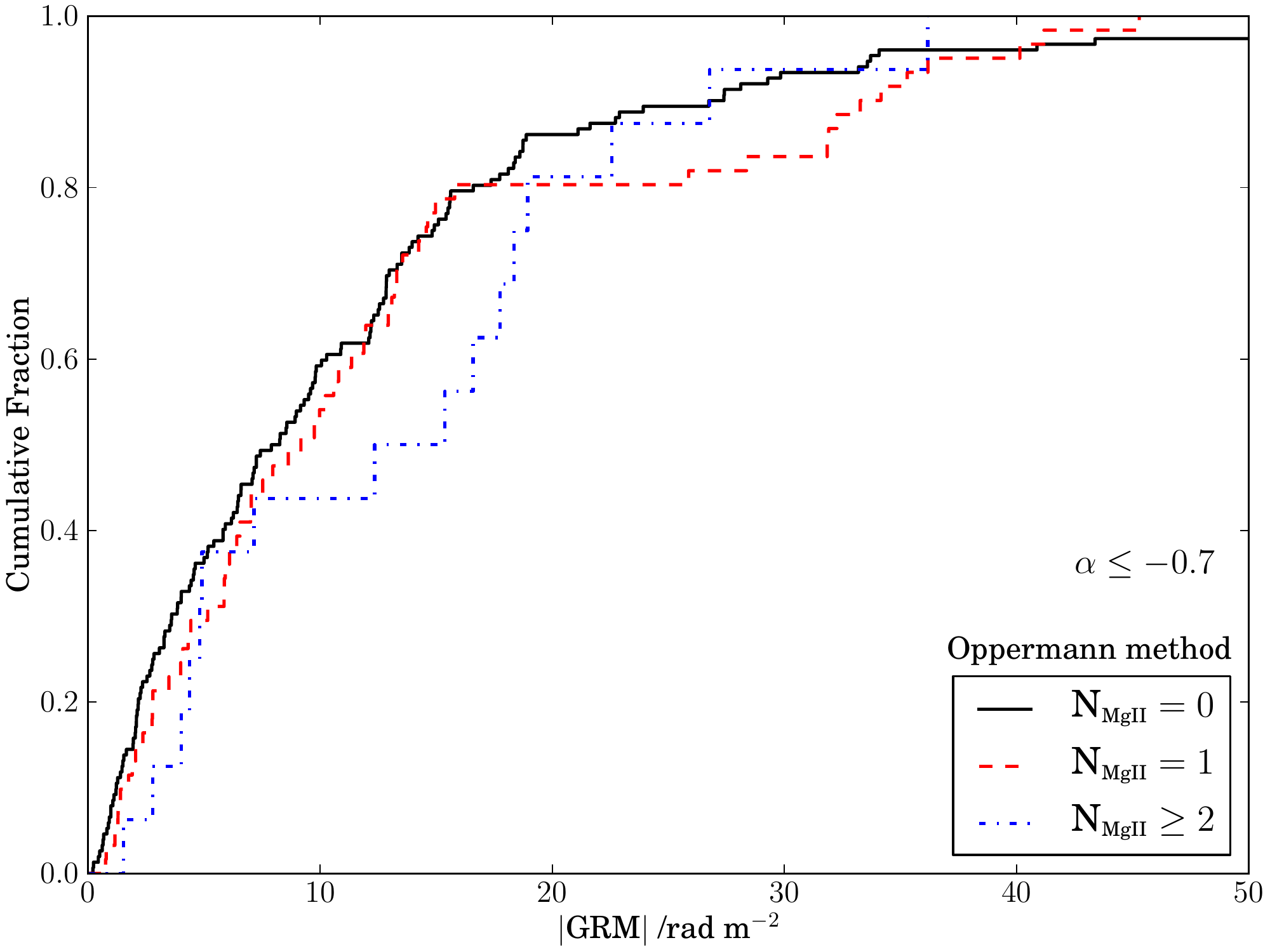}
\caption{ECDFs of the absolute value of the Galactic foreground RMs as calculated using two different algorithms: (i) Left column: the `mean RM' method, (ii) Right column: the \citet{2014arXiv1404.3701O} method. The top panels show the flat-spectrum sources only, while the bottom panels show the steep-spectrum sources only. The black solid lines show the sources without MgII absorption along the line of sight, the red dashed lines show the sources with $1$ absorbing system, and the blue dotted lines show the sources with $\ge2$ absorbing systems. There is no significant difference between any of the data.}
\label{GRMFarnes}
\end{figure*}

The ECDFs of the GRM as calculated using the `mean RM' algorithm, for flat- and steep-sources with $N_{\textrm{MgII}}=0$, 1, and 2 respectively, are shown in Fig.~\ref{GRMFarnes}. The difference in GRM of flat-spectrum sources with $N_{\textrm{MgII}}=0$ versus 1, and also $N_{\textrm{MgII}}=0$ versus 2 absorbers, gives $p$-values of 19\% and 69\% respectively. The difference in GRM of steep-spectrum sources with $N_{\textrm{MgII}}=0$ versus 1, and also $N_{\textrm{MgII}}=0$ versus 2 absorbers, gives $p$-values of 68\% and 11\% respectively. The difference in GRM of flat- and steep-spectrum sources gives a $p$-value of 79\%.

The ECDFs of the GRM as calculated using the \citet{2014arXiv1404.3701O} algorithm, for flat- and steep-sources with $N_{\textrm{MgII}}=0$, 1, and 2 respectively, are also shown in Fig.~\ref{GRMFarnes}. The difference in GRM of flat-spectrum sources with $N_{\textrm{MgII}}=0$ versus 1, and also $N_{\textrm{MgII}}=0$ versus 2 absorbers, gives $p$-values of 60\% and 41\% respectively. The difference in GRM of steep-spectrum sources with $N_{\textrm{MgII}}=0$ versus 1, and also $N_{\textrm{MgII}}=0$ versus 2 absorbers, gives $p$-values of 74\% and 23\% respectively. The difference in GRM of flat- and steep-spectrum sources gives a $p$-value of 53\%.

There is no detectable difference for the GRM of different sources in our sample. This is also independent of the reconstruction algorithm used to calculate the Galactic foreground. Any difference due to the GRM is therefore unable to recreate our result presented in Section~\ref{flatsteepRM}. This is consistent with the Galactic foreground not being responsible for flat-spectrum sources with intervening absorbers having increased RM. 

%==============================================================================%
\section{Discussion}
\label{discussion}
\subsection{Magnetic Fields in Intervening Galaxies}
There are several apparent correlations between variables in Section~\ref{analysis} and~\ref{subsamples} with $p$-values at the $\approx5$\% to 10\% level. While these correlations may be real, one should remain suspicious. As we use $p$-values, any significance is subjective. Such effects, if they exist at all, must be very weak, and we believe such a low significance level to be most consistent with the null hypothesis being true, i.e.\ there is no connection between the two variables \citep[e.g.][]{johnson2013}. On this basis, we have only two statistically significant results, (i) the flat-spectrum sources are consistent with intervening MgII absorption systems increasing the measured RM at 1.4~GHz towards background quasars, while the same correlation is not seen for steep-spectrum sources (as discussed in Section~\ref{flatsteepRM}), and (ii) both the redshifts of flat- and steep-spectrum sources are consistent with lines of sight with higher numbers of MgII absorbers tending to be located at significantly higher redshift (as discussed in Section~\ref{flatsteepredshift}). 

One could argue that such results are caused by the Galactic foreground, but we note that this is inconsistent with our observational findings (see Section~\ref{foreground}). One could also argue that such results are contrary to previously suggested partial coverage models (which imply no connection between RM and MgII absorption at 1.4~GHz; see Section~\ref{introduction}), but we note that although such models are useful for parameterizing the run of polarized fraction with wavelength, they do not describe a physical depolarization model (see Appendix~\ref{theory}). We hypothesize that such partial coverage models are actually the result of probing different emitting regions within the source at different observational frequencies, rather than the effects of any foreground Faraday screen. Importantly, this suggests that in many cases the polarized fraction, which is typically used to estimate the degree of order of magnetic fields, is not a physically meaningful quantity for an \emph{unresolved} source (see Appendix~\ref{theory}).

Nevertheless, as discussed in Section~\ref{combowavelengths}, flat-spectrum sources ($\alpha\approx0$) can be used as a proxy for a source from which the emission is dominated by the core-region, while steep-spectrum sources ($\alpha\approx-0.7$) are dominated by emission from the region of the lobes and jets. By applying this interpretation, let us therefore consider the correlation between MgII absorption and RM in flat-spectrum, aka core-dominated, sources. In the event that no correlation exists, we would expect to have detected a signal at least this strong for just $\approx$1 in 2,250 experiments -- making the result equivalent to a 3.5$\sigma$ event from a normally distributed process. We have therefore either observed a low probability event, or it must be true that there is a connection between RM and MgII absorption in core-dominated sources, while not in jet/lobe-dominated sources. Such evidence suggests that the spectral index is important for discriminating between core- and lobe-dominated sources, and is a reasonable proxy for matching lines of sight at different wavelengths.

As the intervening MgII absorbers are identified towards quasar cores at optical wavelengths, the intervenors can only be said to be obscuring the core at radio wavelengths. An optically selected MgII absorber does not provide an indication of the presence of intervenors along the line of sight towards the lobes/jets. We therefore form three conclusions from Fig.~\ref{ecdfs_cores_lobes}: 
\begin{compactenum}[i]
\item The observed difference in RM between flat-spectrum sources with and without absorption arises due to intervening magnetized plasma in the absorbing systems -- the flat-spectrum of the source ensures that we probe the same line of sight towards the background quasar independently of projection and resolution effects,
\item There is no difference in RM between steep-spectrum sources with and without absorption as we are usually probing different lines of sight at optical and radio wavelengths -- this effect is particularly important at longer radio wavelengths where the steep-spectrum lobes/jets tend to dominate the radio emission,
\item Any difference detected between the flat- and steep-spectrum sources without absorbers could be due to two effects. Firstly, the steep-spectrum sources are only nominally identified as having no absorption. In reality, we likely have not accurately identified the same optical and radio sight line for the steep-spectrum sources. Secondly, the very high RM components of the core may have depolarized at 1.4~GHz, so that only lower RM components are still observable.
\end{compactenum}

To provide a quantitative estimate of the excess RM associated with intervening galaxies, we assume that the correlation between MgII absorption and RM for flat-spectrum sources has an entirely physical origin. We calculate the median RMs in order to ensure robustness, and as a simplifying assumption use Gaussian statistics to calculate the $1\sigma$ uncertainties. Using the difference observed in flat-spectrum sources (as shown in Fig.~\ref{ecdfs_cores_lobes}), we therefore obtain RM[N$_{\text{MgII}}$=0] $=11.4\pm2.2$~rad~m$^{-2}$, RM[N$_{\text{MgII}}$=1] $=18.3\pm2.6$~rad~m$^{-2}$, and RM[N$_{\text{MgII}}\ge2$] $=25.4\pm3.3$~rad~m$^{-2}$ for lines of sight with 0, 1, and $\ge$2 absorption lines respectively. The simplest estimate of the intervening contribution is given by RM[N$_{\text{MgII}}$=1]$-$RM[N$_{\text{MgII}}$=0], so that our data suggests the excess RM associated with a typical intervening system is $6.9\pm1.7$~rad~m$^{-2}$ in the observing frame. This is consistent with previous estimates of excess extragalactic contributions to the RM \citep{2012arXiv1209.1438H}.

For lines of sight with just one absorbing system, the median redshift of the intervening galaxies is $0.87\pm0.06$. Therefore assuming that the Faraday rotation is a linear function with $\lambda^2$, this implies an RM contribution of the order $24\pm6$~rad~m$^{-2}$ in the source rest-frame for a typical intervening cloud of magnetized plasma. This is generally lower than previous estimates obtained at higher radio frequencies, that have estimated rest-frame contributions from $115^{+45}_{-30}$~rad~m$^{-2}$ \citep{2008ApJ...676...70K} to $140^{+80}_{-50}$~rad~m$^{-2}$ \citep{2008Natur.454..302B}. Our estimate improves upon these earlier works as we have both higher statistical significance and have also been able to separate the source contributions based on the total intensity spectral index, although it may also imply some Faraday complexity (see Section~\ref{introduction}).

Following the model presented in \citet{2008Natur.454..302B}, and assuming that MgII absorbing systems with rest-frame equivalent widths between 0.3 to 0.6~$\AA$ are associated with galaxies with a neutral-hydrogen column density of 10$^{19}$~cm$^{-2}$ and a hydrogen ionization fraction of 0.90, we estimate that the typical magnetic field strength associated with each of the intervening systems is $\bar{B}=1.8\pm0.4$~$\upmu$G. Consequently our data are consistent with, and provide the strongest statistical indication to date for, the idea that magnetic fields of substantial strength and coherence were present in normal galaxies in the distant Universe \citep[e.g.][]{1982ApJ...263..518K,1984ApJ...279...19W,1990ApJ...355L..31K,1991MNRAS.248...58W,1995ApJ...445..624O,2008Natur.454..302B,2008ApJ...676...70K,2012ApJ...761..144B,2013ApJ...772L..28B,2013MNRAS.434.3566J}.

%
% Tired of this now. Probably going to have a nap.
%
We cannot currently calculate any physical quantities from the steep-spectrum sources, as we argue that we do not have reliable measurements of whether these sources are truly covered with an absorber. However, as $\approx50$\% of sources are believed to have an intervening absorber \citep{2013ApJ...770..130Z}, one would expect both of the steep-spectrum distributions to sit between the flat-spectrum distributions, which is entirely consistent with the observations (see Fig.~\ref{ecdfs_cores_lobes}).

Our result suggests that different source-components, and consequently different lines of sight, result in the MgII absorption versus RM signal being diluted at 1.4~GHz unless core- and lobe-type sources are considered separately. This divide suggests that either the typical intervenor must be small in angular size relative to the size of the background galaxy, that there is a sharp boundary to the magnetoionic medium in the intervenor, or that the MgII absorbing gas is highly localised within a host galaxy. However, even when separating the sample based on total intensity spectral index, we find no difference in the depolarization of either flat- or steep-sources that have intervening absorption (see Section~\ref{depol_section}). If any depolarization is present due to intervenors, the contribution must be weak. This suggests the magnetic field in the typical intervening galaxy is regular and ordered, at least within the region that is illuminated by background emission.

\subsection{Correlation versus Causation}
\label{correlationcausation}
It is possible to conflate correlation and causation, and so we also examine the possibility that our results could be obtained through systematic effects or confounding variables within our data. There are a number of possible ways in which spurious correlations could be detected in our data, given the presence of some confounding variable. We now explore alternate hypotheses that may explain our finding that our data are consistent with flat-spectrum sources with intervening absorbers having increased RM.

\subsubsection{Evolution of Faraday Rotation with Redshift}
\label{evolvingfaraday}
The most obvious alternate cause for our observed main result would be that the probability that a quasar line of sight intersects an MgII absorber increases as a function of $z$, and that some effect unrelated to the intervenors causes $|\text{RM}|\propto z$. The former is observed in our data as shown in Fig.~\ref{ecdfs_redshift}. Consequently any other property that causes the |RM| to scale positively with $z$ will also manifest as a correlation between RM strength and the number of intervenors. This evolution in |RM| may be due to change in either the integrated magnetic field strength, the electron density, or both along the line of sight.\footnote{Any evolution can in principle be detected, with the exception of special cases where $B_{\parallel}$, $n_{e}$, and the $(1+z)^{-2}$ dilution factor evolve in such a way that the observed-frame RM remains approximately constant. The physics of such models and their applicability to our data, particularly when adding additional intervening contributions to the RM that are each located at different redshifts, is beyond the scope of this paper.}

In such cases, the distribution in $z$ of the sample will determine the magnitude of the spurious correlation, i.e.\ a small range or a uniform distribution in $z$ yields a weak correlation, while conversely a wide range or a highly non-uniform distribution in $z$ yields a strong correlation. To test this, one would ideally resample the data by redshift-binning the flat- and steep-spectrum sources into equal bins, thereby providing a uniform distribution as a function of redshift. We are unable to do this and form firm conclusions as we increasingly fall into the realm of small-sample statistics. However, we are still able to discard such a possibility, and note that a connection between RM and the number of MgII absorbers (as in Fig.~\ref{ecdfs_cores_lobes}) is not detected for the steep-spectrum sources, despite both the flat- and steep-spectrum sources showing a similar relation between the number of absorbers as a function of $z$ (as in Fig.~\ref{ecdfs_redshift}). A KS-test comparing the distribution of redshifts for flat- versus steep-spectrum sources yields a $p$-value of 7.6\% for zero absorbers, 81\% for one absorber, and 99\% for two or more absorbers. Such low significance levels are most consistent with the null hypothesis being true, i.e.\ there is no quantitative difference between the redshift distributions of the flat- and steep-spectrum sources. This suggests a longer line of sight towards a source is not in itself responsible for an increase in RM within our data, and that while sources with more absorbers in our sample are located at higher redshifts, this is equally true for both the flat- and steep-spectrum sources. A spurious correlation that arises due to the sources with absorbers being located further away is therefore not consistent with the observed difference between flat- and steep-spectrum sources. 

To explain the observed difference between flat- and steep-spectrum sources, given the distribution of our sample with redshift, therefore requires another hypothesis. Consider if the flat-spectrum sources were evolving as a function of $z$, while the steep-spectrum sources were not evolving. In such a scenario, we propose that some mechanism causes flat-spectrum sources to have a higher observed-frame RM at higher redshift. We now discuss the possibilities that could cause the observed-frame RM to appear to be scaling as a function of $z$ in flat-spectrum sources, while not in steep-spectrum sources. These explanations broadly fall into two categories: (i) systematic observational effects, and (ii) astrophysical source evolution. Note that we have already discussed the separate possibility of an increased likelihood of a line of sight intersecting an intervenor at higher redshifts.

First we consider systematic observational effects. Our data contain the typical luminosity--redshift degeneracy that is inherent to a flux-limited survey such as the NVSS, as lower luminosity sources at high $z$ are not detected in a flux-limited survey. This selection-effect is commonly termed Malmquist bias. The observed correlation between the number of MgII absorbers and RM for flat-spectrum sources in our sample, therefore implies a \emph{smaller} RM along the line of sight towards the \emph{fainter} flat-spectrum sources, or conversely a \emph{larger} RM towards the \emph{brighter} flat-spectrum sources. Overall, this requires some mechanism that causes the measured RM at fixed-frequency towards flat-spectrum sources to tend to be smaller for low luminosity sources; these sources then drop below the detection threshold at high $z$ and we perceive a net increasing RM. This would be consistent with relativistic beaming models that state that compact, flat-spectrum radio sources are seen at small viewing angles, with weaker cores being seen at progressively larger viewing angles \citep[e.g.][]{1987MNRAS.228..203S}. This suggests that: (a) those weaker cores at larger angles are poorly sampled at high $z$ within a flux-density limited survey (leading to a luminosity effect), and (b) the line of sight through host galaxies with large viewing angles could possibly be reduced, giving rise to lower values of RM for these cores at larger angles (depending on the distribution of electrons and magnetic field near to the central engine). Note that such a scenario need not affect the RM observed towards steep-spectrum sources. Nevertheless, such an effect would appear to be in contradiction to previous studies which have directly measured the evolution of RM as a function of $z$, using either independent data or the same data as in our sample -- with both analyses finding no significant evolution with either redshift or luminosity \citep{2004ApJ...612..749Z,2012arXiv1209.1438H}. More recently, it has been argued that a correlation may have been found, but no analysis of the statistical significance or of confounding factors is currently available \citep{2014arXiv1407.3909P}.

A second possibility is evolution of the sources themselves. If the RM towards quasar cores increased with $z$, while the RM towards lobes/jets remained approximately constant, this would explain the observed correlation. Note that as the same effect is not seen for lobe-dominated sources, this rules out the additional contribution to the RM being located anywhere along the line of sight, and implies that flat-spectrum sources have a higher RM at high $z$ that originates within the local source environment. This is in direct contradiction to theoretical expectations \citep{1996ARA&A..34..155B}, current observational data \citep[e.g.][]{2012arXiv1209.1438H,xu2}, and the expected strong $(1+z)^{-2}$ cosmological dilution effect which arises due to cosmic expansion \citep[e.g.][]{2006ApJ...641..710L} -- all of which suggest that observed-frame RMs, as a proxy for magnetic fields, should be smaller at earlier epochs.

\begin{figure}
\centering
\includegraphics[clip=true, trim=0cm 0cm 0cm 0cm, width=8.5cm]{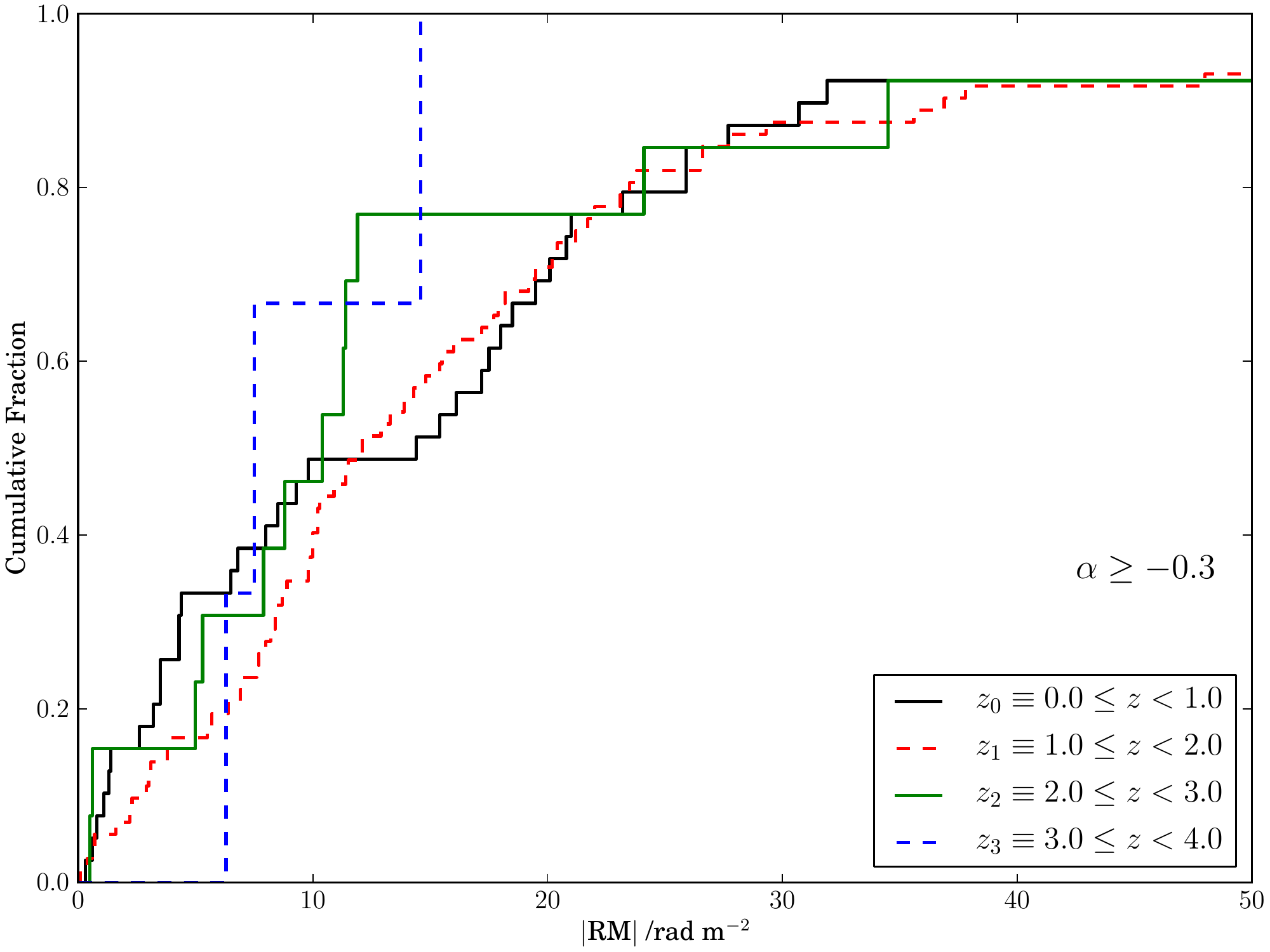}
\bigskip
\includegraphics[clip=true, trim=0cm 0cm 0cm 0cm, width=8.5cm]{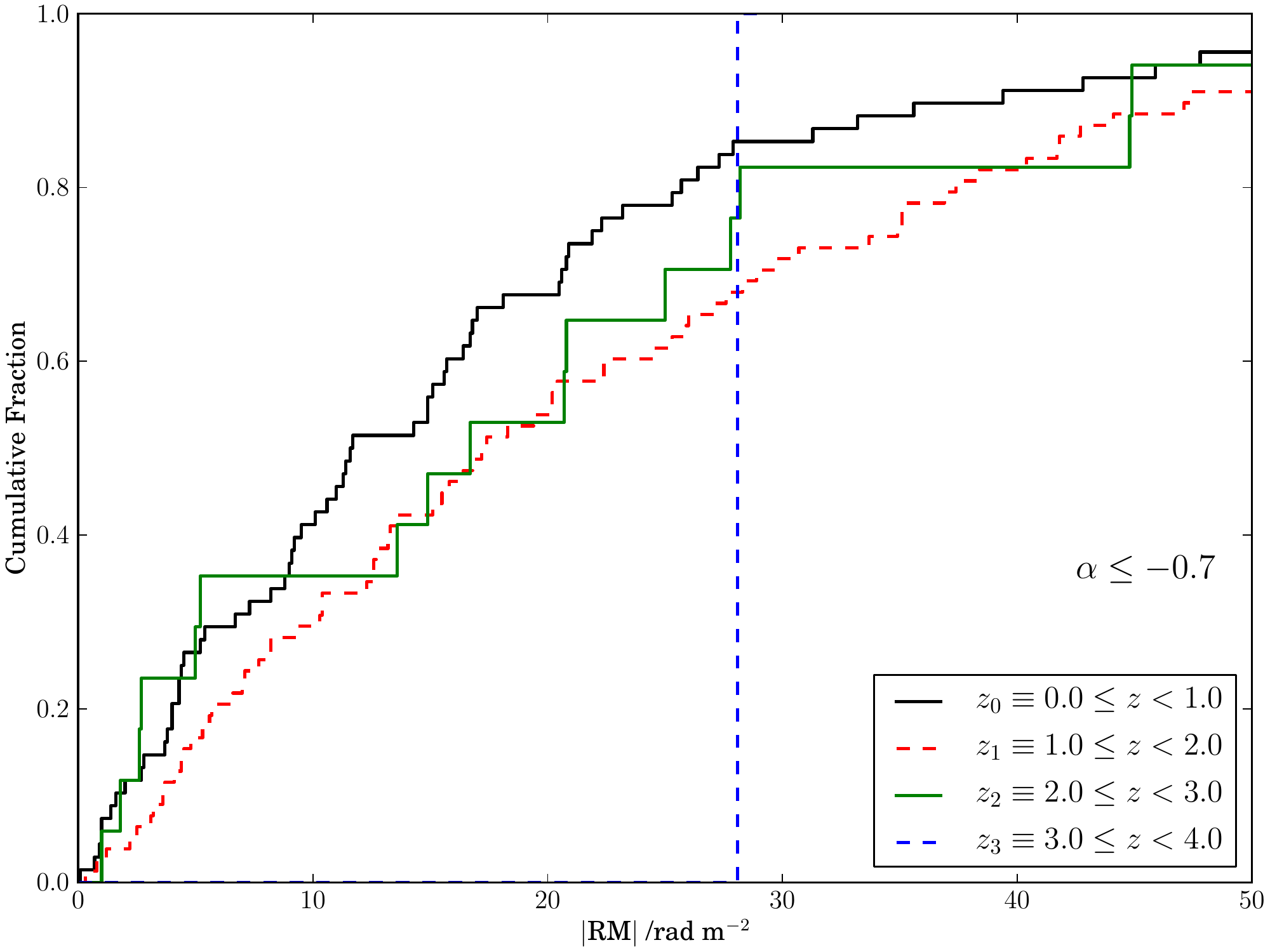}
\caption{ECDFs of the absolute value of the NVSS RMs as a function of the background quasar redshift, $z$. Only sources that have a `clean' line of sight are shown, thereby probing the environment of the background quasars, i.e.\ sources with $N_{\textrm{MgII}}=0$. Both flat- (top panel) and steep-spectrum (bottom panel) sources are shown. The quasars are separated into four redshift bins: $0.0\le z < 1.0$ (black solid line), $1.0\le z < 2.0$ (red dashed line), $2.0\le z < 3.0$ (green solid line), and $3.0\le z < 4.0$ (blue dashed line), that in the main text we refer to as $z_{0}$, $z_{1}$, $z_{2}$, and $z_{3}$ respectively. There is no significant difference between any of the data.}
\label{ecdfs_rmversusz}
\end{figure}

To investigate all of these possibilities, we use the data to look for an evolution of RM versus $z$ in both the flat- and steep-spectrum sources of our sample. We split the data into four bins using the background quasar redshifts: $0.0\le z < 1.0$, $1.0\le z < 2.0$, $2.0\le z < 3.0$, and $3.0\le z < 4.0$, which we shall here refer to as $z_{0}$, $z_{1}$, $z_{2}$, and $z_{3}$ respectively. As it is possible that previous studies of RM versus $z$ have been affected by contributions from intervening sources along the line of sight, we only use sources that have no detected MgII absorber along the line of sight. To date, no study has been able to investigate the evolution of RM versus $z$ for `clean' lines of sight, allowing us to attempt to probe evolution in the local environment of the background quasars themselves. Note that all of our sample are optically identified as quasars in the SDSS (see Section~\ref{data}). Our results are shown in Fig.~\ref{ecdfs_rmversusz}. For flat-spectrum sources, there is a $p$-value of 38\% between $z_{0}$ and $z_{1}$, of 36\% between $z_{0}$ and $z_{2}$, and of 39\% between $z_{0}$ and $z_{3}$. For steep-spectrum sources, there is a $p$-value of 14\% between $z_{0}$ and $z_{1}$, of 75\% between $z_{0}$ and $z_{2}$, and of 23\% between $z_{0}$ and $z_{3}$. There is no statistically significant difference between any of the ECDFs and thus there is no detectable evolution of RM as a function of $z$ in our sample. This is consistent with luminosity effects or source evolution not being responsible for flat-spectrum sources with intervening absorbers having increased RM. One could argue that variations in the GRM (see Section~\ref{foreground}) could mask any RM variation with redshift. While the results presented in Fig.~\ref{ecdfs_rmversusz} cannot rule this out, this scenario still cannot explain how flat-spectrum sources with absorbers have higher RMs, as it would require selection of different GRMs for sources with different numbers of absorbers. Fig.~\ref{GRMFarnes} shows that there is no observable difference in GRM between these same flat-spectrum sources both with and without absorbers, which suggests that the GRM is not affecting our results.

We note that such an analysis cannot be trivially performed for sources with intervening absorbers, as the measured RM then becomes a combination of the RM at the quasar plus then presumably additional components from the multiple absorbers themselves, all of which are located at different redshifts. Our analysis is further complicated by bandwidth depolarization in the NVSS sample for sources with |RM|$\ge350$~rad~m$^{-2}$ \citep{2009ApJ...702.1230T}, as such high RMs could be entirely located at high or low redshift. Nevertheless, this cannot be a significant effect in the RM ECDFs unless a very large fraction of our sources had such an RM -- which is unlikely given our removal of sources at low Galactic latitudes.

\subsubsection{Systematic Observational Effects}
In Section~\ref{evolvingfaraday}, we discussed systematic observational effects such as Malmquist bias that may have generated the observed correlations in our data. We now discuss another systematic effect that could possibly explain our observation that flat-spectrum sources with intervening absorbers having increased RM. This effect is directly related to the signal--to--noise ratio (s/n) of each measurement.

In this alternative hypothesis, we suggest that the more distant sources tend to be fainter in either total or polarized intensity. As the sources with more absorbers are located more distantly in our sample, this could lead to a systematic error. As the NVSS RMs are calculated using two closely-spaced narrow bands \citep{2009ApJ...702.1230T}, a decrease in the s/n could lead to anomalous RM measurements. In this case, the polarized intensity from which the RM is determined, serves as a proxy for the s/n.

\begin{figure*}
\centering
\includegraphics[clip=true, trim=0cm 0cm 0cm 0cm, width=8.5cm]{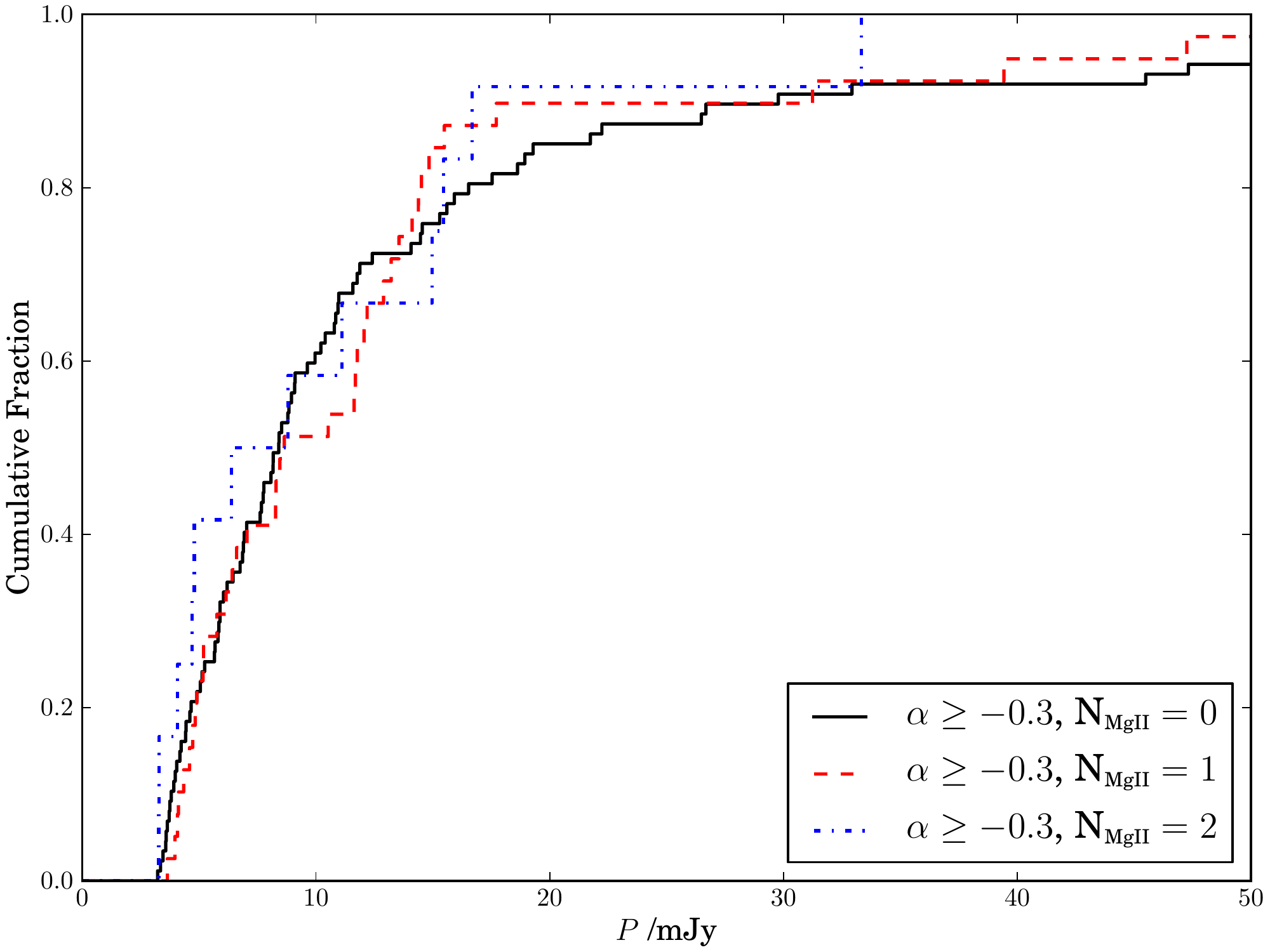}
\includegraphics[clip=true, trim=0cm 0cm 0cm 0cm, width=8.5cm]{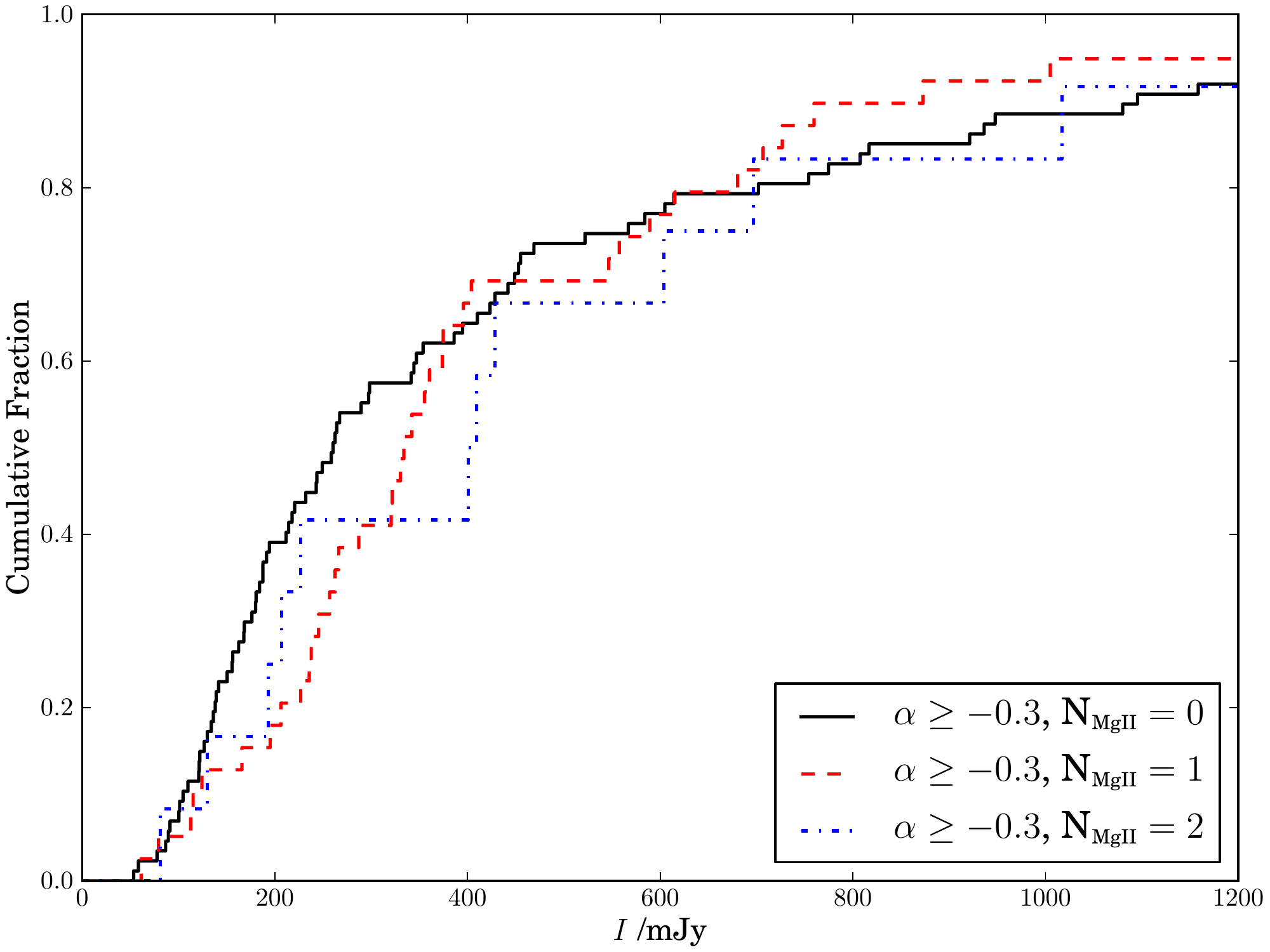}
\bigskip
\includegraphics[clip=true, trim=0cm 0cm 0cm 0cm, width=8.5cm]{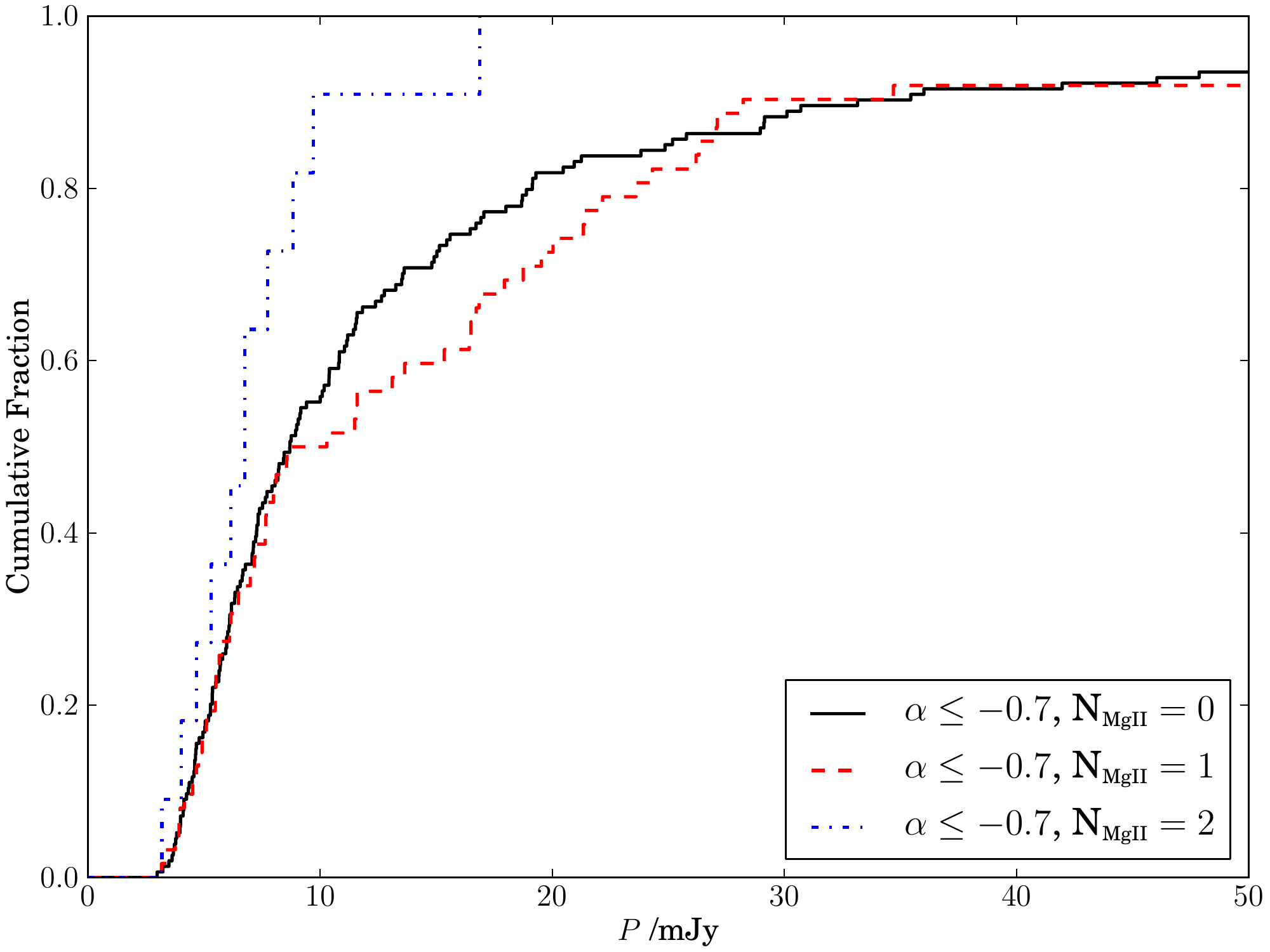}
\includegraphics[clip=true, trim=0cm 0cm 0cm 0cm, width=8.5cm]{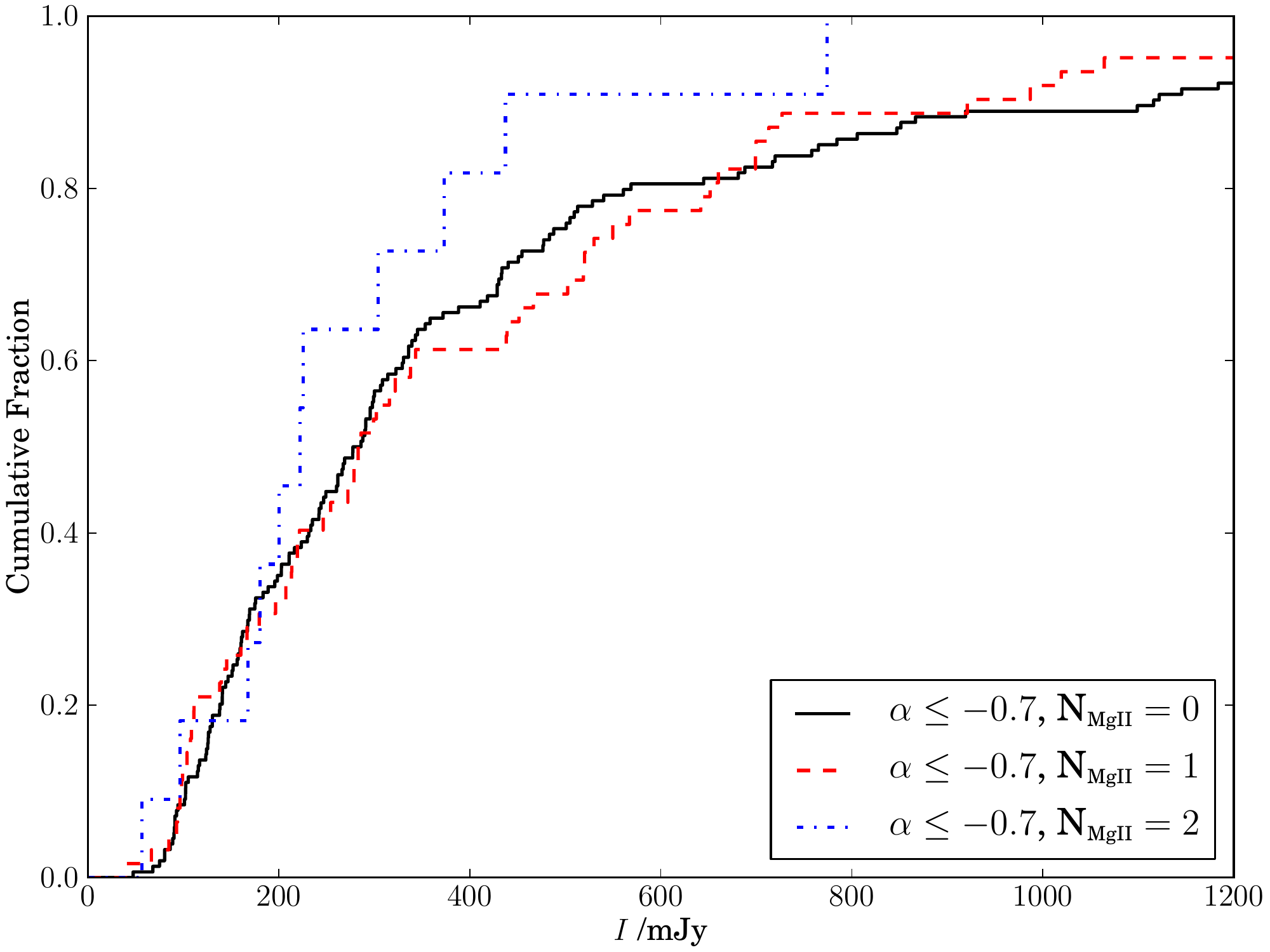}
\caption{ECDFs of the NVSS polarized intensity (left column) and total intensity (right column) for different numbers of MgII absorbers. Both flat-spectrum ($\alpha\ge-0.3$, top row), and steep-spectrum ($\alpha\le-0.7$, bottom row) subsamples are shown. Data are shown for sources with $N_{\textrm{MgII}}=0$ (black solid line), $N_{\textrm{MgII}}=1$ (red dashed line), and $N_{\textrm{MgII}}=2$ (blue dot-dashed line). The polarized and total intensity both serve as a proxy for the s/n of each measurement -- particularly the polarized intensity, from which the NVSS RMs are derived. There is no significant difference between any of the data.}
\label{ecdfs_signalnoiseversusRMz}
\end{figure*}

To investigate this possibility, we plotted the ECDFs of the polarized and total intensity for flat- and steep-spectrum sources with different number of MgII absorbers as shown in Fig.~\ref{ecdfs_signalnoiseversusRMz}. There are a number of different data subsets: (i) For the polarized intensity data from flat-spectum sources, there is a $p$-value of 53\% between sources with zero and one absorbers, and a $p$-value of 69\% between sources with zero and two absorbers. For the polarized intensity data from steep-spectum sources, there is a $p$-value of 35\% between sources with zero and one absorbers, and a $p$-value of 11\% between sources with zero and two absorbers. The $p$-value between the polarized intensity of the flat- and steep-spectrum sources themselves is 91\%, (ii) For the total intensity data from flat-spectum sources, there is a $p$-value of 8.2\% between sources with zero and one absorbers, and a $p$-value of 59\% between sources with zero and two absorbers. For the total intensity data from steep-spectum sources, there is a $p$-value of 80\% between sources with zero and one absorbers, and a $p$-value of 50\% between sources with zero and two absorbers. The $p$-value between the total intensity of the flat- and steep-spectrum sources themselves is 96\%. There is therefore no statistically significant difference in either the polarized or total intensity of sources with different numbers of absorbers in our sample -- our main results are therefore not caused by the effects of s/n.

%\newpage
\subsubsection{k-corrected Polarized Quantities}
\label{toymodels}
All of the aforementioned possibilities neglect the necessity of $k$-corrections to observed polarization quantities. It is possible that wavelength-dependent polarization structure in the nucleus can mimic Faraday rotation -- particularly through the combined interplay of synchrotron self-absorption and depolarization within a compact emitting region \citep[e.g.][]{1988BAAS...20..971O}. Broadband observations may therefore be able to detect Faraday complexity in flat-spectrum sources (i.e.\ a non-linear relationship between polarization angle and $\lambda^2$). There is already some tentative evidence to support this \citep{osullivanpaper,2012MNRAS.421.3300O,farnescatalogue}. Note that it has been previously suggested that sight lines with intervening systems that exhibit Faraday complexity are also associated with low fractional polarization \citep{2012ApJ...761..144B}. It could be attempted to explain this as a selection effect, with the high-frequency RMs selecting for the flat-spectrum component of the source, and the 1.4~GHz RMs selecting for the steep-spectrum component -- thereby generating a pseudo Faraday complexity that arises from the sampling of different emitting regions. Nevertheless, this does not explain why these same sources have a low fractional polarization. However, flat-spectrum NVSS sources have also been shown to have a lower median fractional polarization than steep-spectrum sources \citep{2002A&A...396..463M,2014ApJ...787...99S}. We therefore suggest the alternative hypothesis that Faraday complex sources may be intrinsically flat-spectrum, and that splitting the NVSS sample by fractional polarization, as done by \citet{2012ApJ...761..144B}, selects these flat-spectrum sources. In such cases, as the NVSS RM is measured at fixed-frequency, we would be sampling different regions of this polarization structure at different redshifts, as seen in the source rest-frame. In such a case, the necessary $k$-correction would not just be equivalent to a cosmological dilution factor of $(1+z)^2$, but would rather be a consequence of redshifting a curvilinear run of polarization angle versus $\lambda^2$, while observing with a fixed narrow bandwidth \citep{farnescatalogue}.

We could therefore consider a convoluted toy model that requires $k$-corrections to the run of polarization angle versus $\lambda^2$ in a Faraday complex source. Given the narrow bandwidth used to derive the 1.4~GHz RMs in our sample, the data would have to probe progressively higher rest-frame frequencies at high $z$, which would correspond to the region closer to the central engine, which could have undergone less depolarization at high frequency. In turn, this region closer to the central engine could correspond to a greater pathlength through the source environment, which could possibly lead to larger Faraday rotation. Such an observational effect would not affect the lobe-dominated (aka steep-spectrum) sources, that have ordered magnetic fields on large scales and are optically-thin (leading to simple Faraday rotation, and a linear run of polarization angle with $\lambda^2$). While intriguing, our data are inconsistent with this proposed model for a number of reasons. Previous studies have shown that opacity effects are important in the run of polarized fraction SEDs, allowing flat-spectrum sources to repolarize \citep{farnescatalogue}. This is hard to reconcile with the requirement of depolarization in this proposed alternative hypothesis. The proposed toy model also contradicts the observation that weak MgII absorbers are not correlated with RM along the line of sight \citep{2010ApJ...711..380B}, as this further suggests that there cannot be an additional confounding variable, i.e.\ no evolution of quasar magnetic fields with $z$. One could counter that the nature of weak MgII absorbing systems is still poorly understood \citep[e.g.][]{1999ApJS..120...51C,churchill05,narayanan07,kacprzak08}, or that previous studies have not separated sources based on the spectral index. Regardless, the proposed toy model would still require a strong evolution of the observed-frame RM as a function of $z$ for flat-spectrum sources. As shown in Fig.~\ref{ecdfs_rmversusz}, there is no observed RM evolution in the flat-spectrum sources of our sample. There is only one remaining possibility: that our data are consistent with intervening systems, as traced by MgII absorption, containing regular magnetic fields that increase Faraday rotation along the lines of sight towards distant background quasars.

%==============================================================================%
\section{Conclusions}
\label{conclusion}
We have investigated the current theoretical and observational understanding of Faraday effects originating along the line of sight due to intervening heavy-metal absorbing systems. We have divided a sample of flat- and steep-spectrum radio sources into subsamples both with and without MgII absorption along the line of sight. We have been able to use these samples as a proxy for core- (flat-spectrum) or lobe- (steep-spectrum) dominated sources. This has allowed us to study the same sight line at both optical and radio wavelengths. We find that the core-dominated sample has a larger |RM| when intervening MgII absorbers are present, with a probability of 0.044\% of the increase in |RM| being this large or greater if the data were drawn from the same underlying distribution. Conversely to previous studies, which have found no association between MgII absorption and Faraday rotation at 1.4~GHz, we instead find evidence of an association that is stronger than that which has been presented before at any other observing frequency.

We have considered various alternative effects, including varying luminosity in our essentially flux-limited sample, evolution of magnetic fields with redshift, and other more elaborate possibilities that may cause a spurious correlation. We find that none of them are fully consistent with both our data and our theoretical understanding of cosmic magnetism. The simplest way to explain our observations while remaining consistent with previous observational findings is to require the RM to be increased by additional magnetic fields, or ionised gas, that are associated with intervening MgII absorbing systems along the line of sight. If we assume that the correlation between MgII absorption and RM has an origin entirely due to intervening galaxies, then as a quantitative estimate, our data suggest that a typical absorber provides an additional RM contribution of $6.9\pm1.7$~rad~m$^{-2}$ in the observing frame. At the median redshift of our sample, $z=0.87\pm0.06$, this implies an RM contribution of $24\pm6$~rad~m$^{-2}$ from a typical intervening cloud of magnetized plasma in the source rest-frame. Consequently our data are consistent with, and provide the strongest statistical indication to date for, the idea that coherent magnetic fields of substantial strength ($\bar{B}=1.8\pm0.4$~$\upmu$G) are present in what are presumed to be normal galaxies \citep[e.g.][]{2008ApJ...676...70K}. The possibility that Faraday rotation along the line of sight to a typical quasar could be enhanced by an otherwise essentially invisible population of intervening normal galaxies is an intriguing one. The physical implications of this have been rigorously explored elsewhere, providing constraints for our understanding of galaxy formation and evolution, magnetic field generation, and dynamo mechanisms \citep{1982ApJ...263..518K,1984ApJ...279...19W,1990ApJ...355L..31K,1991MNRAS.248...58W,1995ApJ...445..624O,2008Natur.454..302B,2008ApJ...676...70K,2012ApJ...761..144B,2013ApJ...772L..28B,2013MNRAS.434.3566J}.  

Our data complement previous studies by showing that connections between RM and MgII absorption are still detectable at lower radio frequencies, and that the contribution from intervening systems to the overall Faraday rotation along the line of sight must be weak relative to that from the background quasars. It is also indicates the importance of probing similar lines of sight at optical and radio wavelengths, suggesting that projection effects between cores and lobes have been important contributors to previous studies. However, while our method of using the total intensity spectral index to identify the same line of sight at different wavelengths is a significant primary step, we do not currently have the data available to definitively confirm that the polarized emission is coincident with the total intensity emission. Investigating such potential systematics would likely require full reprocessing of surveys such as the NVSS, or the arrival of next generation surveys such as the Polarization Sky Survey of the Universe's Magnetism (POSSUM) that will be carried out with the Australian Square Kilometre Array Pathfinder (ASKAP) \citep{2010AAS...21547013G}.

The significance of the correlation between intervening absorption lines and RM is currently only at a level equivalent to a $3.5\sigma$ event from a normally distributed process, although we note that we have been unable to calculate either a confidence interval, or the probability of the hypothesis. In future studies, a full Bayesian framework would be useful to further analyse our statistical detection. Our data show that connections between intervening systems and Faraday rotation are difficult to detect, due to the multiple effects that may alter the RM at cosmological distances. Placing the interaction on an even firmer statistical footing will require multiple quantities: larger samples of strong MgII absorbers, higher angular resolution radio data, unambiguous RMs, broadband spectral indices, and improved estimates of the Galactic foreground. Future observations with facilities such as the Square Kilometre Array (SKA) will therefore be important in confirming these results with much greater statistical significance, and for determining the physical properties of the intervening systems themselves, such as improved estimates of the typical magnetic field strength, the physical size, and any redshift dependence of the magnetic field properties. The combination of existing radio morphology classifications \citep{2012arXiv1209.1438H}, other radio surveys such as FIRST \citep{1997ApJ...475..479W}, and measurements of the spectral index (see Section~\ref{combowavelengths}), will also form the foundation of a useful future study. The intervenors themselves could also have implications for an SKA `RM-grid' \citep[e.g.][]{2004NewAR..48.1003G}, as RM measurements from core-dominated sources may have a more complex relation to the magnetic field of the Galactic foreground. This would impede attempts to calculate a residual rotation measure (see Section~\ref{foreground}) using multiple lines of sight within some defined region of sky \citep[e.g.][]{2009ApJ...702.1230T,2012arXiv1209.1438H,2014arXiv1404.3701O}. Broadband measurements of core-dominated sources, combined with reconstructions of the Galactic foreground using simulated data, will be required to investigate such possibilities.

Overall, the new evidence presented here rules out models of partial coverage by inhomogeneous Faraday screens (see Section~\ref{introduction}); the justification for such models has been based on the lack of connection between the number of MgII absorbers and the RM at 1.4~GHz. Taken together with the connection between radio depolarization and total intensity spectral index \citep{farnescatalogue}, our results serve as a reminder of the importance of opacity effects on radio polarization measurements. In combination, these results suggest that depolarization is predominantly occurring in the local environment of the background AGN, while the RM is significantly contributed to by the intervening normal galaxy population. The consequences are important for all future and upcoming radio polarimetric studies.

%==============================================================================%
\section*{Acknowledgements}
We are grateful to the anonymous referee for helpful comments on the original version of the manuscript. J.S.F., S.P.O'S., \& B.M.G. acknowledge the support of the Australian Research Council through grants DP0986386, FS100100033, \& FL100100114 respectively. Parts of this research were conducted within the Australian Research Council Centre of Excellence for All-sky Astrophysics (CAASTRO), through project number CE110001020. The National Radio Astronomy Observatory is a facility of the National Science Foundation operated under cooperative agreement by Associated Universities, Inc. This research has made use of: the SIMBAD database, operated at Centre de Donn\'ees astronomiques de Strasbourg, France; the NASA/IPAC Extragalactic Database (NED) which is operated by the Jet Propulsion Laboratory, California Institute of Technology, under contract with NASA; NASA's Astrophysics Data System Abstract Service; the SDSS, for which funding has been provided by the Alfred P. Sloan Foundation, the Participating Institutions, the National Science Foundation, and the U.S. Department of Energy Office of Science.

%==============================================================================%

%==============================================================================%
\appendix

\section{Theory of Partial Coverage}\label{theory}
The potential presence of partial coverage has previously been inferred from polarized spectral energy distributions (SEDs). For some sources, the polarized SEDs have been described by an equation of the form
\begin{equation}
\Pi(\lambda) = \Pi_{0}\exp{(-2\sigma_{\text{RM}}^2\lambda^4)} \,,
\label{burnbabyburn}
\end{equation}
where $\sigma_{\text{RM}}$ is the RM dispersion of the Faraday screen within a single beam, $\lambda$ is the observing wavelength, $\Pi$ is the fractional polarization, and $\Pi_{0}$ is the fractional polarization at infinite frequency. Such external Faraday depolarization was initially proposed by \citet{1966MNRAS.133...67B} -- a `Burn law'. However, while in some sources the polarized fraction can behave similarly to equation~\ref{burnbabyburn} at progressively shorter wavelengths, it remains unexpectedly constant out to longer wavelengths \citep[e.g.][]{rosetti08,mantovani08}. These SEDs that follow a `Rossetti--Mantovani law' have been explained by assuming that only a fraction of the source is covered by an inhomogeneous Faraday screen \citep[e.g.][]{rosetti08}. In an effort to derive $\sigma_{\text{RM}}$ for these sources, \citet{rosetti08} made an empirical modification to the Burn law so that
\begin{equation}
\Pi(\lambda) = \Pi_{0} \left[ f_{\text{c}}\exp{(-2\sigma_{\text{RM}}^2\lambda^4)} + (1-f_{\text{c}}) \right] \,,
\label{partialcoverage}
\end{equation}
where $f_{\text{c}}$ is interpreted as the covered (depolarizing) fraction of the source, with the uncovered fraction $(1-f_{\text{c}})$ retaining a constant polarized fraction, $(1-f_{\text{c}})\Pi_{0}$, out to arbitrarily long wavelengths. This model has been found to be more successful than the Burn law in reproducing the SED of some sources \citep[e.g.][]{mantovani08,rosetti08,farnescatalogue}. As the existence of such partial coverage SED models and the MgII absorption line studies (see Section~\ref{introduction}) both imply partial coverage, this has been taken together to imply that there must be a link between partial coverage SEDs and MgII absorption lines \citep[e.g.][]{2012ApJ...761..144B,2013ApJ...772L..28B} -- with the inference that intervening MgII host galaxies may be responsible for the partial coverage. Nevertheless, while some polarized SEDs show a similar run in the polarized fraction as a function of wavelength as that predicted by equation~\ref{partialcoverage}, due to sample limitations there is currently no direct evidence available in the literature to connect these same sources to the presence of MgII absorption lines. 

Attempts to justify partial coverage models have previously been made using equation \ref{partialcoverage}, as this explains the functional form of some polarized SEDs \citep[e.g.][]{rosetti08,mantovani08,farnescatalogue}. In addition, \citet{2012ApJ...761..144B} found that Faraday complexity appears to be more commonly observed in weakly polarized sources, which led to the proposal of a toy model that suggests this is related to enhanced depolarization of different source components due to partial coverage. However, flat-spectrum sources are also known to be intrinsically more weakly polarized than their steep-spectrum counterparts \citep[e.g.][]{2002A&A...396..463M,2014ApJ...787...99S}. It is therefore plausible that such effects could be caused by, for example, epoch-dependent variability between observations, or an increased presence of Faraday complexity in the SEDs of flat-spectrum sources. Such possibilities provide a more simple alternative to invoking a partial coverage model.

Irrespective of the presence of an intervenor, the majority of radio sources are known to undergo depolarization at increasing radio wavelengths due to coverage by an inhomogeneous `Faraday screen', i.e.\ a magnetoionic region that is devoid of relativistic particles and that exists somewhere along the line of sight between the observer and the source \citep[e.g.][]{1998MNRAS.299..189S}. Here we assume that these Faraday screens are always inhomogeneous and contain a turbulent or systematically varying regular magnetic field, such that the screen causes depolarization, and not just Faraday rotation. Although the location of these screens along the line of sight cannot be trivially determined \citep[e.g.][]{1966MNRAS.133...67B,1991MNRAS.250..726T,1998MNRAS.299..189S}, more recent data suggests that the most significant predictor of the depolarization properties is the total intensity spectral index \citep{farnescatalogue} -- suggesting that the Faraday screens are located within the local source environment and that radio opacity effects are important for polarization studies.

Following the conventional partial coverage model, consider the case where an intervening galaxy is known to be present and is believed to be partially covering the background quasar. The model of equation~\ref{partialcoverage} makes the critical assumption that while a fraction of the source is partially covered by an inhomogeneous Faraday screen from an intervenor, the other fraction is completely uncovered -- with no covering depolarizing screen whatsoever. How this uncovered portion of the source altogether escapes the effects of Faraday screens, and remains depolarization-free, is not explained. As a typical radio source without an associated intervenor is known to be covered by a depolarizing screen \citep[e.g.][]{farnescatalogue}, similarly, portions of a source without an intervening object should also have a similar screen. One could argue that the screen across the uncovered portion of the source is a non-turbulent Faraday screen, that does not depolarize and which only adds additional Faraday rotation, although there are no suitable candidates for such a physical mechanism in this subset of sources. Analogously, equation~\ref{partialcoverage} states that when the covering fraction tends to zero, the polarized fraction will remain constant at all wavelengths. Such a theory is incompatible with the observational evidence, which shows that there is no realistic expectation of detectable polarization at arbitrarily long wavelengths \citep[e.g.][]{2011MNRAS.418.2336A}. We note that all other depolarization models that are typically available in the literature, such as the Burn law, have all been derived from physical principles \citep[e.g.][]{farnescatalogue}.

In order to be physically justified, any partial coverage model must also allow for depolarization in the uncovered fraction of the source, or explain how the uncovered fraction can become immune to the effects of the inhomogeneous Faraday screens that surround a typical radio source. We therefore extend the partial coverage model to include the effect of an inhomogeneous Faraday screen across the uncovered fraction of a source, such that
\begin{equation}
\Pi(\lambda) = \Pi_{0} \left[ f_{\text{c}}\exp{\left( -2\lambda^4 \left[ \sigma_{\text{interv}}^2+\sigma_{\text{norm}}^2 \right] \right) } + \left( 1-f_{\text{c}} \right) \exp{\left( -2\sigma_{\text{norm}}^2\lambda^4 \right) } \right] \,,
\label{partialcoverage2}
\end{equation}
where $\sigma_{\text{norm}}$ is the RM dispersion in the absence of an intervenor (whether this dispersion originates locally to the source, in the Galaxy, or elsewhere), and $\sigma_{\text{interv}}$ is the screen provided by the intervenor and which is allowed to partially cover a fraction, $f_{\text{c}}$, of the background emitting region, i.e.\ a quasar or radio galaxy. We make the reasonable assumption that the intervening and normal screens are independent and uncorrelated, such that the combined RM dispersion of the two overlapping screens is given by $\sigma_{\Sigma}^2 = \sigma_{\text{interv}}^2+\sigma_{\text{norm}}^2$. We also assume that the background emitting region is optically-thin and that opacity effects are negligible. The functional form of the partial coverage model in equation~\ref{partialcoverage2} is advantageous to previous partial coverage models in that it is physically consistent, and when either $f_{\text{c}}=0$ or $\sigma_{\text{interv}}=0$~rad~m$^{-2}$ we retrieve a conventional Burn law. The original partial coverage model in equation~\ref{partialcoverage} is now only a special case, albeit a non-physical one, when $\sigma_{\text{norm}}=0$~rad~m$^{-2}$.

\begin{figure}
\centering
\includegraphics[clip=true, trim=0cm 0cm 0cm 0cm, width=8.5cm]{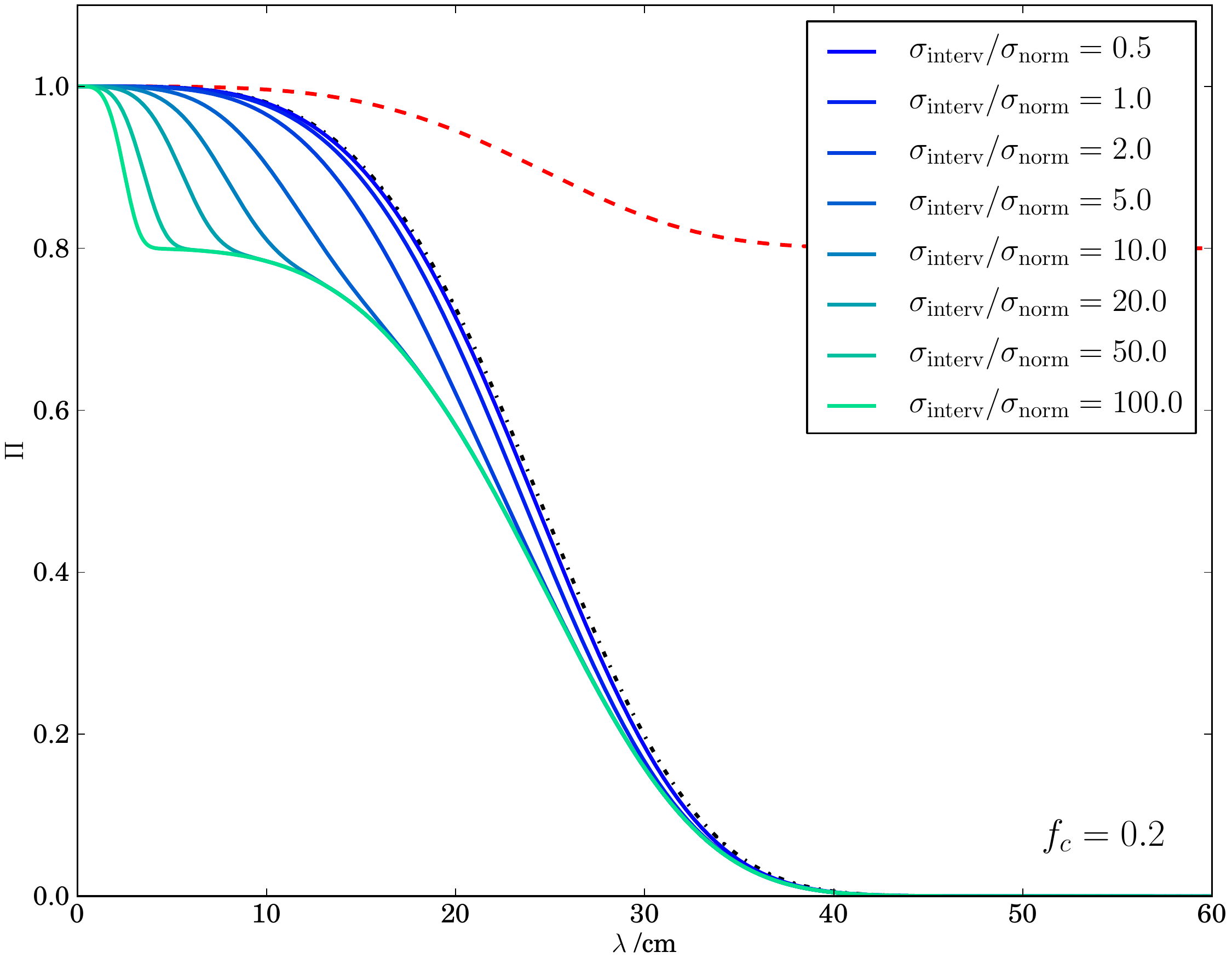}
\includegraphics[clip=true, trim=0cm 0cm 0cm 0cm, width=8.5cm]{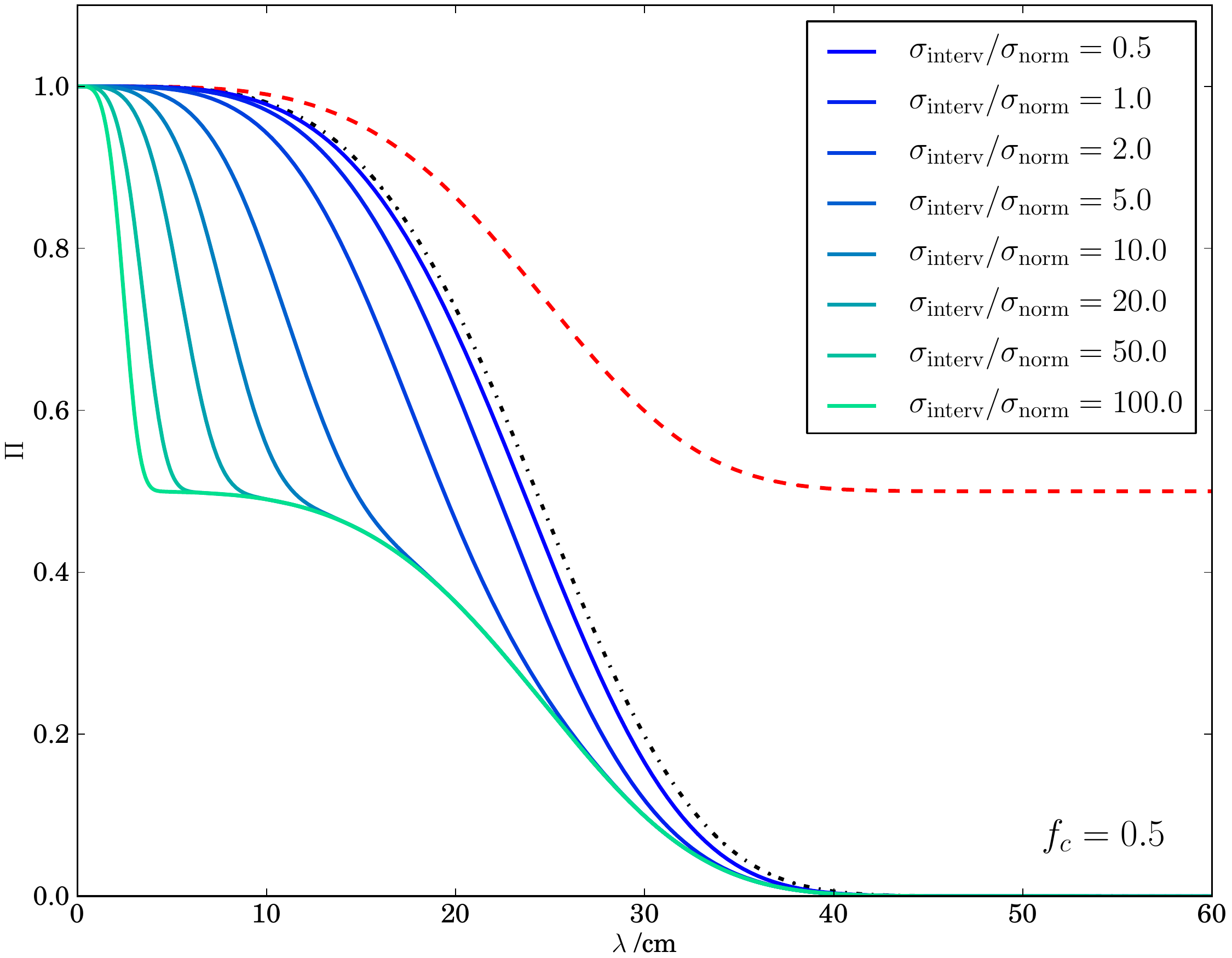}
\includegraphics[clip=true, trim=0cm 0cm 0cm 0cm, width=8.5cm]{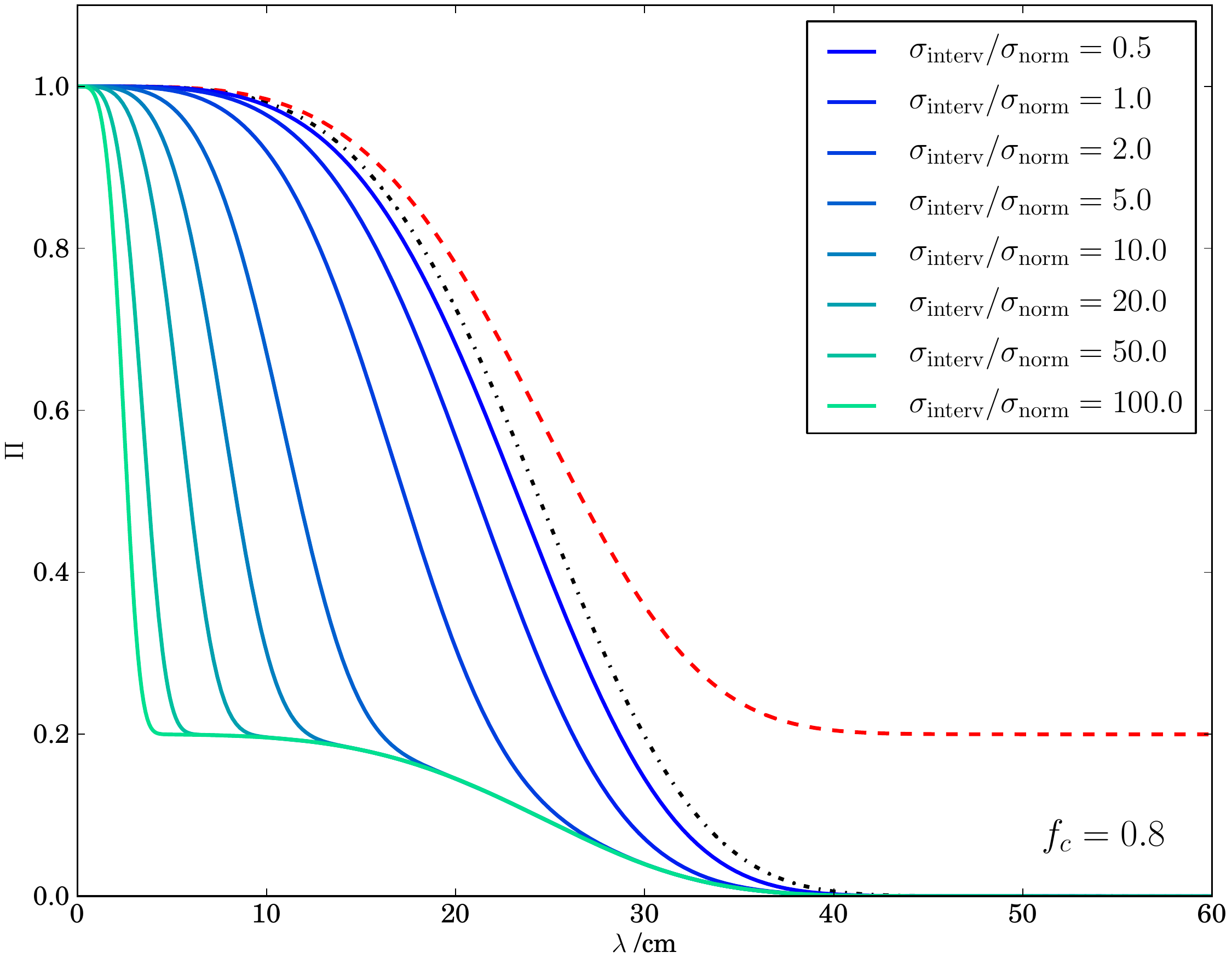}
\includegraphics[clip=true, trim=0cm 0cm 0cm 0cm, width=8.5cm]{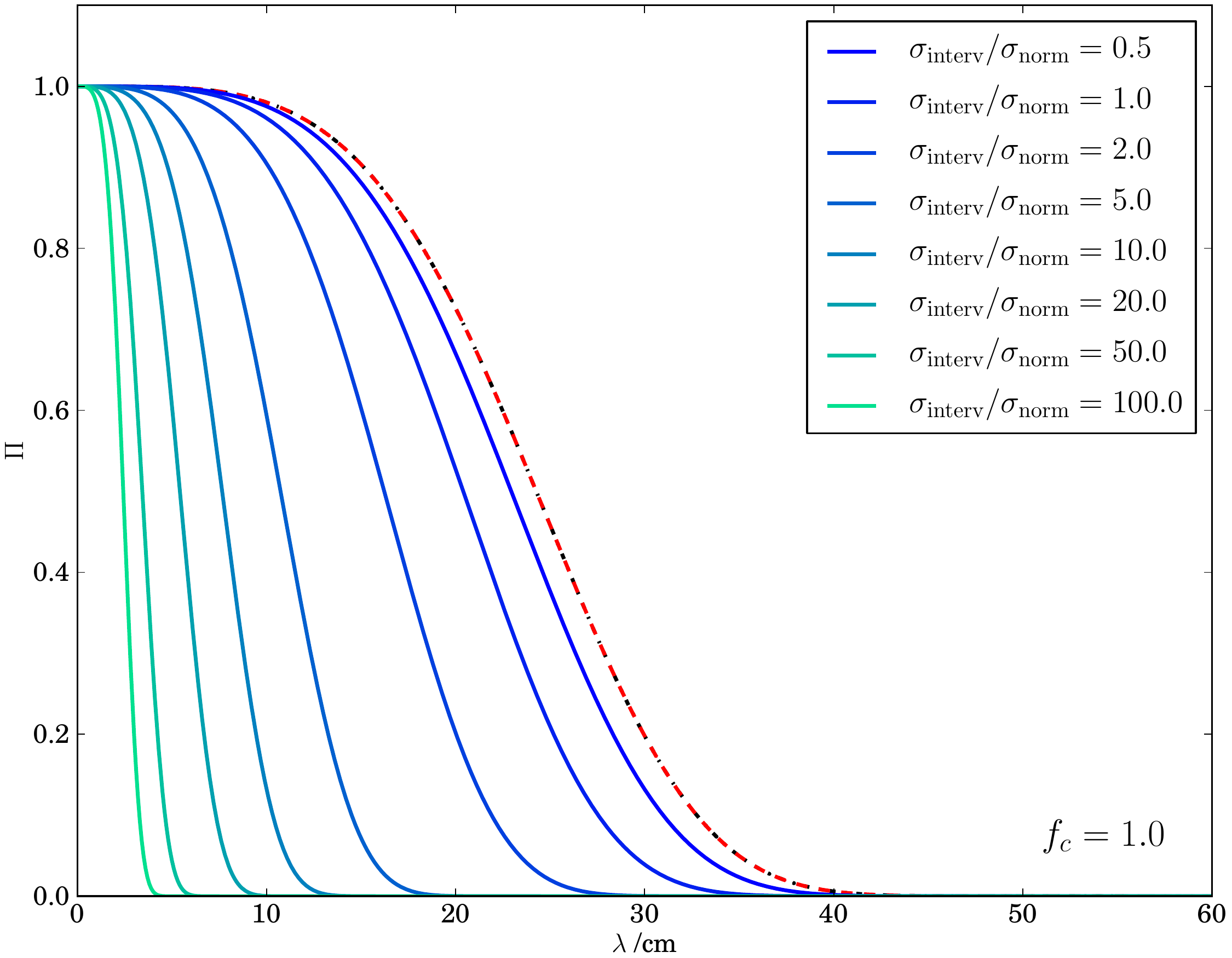}
\caption{A visualisation of polarized SEDs that can arise from partial coverage models, shown as plots of the degree of polarization as a function of wavelength. Such SEDs may arise whenever there are two independent Faraday screens covering different fractions of a source as detailed by equation~\ref{partialcoverage2}. Plots are shown for a covering fraction of $f_{\text{c}}=$0.2 (top left), 0.5 (top right), 0.8 (bottom left), and 1.0 (bottom right). In all plots, the special depolarizing case where there is only an intervening screen across the source, $\sigma_{\text{norm}}=0$~rad~m$^{-2}$ and $\sigma_{\text{interv}}=10$~rad~m$^{-2}$, is shown as a red dashed line. The case where there is no depolarizing intervening screen, $\sigma_{\text{norm}}=10$~rad~m$^{-2}$ and $\sigma_{\text{interv}}=0$~rad~m$^{-2}$, is shown as a black dotted line. The depolarization when both a normal and intervening screen are present is shown by the solid coloured lines (dark to light blue), for several differing contributions from the intervening screen across the covered fraction of the source; in all of these cases the normal screen remains constant. The ratio of the intervening and normal screens, $\sigma_{\text{interv}}/\sigma_{\text{norm}}$, is shown in the legend with darker blue indicating a lower ratio. Even when an infinitesimally small normal screen is present, it is not possible to obtain the constant polarized tail that is typically considered characteristic of `partial coverage' models.}
\label{MgII_Simulation}
\end{figure}

This extended partial coverage model has significant implications for our understanding of partial coverage. It can only recreate the observed constant portion of a polarized SED as a special case, when the normal depolarizing screen that covers a fraction $1-f_{\text{c}}$ of the source is exactly equal to zero, i.e.\ $\sigma_{\text{norm}}=0$~rad~m$^{-2}$. The polarized SEDs that can occur in this depolarization model are shown in Fig.~\ref{MgII_Simulation}. The SEDs show considerable variation depending on the ratio of $\sigma_{\text{interv}}/\sigma_{\text{norm}}$, and exceptionally high-quality observational data would be required to distinguish between these various scenarios. Importantly, for $f_{\text{c}}>0$ and low ratios of $\sigma_{\text{interv}}/\sigma_{\text{norm}}$, the functional form is indistinguishable from the case where $f_{\text{c}}=0$, i.e.\ a Burn law. Furthermore, even for arbitrarily high ratios of $\sigma_{\text{interv}}/\sigma_{\text{norm}}$, the polarized SED will not exhibit the constant polarized tail that is the crucial foundation for a `partial coverage' model. Consequently, while derivation of $\sigma_{\text{RM}}$ using equation~\ref{partialcoverage} may suitably mathematically parameterize the rate of decay of polarization as a function of $\lambda$, it is unlikely that $\sigma_{\text{RM}}$ describes the RM dispersion of a physical Faraday screen. There is therefore little reason to think that the equivalent to the RM dispersion that is derived from equation \ref{partialcoverage} bears any physical relation to the properties of the depolarizing screen across the source.

Although we have shown that SEDs in the form of equation~\ref{partialcoverage} cannot be related to partial coverage, they are still observed \citep{rosetti08,mantovani08}. We therefore also propose an alternative model to explain their origin. It has previously been found that flat-spectrum radio sources typically maintain a relatively constant polarized fraction as a function of $\lambda$, which has been explained as a consequence of multiple optically-thick emission regions in the quasar core \citep{farnescatalogue} -- such sources can maintain an approximately constant polarized fraction as a function of wavelength \citep[e.g.][]{1967ApJ...150..647P,1973MNRAS.161P..31P}. As an extension of this, the polarized SED of an unresolved, compact source may be the superposition of two components: (i) a strongly polarized and strongly depolarizing optically-thin jets/lobes (with $\alpha\approx-0.7$), and (ii) a weakly polarized and weakly depolarizing optically-thick core (with $\alpha\approx0.0$). Such an SED would have a functional form similar to that presented in the original partial coverage model shown in equation~\ref{partialcoverage}, as it would be dominated by the depolarizing jets/lobes at high frequencies and the weakly depolarizing core at low frequencies. Note that this is the inverse of the typical situation in total intensity, where the steep-spectrum jets/lobes dominate the emission at low frequencies. Such a model is falsifiable, as in all cases the approximately constant polarized tail that extends to low frequencies must have a polarized fraction $\le10$\%, which is the maximum degree of polarization for an optically-thick region \citep[e.g.][]{1970ranp.book.....P}. This is consistent with the catalog of \citet{farnescatalogue}, which finds a maximum value for the polarized tail of 5.1\% from their sample of sources that are classified using a partial coverage SED.

However, if correct, this leads to complications for the typical physical understanding of a polarized fraction (which is the ratio of the polarized and total intensity components). In unresolved sources with an SED of `partial coverage' form, our measurements are biased towards the brightest \emph{polarized} intensity within the resolution element seen in projection on the sky. Meanwhile, the brightest \emph{total} intensity for the same source may not correspond to the same emitting region. As the peak polarized and total intensity both sample different regions of the source, this leads to the possibility that a polarized fraction, $\Pi$, may not be at all related to the degree of magnetic field ordering in these unresolved sources. Consequently, at a given frequency a source may be core-dominated in Stokes $I$ and lobe-dominated in $P$, or vice-versa. High resolution and broadband observations of lines of sight with known intervening objects will be necessary to test our predictions.

We therefore highlight how the partial coverage model of equation~\ref{partialcoverage} is both incompatible with observational evidence, and is also empirical rather than physical -- unlike other depolarization laws \citep[e.g.][]{farnescatalogue}. We have adjusted this mathematical model so that it is physically accurate, thereby including the more realistic case where a Faraday screen other than the one provided by the intervenor is also present. In these cases, depolarization from partial coverage can be indistinguishable from a Burn law (see equation~\ref{burnbabyburn}) at all wavelengths, and never has a constant polarized tail even for extreme ratios of $\sigma_{\text{interv}}/\sigma_{\text{norm}}$.

%==============================================================================%
\newpage
\begin{table}
\section{Main Sample Details}\label{sampledeets}
\end{table}
\tabletypesize{\tinyv}
\begin{deluxetable*}{cccccccccccccccc}
\setlength{\tabcolsep}{0.02in} 
\tablewidth{0pt}
\tablecaption{Details of the main sample, listed in order of NVSS Right Ascension. Sources that are used in the `flat' or `steep' spectrum subsamples are indicated by `F' or `S' respectively. All errors are the one sigma uncertainties. The listed redshifts are for the polarized background radio sources.}
\tablehead{
\colhead{\#}  & \colhead{NVSS R.A.} & \colhead{NVSS Dec.} & \colhead{$\alpha$} & \colhead{$\Delta\alpha$} & \colhead{$\chi^2$} & \colhead{Flat/Steep} & \colhead{$N_{\textrm{MgII}}$} & \colhead{RM} & \colhead{$\Delta$RM} & \colhead{$\Pi$} & \colhead{$\Delta\Pi$} & \colhead{$\beta$} & \colhead{$\Delta\beta$} & \colhead{$z$} & \colhead{$\Delta z$}\\
\colhead{}  & \colhead{J2000 ($^\circ$)} & \colhead{J2000 ($^\circ$)} & \colhead{} & \colhead{} & \colhead{} & \colhead{Sub-Sample} & \colhead{} & \colhead{(rad~m$^{-2}$)} & \colhead{(rad~m$^{-2}$)} & \colhead{(\%)} & \colhead{(\%)} & \colhead{} & \colhead{} & \colhead{} & \colhead{}}
\startdata
\tabletypesize{\tinyv}
1 & 00h03m45.18s & -11d08m20.6 & -1.08 & 0.18 & --- & S & 0 & -8.2 & 9.0 & 2.0 & 0.15 & --- & --- & 1.5666 & 0.0003 \\
2 & 00h10m18.25s & +14d33m36.7 & -1.0 & 0.2 & 0.186 & S & 2 & -25.6 & 11.0 & 2.77 & 0.18 & --- & --- & 0.90107 & 0.00027 \\
3 & 00h13m40.31s & +14d57m32.8 & --- & --- & --- & --- & 0 & -19.3 & 19.0 & 8.5 & 0.8 & --- & --- & 0.8241 & 0.00014 \\
4 & 00h32m59.23s & -00d13m18.2 & --- & --- & --- & --- & 1 & -14.9 & 13.0 & 2.2 & 0.19 & --- & --- & 1.5 & 0.0 \\
5 & 00h35m52.91s & -09d11m50.4 & --- & --- & --- & --- & 0 & 4.2 & 12.0 & 5.8 & 0.4 & --- & --- & 1.0021 & 0.0005 \\
6 & 00h35m55.49s & +15d53m16.7 & -0.88 & 0.18 & 0.339 & S & 0 & -33.7 & 9.0 & 1.51 & 0.13 & --- & --- & 1.16279 & 0.00019 \\
7 & 00h37m58.31s & +24d07m11.6 & -0.64 & 0.05 & 0.198 & --- & 0 & 33.4 & 2.5 & 2.4 & 0.13 & -1.15 & 0.25 & 0.0 & 9e-05 \\
8 & 00h43m23.62s & -00d15m51.1 & -1.12 & 0.18 & --- & S & 0 & 16.7 & 10.0 & 4.22 & 0.25 & --- & --- & 2.7929 & 0.0014 \\
\enddata
\label{tab:deets}
\end{deluxetable*}

\tabletypesize{\tinyv}
\begin{deluxetable*}{cccccccccccccccc}
\setlength{\tabcolsep}{0.02in} 
\tablewidth{0pt}
\tablecaption{Continued from Table~\ref{tab:deets}.}
\tablehead{
\colhead{\#}  & \colhead{NVSS R.A.} & \colhead{NVSS Dec.} & \colhead{$\alpha$} & \colhead{$\Delta\alpha$} & \colhead{$\chi^2$} & \colhead{Flat/Steep} & \colhead{$N_{\textrm{MgII}}$} & \colhead{RM} & \colhead{$\Delta$RM} & \colhead{$\Pi$} & \colhead{$\Delta\Pi$} & \colhead{$\beta$} & \colhead{$\Delta\beta$} & \colhead{$z$} & \colhead{$\Delta z$}\\
\colhead{}  & \colhead{J2000 ($^\circ$)} & \colhead{J2000 ($^\circ$)} & \colhead{} & \colhead{} & \colhead{} & \colhead{Sub-Sample} & \colhead{} & \colhead{(rad~m$^{-2}$)} & \colhead{(rad~m$^{-2}$)} & \colhead{(\%)} & \colhead{(\%)} & \colhead{} & \colhead{} & \colhead{} & \colhead{}}
\startdata
\tabletypesize{\tinyv}
82 & 08h48m56.88s & +08d01m27.1 & -0.84 & 0.17 & 0.061 & S & 0 & 67.0 & 10.0 & 2.76 & 0.18 & --- & --- & 0.9577 & 0.00012 \\
83 & 08h50m51.82s & +15d22m15.3 & -0.74 & 0.15 & 0.601 & S & 1 & 49.9 & 1.1 & 12.0 & 0.4 & --- & --- & 2.0144 & 0.0003 \\
84 & 08h51m28.95s & +60d03m20.2 & -0.71 & 0.14 & 0.410 & S & 0 & -14.3 & 9.0 & 5.9 & 0.3 & --- & --- & 0.54264 & 0.00019 \\
85 & 08h52m00.25s & +02d29m34.1 & -1.2 & 0.3 & --- & S & 0 & -8.2 & 7.0 & 5.8 & 0.29 & --- & --- & 1.1767 & 0.0022 \\
86 & 08h52m05.20s & +28d33m59.7 & 0.36 & 0.07 & 2.910 & F & 0 & -0.4 & 6.0 & 4.48 & 0.21 & --- & --- & 1.2819 & 0.0006 \\
87 & 08h53m02.75s & +20d04m21.6 & -0.57 & 0.11 & 0.608 & --- & 0 & 23.8 & 10.0 & 5.25 & 0.29 & --- & --- & 1.9201 & 0.0005 \\
88 & 08h54m11.17s & -00d19m29.9 & -0.78 & 0.18 & --- & S & 0 & 38.4 & 4.0 & 3.11 & 0.15 & --- & --- & 1.3893 & 0.0006 \\
89 & 08h54m35.08s & +07d20m24.3 & 0.68 & 0.14 & 0.248 & F & 0 & 83.0 & 8.0 & 5.71 & 0.28 & --- & --- & 0.65507 & 0.00019 \\
\enddata
\label{tab:deets2}
\end{deluxetable*}

\tabletypesize{\tinyv}
\begin{deluxetable*}{cccccccccccccccc}
\setlength{\tabcolsep}{0.02in} 
\tablewidth{0pt}
\tablecaption{Continued from Table~\ref{tab:deets}.}
\tablehead{
\colhead{\#}  & \colhead{NVSS R.A.} & \colhead{NVSS Dec.} & \colhead{$\alpha$} & \colhead{$\Delta\alpha$} & \colhead{$\chi^2$} & \colhead{Flat/Steep} & \colhead{$N_{\textrm{MgII}}$} & \colhead{RM} & \colhead{$\Delta$RM} & \colhead{$\Pi$} & \colhead{$\Delta\Pi$} & \colhead{$\beta$} & \colhead{$\Delta\beta$} & \colhead{$z$} & \colhead{$\Delta z$}\\
\colhead{}  & \colhead{J2000 ($^\circ$)} & \colhead{J2000 ($^\circ$)} & \colhead{} & \colhead{} & \colhead{} & \colhead{Sub-Sample} & \colhead{} & \colhead{(rad~m$^{-2}$)} & \colhead{(rad~m$^{-2}$)} & \colhead{(\%)} & \colhead{(\%)} & \colhead{} & \colhead{} & \colhead{} & \colhead{}}
\startdata
\tabletypesize{\tinyv}
163 & 10h07m18.06s & +22d51m27.5 & -0.64 & 0.13 & 0.002 & --- & 1 & 12.2 & 10.0 & 9.8 & 0.6 & --- & --- & 2.3061 & 0.0005 \\
164 & 10h07m41.51s & +13d56m29.3 & -0.085 & 0.017 & 0.001 & F & 0 & 11.9 & 2.3 & 2.85 & 0.14 & --- & --- & 2.71533 & 0.00018 \\
165 & 10h08m41.56s & +36d23m22.7 & -0.61 & 0.12 & 1.204 & --- & 1 & 26.9 & 11.0 & 8.7 & 0.5 & --- & --- & 3.1255 & 0.0008 \\
166 & 10h09m43.05s & +05d29m48.5 & -0.92 & 0.18 & 2.508 & S & 1 & -18.8 & 14.0 & 4.5 & 0.3 & --- & --- & 0.94257 & 0.00014 \\
167 & 10h10m27.29s & +41d32m21.7 & --- & --- & --- & --- & 0 & -2.3 & 0.7 & 5.61 & 0.24 & --- & --- & 0.0 & 2.8e-05 \\
168 & 10h10m27.63s & +41d32m36.6 & --- & --- & --- & --- & 0 & 0.7 & 0.9 & 4.85 & 0.19 & --- & --- & 0.0 & 2.8e-05 \\
169 & 10h11m00.31s & +32d57m18.4 & -0.9 & 0.18 & 1.187 & S & 1 & -7.4 & 17.0 & 3.8 & 0.3 & --- & --- & 0.90034 & 0.00018 \\
170 & 10h11m35.26s & +00d57m47.1 & -0.97 & 0.19 & 2.139 & S & 0 & -1.2 & 13.0 & 3.12 & 0.24 & --- & --- & 1.0717 & 0.0009 \\
\enddata
\label{tab:deets3}
\end{deluxetable*}

\tabletypesize{\tinyv}
\begin{deluxetable*}{cccccccccccccccc}
\setlength{\tabcolsep}{0.02in} 
\tablewidth{0pt}
\tablecaption{Continued from Table~\ref{tab:deets}.}
\tablehead{
\colhead{\#}  & \colhead{NVSS R.A.} & \colhead{NVSS Dec.} & \colhead{$\alpha$} & \colhead{$\Delta\alpha$} & \colhead{$\chi^2$} & \colhead{Flat/Steep} & \colhead{$N_{\textrm{MgII}}$} & \colhead{RM} & \colhead{$\Delta$RM} & \colhead{$\Pi$} & \colhead{$\Delta\Pi$} & \colhead{$\beta$} & \colhead{$\Delta\beta$} & \colhead{$z$} & \colhead{$\Delta z$}\\
\colhead{}  & \colhead{J2000 ($^\circ$)} & \colhead{J2000 ($^\circ$)} & \colhead{} & \colhead{} & \colhead{} & \colhead{Sub-Sample} & \colhead{} & \colhead{(rad~m$^{-2}$)} & \colhead{(rad~m$^{-2}$)} & \colhead{(\%)} & \colhead{(\%)} & \colhead{} & \colhead{} & \colhead{} & \colhead{}}
\startdata
\tabletypesize{\tinyv}
244 & 11h20m23.16s & +54d04m27.5 & -0.37 & 0.07 & 3.835 & --- & 0 & 18.7 & 5.0 & 3.9 & 0.18 & --- & --- & 0.92216 & 0.00018 \\
245 & 11h21m29.82s & +12d36m20.0 & -0.77 & 0.15 & 0.846 & S & 0 & 25.7 & 9.0 & 1.84 & 0.15 & --- & --- & 0.6836 & 0.0013 \\
246 & 11h23m38.15s & +05d20m39.0 & -0.4 & 0.08 & 0.702 & --- & 0 & 4.3 & 8.0 & 6.2 & 0.3 & --- & --- & 2.17932 & 0.00026 \\
247 & 11h25m53.70s & +26d10m20.1 & 0.2 & 0.04 & 0.017 & F & 0 & -11.4 & 2.6 & 2.36 & 0.13 & --- & --- & 2.349 & 0.00014 \\
248 & 11h26m28.23s & +39d18m43.0 & -0.77 & 0.15 & 0.994 & S & 0 & 25.9 & 13.0 & 2.87 & 0.23 & --- & --- & 1.5 & 0.0 \\
249 & 11h26m56.73s & +28d46m13.9 & -0.87 & 0.17 & 0.985 & S & 0 & -12.7 & 5.0 & 4.5 & 0.2 & --- & --- & 1.3978 & 0.0009 \\
250 & 11h26m57.66s & +45d16m07.3 & -0.0092 & 0.0018 & 2.733 & F & 1 & 8.3 & 14.0 & 1.01 & 0.12 & --- & --- & 1.81384 & 0.00026 \\
251 & 11h27m02.46s & +02d31m08.9 & -0.42 & 0.08 & 0.774 & --- & 1 & 8.9 & 4.0 & 6.59 & 0.26 & --- & --- & 1.65366 & 0.00022 \\
\enddata
\label{tab:deets4}
\end{deluxetable*}

\tabletypesize{\tinyv}
\begin{deluxetable*}{cccccccccccccccc}
\setlength{\tabcolsep}{0.02in} 
\tablewidth{0pt}
\tablecaption{Continued from Table~\ref{tab:deets}.}
\tablehead{
\colhead{\#}  & \colhead{NVSS R.A.} & \colhead{NVSS Dec.} & \colhead{$\alpha$} & \colhead{$\Delta\alpha$} & \colhead{$\chi^2$} & \colhead{Flat/Steep} & \colhead{$N_{\textrm{MgII}}$} & \colhead{RM} & \colhead{$\Delta$RM} & \colhead{$\Pi$} & \colhead{$\Delta\Pi$} & \colhead{$\beta$} & \colhead{$\Delta\beta$} & \colhead{$z$} & \colhead{$\Delta z$}\\
\colhead{}  & \colhead{J2000 ($^\circ$)} & \colhead{J2000 ($^\circ$)} & \colhead{} & \colhead{} & \colhead{} & \colhead{Sub-Sample} & \colhead{} & \colhead{(rad~m$^{-2}$)} & \colhead{(rad~m$^{-2}$)} & \colhead{(\%)} & \colhead{(\%)} & \colhead{} & \colhead{} & \colhead{} & \colhead{}}
\startdata
\tabletypesize{\tinyv}
325 & 12h30m53.90s & +39d30m15.4 & -0.78 & 0.17 & --- & S & 1 & 5.4 & 4.0 & 7.7 & 0.3 & --- & --- & 2.2 & 0.0 \\
326 & 12h32m12.68s & +33d55m41.0 & -0.88 & 0.18 & 0.410 & S & 0 & -5.2 & 4.0 & 6.0 & 0.26 & --- & --- & 2.4808 & 0.0004 \\
327 & 12h32m34.35s & +51d41m09.5 & -0.24 & 0.05 & 0.849 & F & 0 & 20.4 & 12.0 & 8.3 & 0.6 & --- & --- & 1.938 & 0.001 \\
328 & 12h32m56.56s & +57d22m14.0 & -0.68 & 0.14 & 0.180 & --- & 1 & -15.3 & 11.0 & 4.34 & 0.29 & --- & --- & 2.1171 & 0.0003 \\
329 & 12h33m54.40s & +48d20m52.1 & -0.53 & 0.11 & 3.825 & --- & 0 & -4.1 & 14.0 & 8.0 & 0.6 & --- & --- & 1.0 & 0.0 \\
330 & 12h34m25.67s & +24d31m44.1 & -0.75 & 0.15 & 0.075 & S & 1 & -10.2 & 7.0 & 1.45 & 0.12 & --- & --- & 1.33867 & 0.00013 \\
331 & 12h34m31.69s & +64d55m55.0 & -0.44 & 0.09 & 3.772 & --- & 1 & 33.6 & 14.0 & 4.5 & 0.3 & --- & --- & 3.0321 & 0.00014 \\
332 & 12h34m53.79s & +67d45m50.1 & -0.34 & 0.07 & 3.811 & --- & 0 & 74.2 & 17.0 & 2.64 & 0.24 & --- & --- & 1.36713 & 0.00022 \\
\enddata
\label{tab:deets5}
\end{deluxetable*}

\tabletypesize{\tinyv}
\begin{deluxetable*}{cccccccccccccccc}
\setlength{\tabcolsep}{0.02in} 
\tablewidth{0pt}
\tablecaption{Continued from Table~\ref{tab:deets}.}
\tablehead{
\colhead{\#}  & \colhead{NVSS R.A.} & \colhead{NVSS Dec.} & \colhead{$\alpha$} & \colhead{$\Delta\alpha$} & \colhead{$\chi^2$} & \colhead{Flat/Steep} & \colhead{$N_{\textrm{MgII}}$} & \colhead{RM} & \colhead{$\Delta$RM} & \colhead{$\Pi$} & \colhead{$\Delta\Pi$} & \colhead{$\beta$} & \colhead{$\Delta\beta$} & \colhead{$z$} & \colhead{$\Delta z$}\\
\colhead{}  & \colhead{J2000 ($^\circ$)} & \colhead{J2000 ($^\circ$)} & \colhead{} & \colhead{} & \colhead{} & \colhead{Sub-Sample} & \colhead{} & \colhead{(rad~m$^{-2}$)} & \colhead{(rad~m$^{-2}$)} & \colhead{(\%)} & \colhead{(\%)} & \colhead{} & \colhead{} & \colhead{} & \colhead{}}
\startdata
\tabletypesize{\tinyv}
406 & 13h47m40.96s & +58d12m42.8 & -0.85 & 0.17 & 0.434 & S & 1 & 2.7 & 6.0 & 0.96 & 0.11 & --- & --- & 2.04448 & 0.00022 \\
407 & 13h47m51.00s & +28d36m24.8 & -0.83 & 0.17 & 2.853 & S & 0 & -5.2 & 5.0 & 5.33 & 0.24 & --- & --- & 0.0 & 5e-05 \\
408 & 13h49m34.72s & +53d41m17.2 & -0.45 & 0.09 & 0.063 & --- & 0 & -13.3 & 7.0 & 0.8 & 0.11 & --- & --- & 0.97861 & 0.00015 \\
409 & 13h50m15.25s & +38d12m06.4 & -0.5 & 0.1 & 3.802 & --- & 1 & -0.5 & 14.0 & 2.91 & 0.23 & --- & --- & 1.3927 & 0.0003 \\
410 & 13h50m32.38s & +38d59m22.6 & -0.5 & 0.1 & 0.499 & --- & 0 & -9.4 & 13.0 & 3.17 & 0.21 & --- & --- & 1.59401 & 0.00023 \\
411 & 13h50m52.71s & +30d34m53.6 & -0.067 & 0.013 & 0.150 & F & 1 & 27.7 & 12.0 & 1.79 & 0.14 & --- & --- & 0.71195 & 0.00018 \\
412 & 13h50m54.16s & +05d21m56.0 & -0.8 & 0.29 & --- & S & 0 & 8.2 & 6.0 & 9.4 & 0.4 & --- & --- & 0.0 & 2.7e-05 \\
413 & 13h51m16.10s & +22d11m10.3 & -0.92 & 0.18 & 0.027 & S & 1 & 19.7 & 13.0 & 4.2 & 0.3 & --- & --- & 1.5745 & 0.0004 \\
\enddata
\label{tab:deets6}
\end{deluxetable*}

\tabletypesize{\tinyv}
\begin{deluxetable*}{cccccccccccccccc}
\setlength{\tabcolsep}{0.02in} 
\tablewidth{0pt}
\tablecaption{Continued from Table~\ref{tab:deets}.}
\tablehead{
\colhead{\#}  & \colhead{NVSS R.A.} & \colhead{NVSS Dec.} & \colhead{$\alpha$} & \colhead{$\Delta\alpha$} & \colhead{$\chi^2$} & \colhead{Flat/Steep} & \colhead{$N_{\textrm{MgII}}$} & \colhead{RM} & \colhead{$\Delta$RM} & \colhead{$\Pi$} & \colhead{$\Delta\Pi$} & \colhead{$\beta$} & \colhead{$\Delta\beta$} & \colhead{$z$} & \colhead{$\Delta z$}\\
\colhead{}  & \colhead{J2000 ($^\circ$)} & \colhead{J2000 ($^\circ$)} & \colhead{} & \colhead{} & \colhead{} & \colhead{Sub-Sample} & \colhead{} & \colhead{(rad~m$^{-2}$)} & \colhead{(rad~m$^{-2}$)} & \colhead{(\%)} & \colhead{(\%)} & \colhead{} & \colhead{} & \colhead{} & \colhead{}}
\startdata
\tabletypesize{\tinyv}
487 & 14h59m45.21s & +33d01m04.5 & -0.77 & 0.15 & 0.048 & S & 0 & 12.3 & 9.0 & 4.84 & 0.27 & --- & --- & 1.0 & 1.0 \\
488 & 15h00m07.18s & +56d36m03.1 & -0.92 & 0.18 & 0.379 & S & 0 & -2.0 & 4.0 & 5.72 & 0.22 & --- & --- & 0.8849 & 0.0003 \\
489 & 15h00m27.08s & +45d09m02.7 & -0.92 & 0.18 & 0.806 & S & 0 & -20.4 & 7.0 & 6.0 & 0.3 & --- & --- & 1.2039 & 0.0004 \\
490 & 15h01m24.56s & +56d19m49.4 & -0.47 & 0.09 & 1.229 & --- & 2 & -5.2 & 5.0 & 5.56 & 0.25 & --- & --- & 1.466 & 0.0003 \\
491 & 15h04m25.03s & +10d29m38.5 & 0.22 & 0.04 & 0.001 & F & 0 & 0.7 & 1.5 & 3.58 & 0.15 & 0.43 & 0.14 & 1.8394 & 0.0006 \\
492 & 15h04m26.71s & +28d54m30.6 & -0.63 & 0.13 & 0.083 & --- & 1 & -1.9 & 4.0 & 2.72 & 0.14 & --- & --- & 2.28243 & 0.00013 \\
493 & 15h04m31.13s & +47d41m49.4 & -0.87 & 0.17 & 0.185 & S & 1 & 0.2 & 5.0 & 5.02 & 0.24 & --- & --- & 0.8238 & 0.0008 \\
494 & 15h05m06.46s & +03d26m30.4 & 0.68 & 0.14 & 1.115 & F & 0 & 25.9 & 8.0 & 2.24 & 0.15 & --- & --- & 0.0 & 9e-05 \\
\enddata
\label{tab:deets7}
\end{deluxetable*}

\tabletypesize{\tinyv}
\begin{deluxetable*}{cccccccccccccccc}
\setlength{\tabcolsep}{0.02in} 
\tablewidth{0pt}
\tablecaption{Continued from Table~\ref{tab:deets}.}
\tablehead{
\colhead{\#}  & \colhead{NVSS R.A.} & \colhead{NVSS Dec.} & \colhead{$\alpha$} & \colhead{$\Delta\alpha$} & \colhead{$\chi^2$} & \colhead{Flat/Steep} & \colhead{$N_{\textrm{MgII}}$} & \colhead{RM} & \colhead{$\Delta$RM} & \colhead{$\Pi$} & \colhead{$\Delta\Pi$} & \colhead{$\beta$} & \colhead{$\Delta\beta$} & \colhead{$z$} & \colhead{$\Delta z$}\\
\colhead{}  & \colhead{J2000 ($^\circ$)} & \colhead{J2000 ($^\circ$)} & \colhead{} & \colhead{} & \colhead{} & \colhead{Sub-Sample} & \colhead{} & \colhead{(rad~m$^{-2}$)} & \colhead{(rad~m$^{-2}$)} & \colhead{(\%)} & \colhead{(\%)} & \colhead{} & \colhead{} & \colhead{} & \colhead{}}
\startdata
\tabletypesize{\tinyv}
568 & 16h31m45.29s & +11d56m03.3 & -0.5 & 0.1 & 0.715 & --- & 1 & 39.9 & 2.0 & 2.37 & 0.12 & -0.3 & 1.1 & 1.7875 & 0.0003 \\
569 & 16h36m16.64s & +17d35m08.3 & -0.76 & 0.15 & 0.003 & S & 1 & 65.5 & 6.0 & 4.77 & 0.21 & --- & --- & 1.9081 & 0.0004 \\
570 & 16h36m38.21s & +21d12m55.5 & 0.041 & 0.008 & 0.143 & F & 2 & 54.0 & 13.0 & 1.15 & 0.13 & --- & --- & 1.80168 & 0.00029 \\
571 & 16h39m49.80s & +24d43m34.5 & -0.92 & 0.18 & 1.028 & S & 3 & 61.4 & 14.0 & 2.55 & 0.21 & --- & --- & 1.5866 & 0.00026 \\
572 & 16h39m56.04s & +47d05m24.0 & -0.5 & 0.1 & 3.809 & --- & 1 & 27.9 & 7.0 & 4.23 & 0.21 & --- & --- & 0.86048 & 0.00021 \\
573 & 16h44m51.10s & +37d30m27.4 & --- & --- & --- & --- & 0 & 23.3 & 13.0 & 4.4 & 0.3 & --- & --- & 0.0 & 6e-05 \\
574 & 16h44m55.26s & +37d29m44.2 & --- & --- & --- & --- & 0 & 11.1 & 14.0 & 6.1 & 0.5 & --- & --- & 0.0 & 6e-05 \\
575 & 16h48m29.29s & +41d04m05.8 & -0.29 & 0.06 & 3.184 & F & 0 & 3.2 & 13.0 & 2.11 & 0.17 & --- & --- & 0.8516 & 0.0001 \\
\enddata
\label{tab:deets8}
\end{deluxetable*}

\clearpage

%==============================================================================%


\begin{thebibliography}{}
\bibitem[Abazajian et al.(2009)]{2009ApJS..182..543A}
Abazajian, K.~N., Adelman--McCarthy, J.~K., Ag\"{u}eros, M.~A., et al., 2009, \apjs, 182, 543. 

\bibitem[Arshakian \& Beck(2011)]{2011MNRAS.418.2336A} 
Arshakian, T.~G., Beck, R., 2011, \mnras, 418, 2336.

\bibitem[Barton \& Cooke(2009)]{2009AJ....138.1817B} 
Barton, E.~J., Cooke, J., 2009, \aj, 138, 1817.

\bibitem[Beck et al.(1996)]{1996ARA&A..34..155B}
Beck, R., Brandenburg, A., Moss, D., Shukurov, A., Sokoloff, D., 1996, \araa, 34, 155. 

\bibitem[Becker et al.(1991)]{1991ApJS...75....1B} Becker, R.~H., White, 
R.~L., \& Edwards, A.~L.\ 1991, \apjs, 75, 1. 

\bibitem[Bernet et al.(2008)]{2008Natur.454..302B} Bernet, M.~L., Miniati, 
F., Lilly, S.~J., Kronberg, P.~P., 
\& Dessauges-Zavadsky, M.\ 2008, \nat, 454, 302. 

\bibitem[Bernet et al.(2010)]{2010ApJ...711..380B} Bernet, M.~L., Miniati, 
F., \& Lilly, S.~J.\ 2010, \apj, 711, 380.

\bibitem[Bernet et al.(2012)]{2012ApJ...761..144B} Bernet, M.~L., Miniati, 
F., \& Lilly, S.~J.\ 2012, \apj, 761, 144.

\bibitem[Bernet et al.(2013)]{2013ApJ...772L..28B} 
Bernet, M.~L., Miniati, F., \& Lilly, S.~J.\ 2013, \apj, 772, 28. 

\bibitem[Bordoloi et al.(2014)]{2014ApJ...784..108B} 
Bordoloi, R., Lilly, S.~J., Kacprzak, G.~G., Churchill, C.~W., 2014, \apj, 784, 108.

\bibitem[Burn(1966)]{1966MNRAS.133...67B} Burn, B.~J.\ 1966, \mnras, 133, 
67. 

\bibitem[Churchill \& Charlton(1999)]{1999AAS...195.5201C} 
Churchill, C., \& Charlton, J., 1999, BAAS, 31, 1451.

\bibitem[Churchill et al.(1999)]{1999ApJS..120...51C} 
Churchill, C.~W., Rigby, J.~R., Charlton, J.~C., Vogt, S.~S., 1999, \apjs, 120, 51.

\bibitem[Churchill et al.(2005)]{churchill05} 
Churchill, C.~W., Kacprzak, G.~G., \& Steidel, C.~C., 2005, in IAU Colloq. 199, `Probing Galaxies through Quasar Absorption Lines, ed. P.~R. Williams, C.~G. Shu, \& B. Menard (Cambridge: Cambridge Univ. Press), 24.

\bibitem[Condon et al.(1998)]{1998AJ....115.1693C} Condon, J.~J., Cotton, 
W.~D., Greisen, E.~W., et al.\ 1998, \aj, 115, 1693. 

\bibitem[Douglas et al.(1996)]{1996AJ....111.1945D} Douglas, J.~N., Bash, 
F.~N., Bozyan, F.~A., Torrence, G.~W., \& Wolfe, C.\ 1996, \aj, 111, 1945. 

\bibitem[Farnes et al.(2014)]{farnescatalogue} 
Farnes, J.~S., Gaensler, B.~M., Carretti, E., 2014, \apjs, 212, 15. 

\bibitem[Gaensler et al.(2004)]{2004NewAR..48.1003G} Gaensler, B.~M., Beck, 
R., \& Feretti, L.\ 2004, New A Rev., 48, 1003. 

\bibitem[Gaensler et al.(2010)]{2010AAS...21547013G} 
Gaensler, B.~M., Landecker, T.~L., Taylor, A.~R., \& POSSUM Collaboration, 2010, BAAS, 42, 515.

\bibitem[Gregory et al.(1996)]{1996ApJS..103..427G} Gregory, P.~C., Scott, 
W.~K., Douglas, K., \& Condon, J.~J.\ 1996, \apjs, 103, 427. 

\bibitem[Hammond et al.(2012)]{2012arXiv1209.1438H} Hammond, A.~M., 
Robishaw, T., \& Gaensler, B.~M.\ 2012, preprint (arXiv:1209.1438v3).

\bibitem[Johnson(2013)]{johnson2013}
Johnson, V.~E., 2013, Proc. Nat. Acad. Sci. 110, 19313.

\bibitem[Jones et al.(2010)]{2010ApJ...715.1497J} Jones, T.~M., Misawa, T., 
Charlton, J.~C., Mshar, A.~C., \& Ferland, G.~J.\ 2010, \apj, 715, 1497. 

\bibitem[Joshi \& Chand(2013)]{2013MNRAS.434.3566J} 
Joshi, R., \& Chand, H.,\ 2013, \mnras, 434, 3566.

\bibitem[Kacprzak et al.(2008)]{kacprzak08} 
Kacprzak, G.~G., Churchill, C.~W., Steidel, C.~C., \& Murphy, M.~T., 2008, \aj, 135, 922.

\bibitem[Klein et 
al.(2003)]{2003A&A...406..579K} Klein, U., Mack, K.-H., Gregorini, L., \& Vigotti, M.\ 2003, \aap, 406, 579. 

\bibitem[Kronberg \& Perry(1982)]{1982ApJ...263..518K}
Kronberg, P.~P., Perry, J.~J., 1982, \apj, 263, 518.

\bibitem[Kronberg et al.(1990)]{1990ApJ...355L..31K}
Kronberg, P.~P., Perry, J.~J., Zukowski, E.~L.~H., 1990, \apj, 355, 31.

\bibitem[Kronberg et al.(2008)]{2008ApJ...676...70K}
Kronberg, P.~P., Bernet, M.~L., Miniati, F., Lilly, S.~J., Short, M.~B., Higdon, D.~M., 2008, \apj, 676, 70.

\bibitem[Kulsrud \& Zweibel(2008)]{2008RPPh...71d6901K} 
Kulsrud, R.~M., Zweibel, E.~G., 2008, Rep. Prog. Phys., 71, 046901.

\bibitem[Longair(2011)]{2011hea..book.....L} 
Longair, M.~S.\ 2011, High Energy Astrophysics, Cambridge, UK: Cambridge University Press, 2011.  

\bibitem[L\'{o}pez(2006)]{2006ApJ...641..710L} 
L\'{o}pez, E.~D., 2006, \apj, 641, 710. 

\bibitem[Mantovani et al.(2009)]{mantovani08} 
Mantovani, F., Mack, K.-H., Montenegro-Montes, F.~M., Rossetti, A., \& Kraus, A.\ 2009, \aap, 502, 61.

\bibitem[Massardi et al.(2013)]{2013arXiv1309.2527M} 
Massardi, M., Burke-Spolaor, S.~G., Murphy, T., et al., 2013, \mnras, 436, 2915.

\bibitem[Mesa et al.(2002)]{2002A&A...396..463M} 
Mesa, D., Baccigalupi, C., De Zotti, G., et al., 2002, \aap, 396, 463.

\bibitem[Murphy et al.(2010)]{2010MNRAS.402.2403M} Murphy, T., Sadler, 
E.~M., Ekers, R.~D., et al.\ 2010, \mnras, 402, 2403.

\bibitem[Narayanan et al.(2007)]{narayanan07} 
Narayanan, A., Misawa, T., Charlton, J.~C., \& Kim, T.~S., 2007, \apj, 660, 1093.

\bibitem[O'Dea(1988)]{1988BAAS...20..971O} 
O'Dea, C.~P., 1988, \baas, 20, 971. 

\bibitem[O'Sullivan et al.(2009)]{osullivanpaper} 
O'Sullivan, S.~P., Gabuzda, D.~C., 2009, \mnras, 393, 429.

\bibitem[O'Sullivan et al.(2012)]{2012MNRAS.421.3300O} 
O'Sullivan, S.~P., Brown, S., Robishaw, T., et al.\ 2012, \mnras, 421, 3300.

\bibitem[Oppermann et al.(2012)]{2012A&A...542A..93O} 
Oppermann, N., Junklewitz, H., Robbers, G., et al.\ 2012, \aap, 542, 93.

\bibitem[Oppermann et al.(2014)]{2014arXiv1404.3701O} 
Oppermann, N., Junklewitz, H., Greiner, M., et al.\ 2014, \aap, submitted (arXiv:1404.3701).

\bibitem[Oren \& Wolfe(1995)]{1995ApJ...445..624O}
Oren, A.~L., Wolfe, A.~M., 1995, \apj, 445, 624.

\bibitem[Pacholczyk \& Swihart(1967)]{1967ApJ...150..647P}
Pacholczyk, A.~G., \& Swihart, T.~L., 1967, \apj, 150, 647.

\bibitem[Pacholczyk(1970)]{1970ranp.book.....P}
Pacholczyk, A.~G., 1970, `Radio astrophysics. Nonthermal processes in galactic and extragalactic sources', San Francisco: Freeman, 1970.

\bibitem[Pacholczyk \& Gregory(1973)]{1973MNRAS.161P..31P}
Pacholczyk, A.~G., \& Gregory, S.~A., 1973, \mnras, 161, 31.

\bibitem[Perry et al.(1993)]{1993ApJ...406..407P}
Perry, J.~J., Watson, A.~M., Kronberg, P.~P., 1993, \apj, 406, 407.

\bibitem[Pshirkov et al.(2014)]{2014arXiv1407.3909P}
Pshirkov, M.~S., Tinyakov, P.~G., Urban, F.~R., 2014, preprint (arXiv:1407.3909).

\bibitem[Rengelink et 
al.(1997)]{1997A&AS..124..259R} Rengelink, R.~B., Tang, Y., de Bruyn, A.~G., et al.\ 1997, \aaps, 124, 259. 

\bibitem[Rigby et al.(1998)]{1998AAS...193.0405R} 
Rigby, J.~R., Churchill, C.~W., Charlton, J.~C., 1998, \baas, 30, 1248.

\bibitem[Rossetti et al.(2008)]{rosetti08} 
Rossetti, A., Dallacasa, D., Fanti, C., Fanti, R., \& Mack, K.-H.\ 2008, \aap, 487, 865.

\bibitem[Saikia et al.(1987)]{1987MNRAS.228..203S} 
Saikia, D.~J., Salter, C.~J., Neff, S.~G., Gower, A.~C., Sinha, R.~P., \& Swarup, G., 1987, MNRAS, 228, 203.

\bibitem[Simard-Normandin et 
al.(1980)]{1980A&AS...40..319S} Simard-Normandin, M., Kronberg, P.~P., \& Neidhoefer, J.\ 1980, \aaps, 40, 319. 

\bibitem[Simard-Normandin et 
al.(1981)]{1981A&AS...43...19S} Simard-Normandin, M., Kronberg, P.~P., \& Neidhoefer, J.\ 1981, \aaps, 43, 19. 

\bibitem[Simard-Normandin et 
al.(1982)]{1982A&AS...48..137S} Simard-Normandin, M., Kronberg, P.~P., \& Button, S.\ 1982, \aaps, 48, 137. 

\bibitem[Sokoloff et al.(1998)]{1998MNRAS.299..189S} Sokoloff, D.~D., 
Bykov, A.~A., Shukurov, A., et al.\ 1998, \mnras, 299, 189. 

\bibitem[Steidel \& Sargent(1992)]{1992ApJS...80....1S} 
Steidel, C.~C., Sargent, W.~L.~W., 1992, \apjs, 80, 1.	

\bibitem[Stil et al.(2014)]{2014ApJ...787...99S} 
Stil, J.~M., Keller, B.~W., George, S.~J., Taylor, A.~R., 2014, \apj, 787, 99. 

\bibitem[Tabara \& Inoue(1980)]{1980A&AS...39..379T} 
Tabara, H., \& Inoue, M.\ 1980, \aaps, 39, 379. 

\bibitem[Taylor et al.(2009)]{2009ApJ...702.1230T} Taylor, A.~R., Stil, 
J.~M., \& Sunstrum, C.\ 2009, \apj, 702, 1230. 

\bibitem[Tingay et al.(2003)]{2003PASJ...55..351T} Tingay, S.~J., Jauncey, 
D.~L., King, E.~A., et al.\ 2003, \pasj, 55, 351. 

\bibitem[Tribble(1991)]{1991MNRAS.250..726T} Tribble, P.~C.\ 1991, \mnras, 
250, 726. 

\bibitem[Tripp et al.(1997)]{1997ApJS..112....1T} 
Tripp, T.~M., Lu, L., Savage, B.~D., 1997, \apjs, 112, 1.

\bibitem[Watson \& Perry(1991)]{1991MNRAS.248...58W}
Watson, A.~M., Perry, J.~J., 1991, \mnras, 248, 58.

\bibitem[Welter et al.(1984)]{1984ApJ...279...19W} 
Welter, G.~L., Perry, J.~J., Kronberg, P.~P., 1984, \apj, 279, 19.

\bibitem[White et al.(1997)]{1997ApJ...475..479W} 
White, R.~L., Becker, R.~H., Helfand, D.~J., Gregg, M.~D., 1997, \apj, 475, 479.

\bibitem[Xu \& Han(2014a)]{xu2} 
Xu, J., \& Han, J.~L., 2014a, MNRAS, 442, 3329.

\bibitem[Xu \& Han(2014b)]{2014arXiv1405.5087X} 
Xu, J., \& Han, J.~L., 2014b, Research in Astronomy and Astrophysics, 14, 942.

\bibitem[Zavala \& Taylor(2004)]{2004ApJ...612..749Z} 
Zavala, R.~T., Taylor, G.~B.,\ 2003, \apj, 612, 749.

\bibitem[Zhu 
\& M{\'e}nard(2013)]{2013ApJ...770..130Z} Zhu, G., \& M{\'e}nard, B.\ 2013, \apj, 770, 130. 

\bibitem[Zukowski et 
al.(1999)]{1999A&AS..135..571Z} Zukowski, E.~L.~H., Kronberg, P.~P., Forkert, T., \& Wielebinski, R.\ 1999, \aaps, 135, 571. 

\end{thebibliography}
\end{document}